\documentclass[10pt]{article}

\usepackage{a4wide,amsmath,amssymb,accents,graphicx,mathrsfs,verbatim,pgf}
\usepackage{authblk,array}

\title{\textbf{Electromagnetism, Local Covariance, the Aharonov-Bohm Effect and
Gauss' Law}}
\author[1]{Ko Sanders\thanks{ko\_sanders@hotmail.com}}
\author[2]{Claudio Dappiaggi\thanks{claudio.dappiaggi@unipv.it}}
\author[3]{Thomas-Paul Hack\thanks{hack@dima.unige.it}}
\affil[1]{Enrico Fermi Institute\\
University of Chicago\\
5640 South Ellis Avenue\\
Chicago, IL 60637 - U.S.A.}
\affil[2]{Dipartimento di Fisica \& INFN, Sezione di Pavia\\
Universit\`a degli Studi di Pavia\\
Via Bassi, 6 - I-27100 Pavia - Italy}
\affil[3]{Dipartimento di Matematica\\
Universit\`a degli Studi di Genova\\
Via Dodecaneso 35 - I-16146 Genova Italy}
\date{revised version 25 March 2014}

\newtheorem{definition}{Definition}[section]
\newtheorem{theorem}[definition]{Theorem}
\newtheorem{proposition}[definition]{Proposition}
\newtheorem{corollary}[definition]{Corollary}
\newtheorem{lemma}[definition]{Lemma}
\newenvironment{proof*}{\smallskip\par\noindent\emph{Proof: }
 \ignorespaces}{\hfill$\Box$\smallskip\par\ignorespaces}
\newtheorem{remark}[definition]{Remark}

\newtheorem{example}[definition]{Example}

\newcommand{\map}[3]{\ensuremath{#1\!:\!#2\!\rightarrow\!#3}}

\begin{document}
\maketitle

\begin{abstract}
We quantise the massless vector potential $A$ of electromagnetism in the presence of a classical
electromagnetic (background) current, $j$, in a generally covariant way on arbitrary globally
hyperbolic spacetimes $M$. By carefully following general principles and procedures we clarify a
number of topological issues. First we combine the interpretation of $A$ as a connection on a
principal $U(1)$-bundle with the perspective of general covariance to deduce a physical gauge
equivalence relation, which is intimately related to the Aharonov-Bohm effect. By Peierls' method
we subsequently find a Poisson bracket on the space of local, affine observables of the theory.
This Poisson bracket is in general degenerate, leading to a quantum theory with non-local
behaviour. We show that this non-local behaviour can be fully explained in terms of Gauss' law.
Thus our analysis establishes a relationship, via the Poisson bracket, between the Aharonov-Bohm
effect and Gauss' law -- a relationship which seems to have gone unnoticed so far. Furthermore,
we find a formula for the space of electric monopole charges in terms of the topology of the
underlying spacetime. Because it costs little extra effort, we emphasise the cohomological
perspective and derive our results for general $p$-form fields $A$ ($p<\mathrm{dim}(M)$), modulo
exact fields, for the Lagrangian density $\mathcal{L}=\frac12 dA\wedge*dA+ A\wedge*j$. In
conclusion we note that the theory is not locally covariant, in the sense of
Brunetti-Fredenhagen-Verch. It is not possible to obtain such a theory by dividing out the centre
of the algebras, nor is it physically desirable to do so. Instead we argue that electromagnetism
forces us to weaken the axioms of the framework of local covariance, because the failure of
locality is physically well-understood and should be accommodated.
\end{abstract}

\section{Introduction}\label{Sec_Introduction}

The number of rigorous studies into a quantised, free electromagnetic field system propagating
in a globally hyperbolic spacetime is fairly small and unfortunately these studies have been
plagued by problems, or limitations, which are related to the topological properties of the
background spacetime. The main goal of this paper is to give a new presentation which overcomes
these shortcomings and which fully clarifies all the topological properties of the theory. In
this introduction we will briefly describe the geometric point of view that will be expounded in
the remainder of our paper and we will indicate the problems that previous investigations
encountered and how they will be overcome.

Historically, electromagnetism was described by a field strength $F$ in Minkowski spacetime,
which is a two-form that contains both the electric field strength $E$ and the magnetic field
strength $B$. The Maxwell equations for $F$ entail that it is closed, $dF=0$, and as the topology
of Minkowski spacetime is trivial we may always write $F=dA$, where $A$ is the so-called vector
potential. Instead of using $F$ as a fundamental object, one may use gauge equivalence classes of
vector potentials $A$, where two vector potentials are identified when they give rise to the same
field strength $F$. This means that they differ by a closed, or, equivalently, an exact, one-form
in Minkowski spacetime. (See \cite{B} for a description of electromagnetism in Minkowski
spacetime from the perspective of algebraic quantum field theory.)

When generalising the theory to more general spacetimes one encounters several topological
obstructions. Firstly, not every closed two-form $F$ is exact, so there may not always be a
vector potential. Secondly, two vector potentials that give rise to the same field strength
differ by a closed one-form, but this one-form may not be exact, so the choice of gauge
equivalence relation to be used becomes relevant. This raises the question which of the three
equivalent formulations of electromagnetism in Minkowski spacetime leads to the correct
generalisation in curved spacetimes: the theory based on $F$, or a theory based on $A$ with
either choice of gauge equivalence.

In this paper we will identify connections on a trivial principal $U(1)$-bundle over a spacetime
$M$ with vector potential one-forms $A$. Using general covariance this naturally leads to a gauge
equivalence relation that identifies vector potentials that differ by an exact
one-form.\footnote{Notice that, for any principal $U(1)$-bundle, the associated bundle of
connections is an affine bundle modeled on the space of one-forms on the base manifold
\cite{BDS2}.} This point of view, which essentially coincides with that taken in the standard
model of elementary particles, has emerged in the course of time and incorporates the well-known
Aharonov-Bohm effect \cite{PT,DW}. In order to treat this effect most clearly, we will include in
our description an electromagnetic current, $j$, which is regarded as a given background
structure. The choice of gauge equivalence that we employ can then be motivated by the following
physical considerations. Using the Aharonov-Bohm effect, which is experimentally established
\cite{PT}, we can distinguish vector potentials that differ by a closed one-form which is not
exact. This means that the field strength $F$ itself does not contain enough information to
account for all physical effects and also that a gauge equivalence on $A$ using closed forms,
rather than exact ones, is too crude.

A few early studies in the quantisation of free electromagnetism focussed on some particular
(curved) spacetimes and used methods that are ill-suited for a generally covariant approach.
\cite{Ashtekar} noticed that in Kruskal spacetime there are many inequivalent Hilbert space
representations for free electromagnetism, which are labeled by magnetic and electric charges,
when the field strength $F$ is taken as the basic object. The existence of inequivalent
representations is a circumstance which is now known to hold even for free scalar fields in
general curved spacetimes and which is treated in the modern literature by separating the
construction of the abstract algebra of observables and its representation. In \cite{AshIsh92}
the usual Weyl quantisation in Minkowski spacetime is compared to an interesting proposal for
quantising the holonomies of the electric field together with the magnetic field. The two
approaches are found to be inequivalent, but the holonomy based approach seems to make
essential use of the choice of a Cauchy surface, which makes it doubtful that the approach
can be made generally covariant. We will follow the direction set out in the more recent
literature, starting with \cite{D}, that uses an algebraic approach based on the Weyl
algebra, because it is the most obvious way forward towards a generally covariant theory.

Some of the recent quantisations of free electromagnetism in curved spacetimes were inadequate
for describing the Aharonov-Bohm effect (\cite{DL,DS} and \cite{Hol} Appendix A): they either
took the field strength as its basic object or they identified two vector potentials that differ
by a closed one-form. Other investigations run into problems in the quantisation procedure.
Although the well-posedness of the classical Maxwell equations was not in doubt (see
e.g.\ \cite{P} for $p$-form fields), \cite{D,FP,P} only carry out the quantisation in spacetimes
with a compact Cauchy surface (and \cite{FP} additionally assumes the triviality of a certain de
Rham cohomology group). The reason appears to be that they want to equip the space of spacelike
compact solutions with a non-degenerate symplectic form. This symplectic form gives rise to a
Poisson space of observables, which is quantised \cite{D,P} using (infinitesimal) Weyl algebras.
\cite{Lang} follows a similar path, but without imposing topological restrictions. Although this
quantisation procedure is successful on any spacetime, it does not behave well under embeddings
(cf.\ Remark \ref{Rem_WrongQ}). Alternatively, \cite{DL,DS} consider general spacetimes and
define a (degenerate) pre-symplectic space, which is quantised directly (see also \cite{FP}).
This can lead to algebras with a non-trivial centre, depending on the topology of the underlying
spacetime, which entails that the theory is not locally covariant in the sense of \cite{BFV}.
Indeed, when a spacetime with non-trivial cohomology is embedded into Minkowski spacetime, this
can lead to algebraic embeddings which vanish on the non-trivial centre. Although the lack of
injectivity was completely characterised in these papers, its interpretation remained to be
understood.

\begin{table}[t]
\centering
\begin{tabular}{l|l|l}
Field theoretic object&Geometric interpretation&Support\\
\hline
\hline
field configurations, modulo gauge&ambient kinematic phase space $\mathcal{F}$&general\\
\hline
Euler-Lagrange solutions, modulo gauge&dynamical phase space
manifold $\mathcal{S}$&general\\
\hline
solutions to linearised equations around $A_0$&tangent space
$T_{A_0}\mathcal{S}$&general\\
\hline
observables for the linearised equation&cotangent space
$T^*_{A_0}\mathcal{S}$&compact\\
\hline
Peierls' bracket&Poisson $2$-vector field on $\mathcal{S}$, i.e.\ an&(n.a.)\\
&antisymmetric bilinear form on $T^*\mathcal{S}$.&
\end{tabular}\caption{The geometric interpretation of (classical) phase space,
using a Lagrangian approach and ignoring issues of infinite dimensional topology.
The compact supports arise from a duality. Note that in systems without gauge
symmetry there is an injection $G:T_{A_0}^*\mathcal{S}\rightarrow T_{A_0}\mathcal{S}$,
whose range consists of spatially compact solutions. One may then interpret the
Peierls' bracket as a (densely defined) symplectic form on $T_{A_0}\mathcal{S}$.
In the presence of gauge symmetries, however, $G$ may fail to be injective (see
Remark \ref{Rem_WrongQ} below) and the quantisation schemes based on symplectic and
Poisson structures are no longer equivalent.}
\label{Tab_Geo}
\end{table}

Our presentation differs\footnote{We wish to point out that \cite{Hol} seems to follow the same
quantisation scheme as we do in its study of quantum Yang-Mills theories. However, this paper
does not compute the centre of the quantum algebra for the $U(1)$ case or investigate its
interpretation in this setting. In fact, it only discusses these issues in its Appendix A,
where an alternative quantisation scheme is used.} by using Peierls' method \cite{Pei,Ha} to
directly find a Poisson structure on the space of observables, bypassing the need for a
symplectic form. This procedure fits in a general geometric framework for Lagrangian field
theories \cite{FR12,K12}, whose most salient aspects are indicated in Table \ref{Tab_Geo}, and
the resulting affine Poisson space may be quantised using ideas from deformation quantisation,
in particular Fedosov's quantisation method (cf.\ \cite{Wa}). Whereas the two approaches are
equivalent for the scalar field, where a non-degenerate symplectic form always exists, this is
no longer the case for electromagnetism, due to the gauge symmetry. In order to obtain an
equivalent formulation in terms of the space of classical spacelike compact solutions, one
would have to modify the gauge equivalence of those solutions in a subtle, but very relevant,
way. (This modification was also noted, but not explained, by \cite{FeH} in the case of
linearised gravity (see also \cite{HS}). For spacetimes with compact Cauchy surfaces the two
approaches are equivalent.)

\begin{figure}
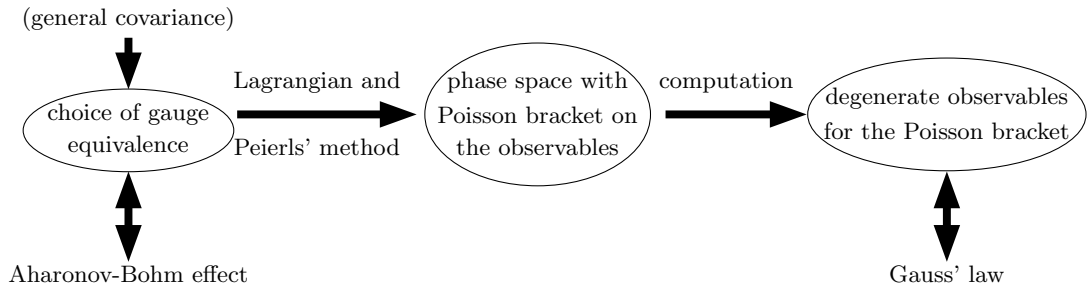

\begin{center}
\begin{pgfpicture}{0cm}{0cm}{14.45cm}{4.25cm}
\color{black}
\pgfputat{\pgfxy(1.7,3.825)}{\pgfbox[center,center]{{\small (general covariance)}}}
\pgfputat{\pgfxy(1.7,0.425)}{\pgfbox[center,center]{{\small Aharonov-Bohm effect}}}
\pgfputat{\pgfxy(1.7,2.55)}{\pgfbox[center,center]{{\small choice of gauge}}}
\pgfputat{\pgfxy(1.7,2.125)}{\pgfbox[center,center]{{\small equivalence}}}
\pgfputat{\pgfxy(7.14,2.975)}{\pgfbox[center,center]{{\small phase space with}}}
\pgfputat{\pgfxy(7.14,2.55)}{\pgfbox[center,center]{{\small Poisson bracket on}}}
\pgfputat{\pgfxy(7.14,2.125)}{\pgfbox[center,center]{{\small the observables}}}
\pgfputat{\pgfxy(12.58,2.7625)}{\pgfbox[center,center]{{\small degenerate observables}}}
\pgfputat{\pgfxy(12.58,2.3375)}{\pgfbox[center,center]{{\small for the Poisson bracket}}}
\pgfputat{\pgfxy(12.58,0.425)}{\pgfbox[center,center]{{\small Gauss' law}}}
\pgfellipse[stroke]{\pgfxy(1.7,2.3375)}{\pgfxy(1.4,0)}{\pgfxy(0,0.55)}
\pgfellipse[stroke]{\pgfxy(7.14,2.55)}{\pgfxy(1.5,0)}{\pgfxy(0,0.95)}
\pgfellipse[stroke]{\pgfxy(12.58,2.55)}{\pgfxy(1.85,0)}{\pgfxy(0,0.75)}
\pgfsetlinewidth{3pt}
\pgfsetendarrow{\pgfarrowtriangle{5pt}}
\pgfxyline(3.145,2.55)(5.27,2.55)
\pgfxyline(8.84,2.55)(10.455,2.55)
\pgfxyline(1.7,3.57)(1.7,3.145)
\pgfsetstartarrow{\pgfarrowtriangle{5pt}}
\pgfxyline(1.7,1.53)(1.7,0.85)
\pgfxyline(12.58,1.53)(12.58,0.85)
\pgfputat{\pgfxy(4.2075,2.975)}{\pgfbox[center,center]{{\small Lagrangian and}}}
\pgfputat{\pgfxy(4.2075,2.125)}{\pgfbox[center,center]{{\small Peierls' method}}}
\pgfputat{\pgfxy(9.6475,2.975)}{\pgfbox[center,center]{{\small computation}}}
\end{pgfpicture}
\end{center}
\caption{A diagrammatic representation of the logical relationship between the Aharonov-Bohm
effect and Gauss' law via the Poisson bracket.}
\label{Fig_ABGauss}
\end{figure}

Carefully computing the Poisson structure by the standard procedure (Peierls' method), we find
a different space of degeneracies than \cite{DL,DS}. Furthermore, we show that there is a
perfectly satisfactory explanation for these degeneracies in the form of Gauss' law. In
particular, the lack of injectivity of algebraic morphisms is only a lack of locality, not of
general covariance, which occurs when observables in a spacetime region $M$ exploit Gauss' law to
measure charges that are located elsewhere in spacetime. (Using classical spacelike compact
solutions without modifying the gauge equivalence, one would not find any degeneracies, but the
theory would not behave well under embeddings.) The logical relationship between the Aharonov-Bohm
effect and Gauss' law that we establish by this procedure is indicated in Figure
\ref{Fig_ABGauss}. In addition to a full clarification of the lack of locality of the quantum
vector potential, our analysis also leads to a formula for the space of electric monopole
charges in terms of the topology of the underlying spacetime. Moreover, with little extra
effort we derive our results for general $p$-form fields $A$, where $p<n$, the spacetime
dimension.

It is not possible to recover locally covariant theories by dividing out the centre of the
algebras, nor is this physically desirable. It is possible to obtain such theories by going to an
off-shell algebra, at the price of losing the dynamics, which is also physically undesirable.
Instead, we argue that one should generalise the axiomatic framework of local covariance, in
order to accommodate the lack of locality of electromagnetism, which is physically
well-understood. What kind of axiomatic restriction should be placed on the (lack of) injectivity
for general spacetime embeddings, if any, remains unclear. For the theories that we consider,
injectivity of morphisms still holds for embeddings of spacetimes with trivial topology (in the
spirit of \cite{FH}). However, a purely topological resolution of this issue seems unlikely,
because other theories (like linearised gravity \cite{FeH}) possess gauge symmetries that are not
related to the spacetime topology alone, but also to the background metric.

In the algebraic approach the construction of the algebras of the theory is only the first step,
which should be followed by a discussion of the class of physical states. This topic, however,
lies outside the scope of our paper, which aims to clarify the topological issues involved in
the classical theory and on preserving them during quantisation. Nevertheless we would like to
remark here that we expect that it should be possible to extend the results of \cite{FP} to
define Hadamard states for our theory on any globally hyperbolic spacetime and to prove the
existence of such states by a deformation argument. Also the construction of Hadamard states
from a bulk-to-boundary correspondence \cite{DS} is expected to remain valid. In addition we
would like to point the interested reader to \cite{FS}, which constructs quasi-free Hadamard
states on a large class of spacetimes with the additional property that a Gupta-Bleuler type
description of the representation remains valid.

We have organised the contents of our paper as follows. In Section \ref{Sec_Classical} we will
describe the (essentially well-known) results on the classical dynamics of the vector potential
and its $p$-form generalisations. The main result is the well-posedness of the initial value
formulation in the presence of a background current, also for distributional field
configurations. In Section \ref{Sec_Poisson} we find the Poisson structure on the classical phase
space, using Peierls' method, and we study its degeneracy, which is related to Gauss' law and the
spacetime topology. Due to the background current the phase space is in general an affine Poisson
space, which will be quantised in Section \ref{Sec_Quantum}. A quantisation of the field strength
can be derived from that of the vector potential. Finally we will show that the theory is not
locally covariant and that the lack of locality may be interpreted in terms of Gauss' law, also
at the quantum level.

\section{Classical Dynamics of $p$-Form Fields}\label{Sec_Classical}

Most of the material that we present here on the classical dynamics of the vector potential and
its $p$-form generalisations is not new, but in view of our later applications it is fitting to
give some results and notations that go beyond standard treatments. This concerns in particular
details on distributional solutions to normally hyperbolic and the Maxwell equations. The
Subsections \ref{SSec_Geo} and \ref{SSec_NHyp} introduce the relevant results and notations from
differential geometry, based on \cite{BT,dR}, and for the Cauchy problem for normally hyperbolic
operators \cite{BGP}. Subsequently we turn to the Cauchy problem for the Maxwell equations, in
Subsection \ref{SSec_Maxwell}, which is a slight generalisation of the work of \cite{P}.

\subsection{Geometric preliminaries}\label{SSec_Geo}

Consider a smooth, $n$-dimensional manifold $\mathcal{M}$, which we assume to be Hausdorff,
connected, oriented and paracompact. We will denote by $\bigwedge^p\mathcal{M}$ the vector
bundle of alternating $p$-linear forms on $T\mathcal{M}$ and the space of their smooth sections,
the $p$-forms on $\mathcal{M}$, will be denoted by $\Omega^p(\mathcal{M})$. The exterior algebra
of differential forms is $\Omega(\mathcal{M})=\oplus_{p=0}^n\Omega^p(\mathcal{M})$, equipped with
the exterior (wedge) product. The exterior derivative
$\map{d}{\Omega(\mathcal{M})}{\Omega(\mathcal{M})}$ maps $p$-forms to $(p+1)$-forms, does not
increase the support and satisfies $d\circ d=0$. A differential form $\alpha$ is called closed
when $d\alpha=0$ and exact when $\alpha=d\beta$ for some differential form $\beta$. The space of
closed $p$-forms will be denoted by $\Omega_d^p(\mathcal{M})$. Corresponding spaces of compactly
supported forms are indicated with a subscript $0$ and we may define the de Rham cohomology
groups of $\mathcal{M}$
\[
H^p(\mathcal{M}):=\frac{\Omega^p_d(\mathcal{M})}{d\Omega^{p-1}(\mathcal{M})}\quad
H^p_0(\mathcal{M}):=\frac{\Omega^p_{0,d}(\mathcal{M})}{d\Omega^{p-1}_0(\mathcal{M})}.
\]

The orientation of $\mathcal{M}$ allows us to define integration as a linear map
$\map{\int_{\mathcal{M}}}{\Omega^n_0}{\mathbb{R}}$ and there is a bilinear map
\[
\map{(\, ,)}{\Omega^p(\mathcal{M})\otimes\Omega^{n-p}_0(\mathcal{M})}{\mathbb{R}}\quad
(\alpha,\beta):=\int_{\mathcal{M}}\alpha\wedge\beta.
\]
By Stokes' Theorem we have $(d\alpha,\beta)=(-1)^{p+1}(\alpha,d\beta)$ if $\alpha$ is a
$p$-form. Moreover, the pairing $(\, ,)$ gives rise to the following isomorphism, known as
Poincar\'e duality:
\[
H^p_0(\mathcal{M})^*\simeq H^{n-p}(\mathcal{M}).
\]
When the de Rham-cohomology groups are finite dimensional we also have
$H^p_0(\mathcal{M})\simeq H^{n-p}(\mathcal{M})^*$.

\begin{example}\label{Ex_Exactness}
It is important to note that for a compactly supported, closed form
$\alpha\in\Omega^p_{0,d}(\mathcal{M})$ the fact that $[\alpha]=0\in H^p_0(\mathcal{M})$ trivially
implies that $[\alpha]=0\in H^p(\mathcal{M})$, but the converse is generally not true. A typical
example in $\mathbb{R}$ is the form $\alpha:=f(r)dr$ with $f\in C_0^{\infty}(\mathbb{R})$. We
always have $\alpha=d\beta$, where $\beta$ is the function $\beta(r):=\int_{-\infty}^rf(s)ds$,
which vanishes in a neighbourhood of $r=-\infty$ and is constant in a neighbourhood of
$r=\infty$. $\beta$ is compactly supported if and only if $\int f=0$.
\end{example}

We denote by $\mathcal{D}^p(\mathcal{M}):=\Omega^{n-p}_0(\mathcal{M})'$ the space of
distribution densities with values in the dual vector bundle of
$\bigwedge^{n-p}\mathcal{M}$.\footnote{In this paper the term distribution is always meant in
the sense of analysis, because the differential geometric notion of distribution (as it occurs
e.g.\ in the formulation of Frobenius' Theorem) will not be needed explicitly. In the literature,
however, distributional sections of $\bigwedge^p\mathcal{M}$ are often called currents, in order
to avoid confusion. The term current was introduced by de Rham (cf.\ \cite{dR}) because, in the
setting of electromagnetism, such objects can be interpreted as electromagnetic currents.
Ironically, in this paper we will mostly consider smooth electromagnetic background currents.}
The pairing $(\, ,)$ can be used to construct a canonical embedding of $\Omega^p(\mathcal{M})$
into $\mathcal{D}^p(\mathcal{M})$, given by $\alpha\mapsto (\alpha,.)$. Differential operators on
distributions and exterior products with smooth forms are to be understood by duality in terms of
the pairing $(\, ,)$. As for smooth differential forms we define a distributional differential
form $\alpha$ to be closed, respectively exact, when $d\alpha=0$, respectively $\alpha=d\beta$.
Compactly supported and closed distribution densities are indicated by the same subscripts as in
the smooth case.

An exact distributional form $\alpha=d\beta\in\mathcal{D}^p(\mathcal{M})$ vanishes on all closed
$\gamma\in\Omega^{n-p}_{0,d}(\mathcal{M})$, because $(\alpha,\gamma)=(-1)^p(\beta,d\gamma)=0$.
That the converse is also true is a result of de Rham (\cite{dR} Sec.\ 22, 23, in particular
Theorem 17'):
\begin{theorem}\label{Thm_dR}
$\alpha\in\mathcal{D}^p_d(\mathcal{M})$ is in $d\mathcal{D}^{p-1}(\mathcal{M})$ if and only if
$(\alpha,\gamma)=0$ for all $\gamma\in\Omega^{n-p}_{0,d}(\mathcal{M})$.
$\alpha\in\mathcal{D}^p_{0,d}(\mathcal{M})$ is in $d\mathcal{D}^{p-1}_0(\mathcal{M})$ if and only
if $(\alpha,\gamma)=0$ for all $\gamma\in\Omega^p_d(\mathcal{M})$.
\end{theorem}
Consequently, the cohomology groups for distributional $p$-forms, which are defined by
\[
(H')^p(\mathcal{M}):=\frac{\mathcal{D}^p_d(\mathcal{M})}{d\mathcal{D}^{p-1}(\mathcal{M})}\quad
(H')^p_0(\mathcal{M}):=\frac{\mathcal{D}^p_{0,d}(\mathcal{M})}{d\mathcal{D}^{p-1}_0(\mathcal{M})}
\]
can be identified with those for smooth $p$-forms as follows (cf.\ \cite{dR} Theorem 14):
\[
(H')^p(\mathcal{M})\simeq H^{n-p}_0(\mathcal{M})^*\simeq H^p(\mathcal{M}),\quad
(H')^p_0(\mathcal{M})\simeq H^p_0(\mathcal{M}).
\]

As a further piece of notation we consider a smoothly embedded, oriented submanifold
$\Sigma\subset\mathcal{M}$, so that the vector bundles $\bigwedge^p\mathcal{M}$ restrict to
vector bundles $\bigwedge^p\mathcal{M}|_{\Sigma}$. In general these restricted bundles cannot be
canonically identified with $\bigwedge^p\Sigma$, except for $p=0$, as can be seen by considering
their dimensions. With some abuse of notation we will write
$\Omega^p(\mathcal{M})|_{\Sigma}$ for smooth sections of $\bigwedge^p\mathcal{M}|_{\Sigma}$ over
$\Sigma$, and similarly for the case of compactly supported sections and distribution densities
on $\Sigma$. Note that the restriction of a density from $\mathcal{M}$ to $\Sigma$ incurs an
additional factor, when compared to ordinary sections, due to the change in volume form.

By a spacetime $M=(\mathcal{M},g)$ we mean an $n$-dimensional manifold $\mathcal{M}$ as above
with $n\ge 2$, endowed with a smooth pseudo-Riemannian metric $g$ of signature $+-\ldots-$. We
will assume that $M$ is globally hyperbolic, which means by definition that it admits a Cauchy
surface. The latter is a subset which is intersected exactly once by each inextendible timelike
curve. A globally hyperbolic spacetime can be foliated by smooth, spacelike Cauchy surfaces
\cite{BS}. In the remainder of our paper we will only consider Cauchy surfaces which are
spacelike and smooth.

It will occasionally be useful to consider forms whose support properties are related to the
Lorentzian geometry of a spacetime $M$ as follows \cite{S,Fri,BGP,FeH}:
\begin{eqnarray}
\Omega_{sc}^p(M)&=&\left\{\alpha\in\Omega^p(M)|\
\mathrm{supp}(\alpha)\subset J(K)\mathrm{\ for\ some\ compact\ } K\subset M
\right\}\nonumber\\
\Omega_{tc}^p(M)&=&\left\{\alpha\in\Omega^p(M)|\
\mathrm{supp}(\alpha)\subset J^+(\Sigma^-)\cap J^-(\Sigma^+)
\mathrm{\ for\ two\ Cauchy\ surfaces\ } \Sigma^{\pm}\subset M\right\}.\nonumber
\end{eqnarray}
The subscripts ``sc'' and ``tc'' stand for spacelike compact and timelike compact,
respectively. We note that $\Omega_{sc}^p(M)\cap\Omega_{tc}^p(M)=\Omega^p_0(M)$, by global
hyperbolicity. Distribution densities with timelike, resp.\ spacelike, compact support will be
denoted similarly by $\mathcal{D}^p_{tc}(M)$, resp.\ $\mathcal{D}^p_{sc}(M)$.

In terms of local coordinates and an arbitrary (local) derivative operator $\nabla_a$, the
differential geometric calculus given above can be expressed as follows. A $p$-form $\alpha$
corresponds to a fully anti-symmetric tensor $\alpha_{a_1\cdots a_p}$. We have
$dx^{a_1}\wedge\cdots\wedge dx^{a_p}=p!dx^{[a_1}\otimes\cdots\otimes dx^{a_p]}$, where the
square brackets denote antisymmetrisation as an idempotent operator. The exterior product is
given by
 $(\alpha\wedge\beta)_{a_1\cdots a_{p+q}}=\frac{(p+q)!}{p!q!}\alpha_{[a_1\cdots a_p}
\beta_{a_{p+1}\cdots a_{p+q}]}$ and the exterior derivative takes the form
$(d\alpha)_{a_0\cdots a_p}=(p+1)\nabla_{[a_0}\alpha_{a_1\cdots a_p]}$. The metric volume form
is given by
$(d\mathrm{vol}_g)_{a_1\cdots a_n}=\sqrt{|\det(g_{\mu\nu})|}\epsilon_{a_1\cdots a_n}$,
with the Levi-Civita tensor satisfying $\epsilon_{1\cdots n}=1$.

The metric volume form allows us to define a fibre-preserving linear involution
$\map{*}{\bigwedge^p(M)}{\bigwedge^{n-p}(M)}$, called the Hodge dual. In local coordinates we have
\[
(*\alpha)_{a_{p+1}\cdots a_n}=\frac{\sqrt{|\det(g_{\mu\nu})|}}{p!}
\epsilon_{a_1\cdots a_n}\alpha^{a_1\cdots a_p},
\]
which gives rise to a support preserving map $\map{*}{\Omega^p(M)}{\Omega^{n-p}(M)}$. For any
embedding $\Sigma\subset M$ the Hodge dual $*$ restricts to a map on the restricted vector bundle
$\bigwedge M|_{\Sigma}$ which has the same pointwise properties. In addition, if $\Sigma$ is not
null, we can consider the Hodge dual in the induced metric, which we denote by $*_{\Sigma}$. For
$\alpha,\beta\in\bigwedge^p_x\mathcal{M}$ and $\gamma\in\bigwedge^{n-p}_x\mathcal{M}$ we have the
following identities:
\[
*\!*\alpha=(-1)^{n-1+p(n-p)}\alpha\quad
*\alpha\wedge*\gamma=(-1)^{n-1}\alpha\wedge\gamma\quad
*\alpha\wedge\beta=(-1)^{p(n-p)}\alpha\wedge*\beta
\]
\[
\alpha\wedge*\beta=\beta\wedge*\alpha=\frac{1}{p!}\alpha^{a_1\cdots a_p}\beta_{a_1\cdots a_p}
d\mathrm{vol}_g.
\]
From this it is easy to see that the Hodge dual can be extended to an operation on distributions,
by duality.

The exterior co-derivative is defined by $\delta:=(-1)^p*^{-1}d*$, when acting on $p$-forms. It
defines a linear map $\map{\delta}{\Omega^p(M)}{\Omega^{p-1}(M)}$ which does not increase the
support and in local coordinates it takes the form
\[
(\delta\alpha)_{a_2\cdots a_p}=-\nabla^{a_1}\alpha_{a_1\cdots a_p}.
\]
A differential form $\alpha$ is called co-closed when $\delta\alpha=0$ and co-exact when
$\alpha=\delta\beta$ for some differential form $\beta$. The space of co-closed $p$-forms will be
denoted by $\Omega^p_{\delta}(M)$ and similarly for the distributional and compactly supported
case. Because $\delta\circ\delta=0$ one can also define cohomology groups, but these are easily
seen to be isomorphic to the de Rham cohomology groups, by Hodge duality.
For $\alpha,\beta\in\Omega^p(M)$ and $\gamma\in\Omega^{n-p+1}_0(M)$ we note that
\[
(\alpha,*\beta)=(\beta,*\alpha)\qquad
(\delta\alpha,\gamma)=(-1)^p(\alpha,\delta\gamma)\qquad
(\delta\alpha,*\beta)=-(\alpha,*d\beta),
\]
where the last equality is valid when the supports of $\alpha$ and $\beta$ have a compact intersection.

\subsection{The Cauchy problem for normally hyperbolic operators}\label{SSec_NHyp}

In preparation for the initial value formulation of the Maxwell equations, we first review the
Cauchy problem for a normally hyperbolic operator $P$ acting on sections of a vector bundle
$\mathcal{V}$ over $M$ \cite{BGP}. The example that is of prime importance in this paper is the
de Rham-Laplace-Beltrami operator $\Box:=d\delta+\delta d$, whose action on $A\in\Omega^p(M)$ is
given in local coordinates by (\cite{dR} Sec.\ 26)
\begin{equation}\label{Eqn_BoxLoc}
(\Box A)_{a_1\cdots a_p}=-\nabla^b\nabla_bA_{a_1\cdots a_p}
+pR_{[a_1}^{\phantom{[a_1}b}A_{|b|a_2\cdots a_p]}
-\left(\begin{array}{c}p\\ 2\end{array}\right)
R_{[a_1a_2}^{\phantom{[a_1a_2}bc}A_{|bc|a_3\cdots a_p]}.
\end{equation}
Note that $\Box$ is indeed a normally hyperbolic operator on $\Omega^p(M)$ for any
$p\in\left\{0,\ldots ,n\right\}$ (cf.\ \cite{BGP}).

In general we denote the smooth sections of $\mathcal{V}$ over $M$ by $\Gamma(\mathcal{V})$ and
the distribution densities with values in the dual bundle $\mathcal{V}^*$ by
$\Gamma'(\mathcal{V}^*):=\Gamma_0(\mathcal{V})'$, where the subscript indicates a compact support,
as usual. The duality
\[
(\alpha,A):=\int_M\alpha(A)d\mathrm{vol}_g,\qquad
\alpha\in\Gamma_0(\mathcal{V}^*), A\in\Gamma(\mathcal{V})
\]
allows us to identify $\Gamma(\mathcal{V})$ with a subspace of $\Gamma'(\mathcal{V})$. $P$
has a formally adjoint operator $P^*$ on $\mathcal{V}^*$, which satisfies
$(\alpha,PA)=(P^*\alpha,A)$ and which is also normally hyperbolic. Given $P$ there is a uniquely
associated $P$-compatible connection on $\mathcal{V}$, which we denote by $\nabla_a$
(cf.\ \cite{BGP} Lemma 1.5.5).

If we denote the restriction of the bundle to a Cauchy surface $\Sigma$ by
$\mathcal{V}|_{\Sigma}$, then the main result on the Cauchy problem can be formulated as follows:
\begin{theorem}\label{Thm_WaveCauchy}
Given $j\in\Gamma(\mathcal{V})$ and $A_0,A_1\in\Gamma(\mathcal{V}|_{\Sigma})$, where
$\Sigma\subset M$ is a Cauchy surface with future pointing unit normal vector field $n^a$, there
is a unique $A\in\Gamma(\mathcal{V})$ such that
\[
PA=j,\quad A|_{\Sigma}=A_0,\quad n^a\nabla_aA|_{\Sigma}=A_1,
\]
where $\nabla_a$ is the $P$-compatible connection. $A$ depends continuously on the data
$(j,A_0,A_1)$ (in the usual Fr\'echet topology of smooth sections). Furthermore, $A$ is supported
in $J(K)$ with $K:=\mathrm{supp}(j)\cup\mathrm{supp}(A_0)\cup\mathrm{supp}(A_1)$.
\end{theorem}
For the case of compactly supported data $(j,A_0,A_1)$ (and the test-section topology on
$\Gamma_0(\mathcal{V})$ and $\Gamma_0(\mathcal{V}|_{\Sigma})$) a proof can be found in
\cite{BGP}, Theorems 3.2.11 and 3.2.12. For data with general supports, the proof of existence,
uniqueness and the support property is directly analogous to that of the scalar case, which is
given in \cite{G} Corollary 5. The continuity for general data follows from the compactly
supported case, in light of the support properties of the solution.

Below we will show that the regularity of the initial data is not essential, if one also allows
distributional solutions. First, however, we will establish some useful results concerning
fundamental solutions for $P$. Using Theorem \ref{Thm_WaveCauchy} one can prove the existence of
unique advanced ($-$) and retarded ($+$) fundamental solutions $G^{\pm}$ for $P$. These are
defined as distributional sections of the bundle $\mathcal{V}\boxtimes\mathcal{V}^*$ over
$M^{\times 2}$ and by their support properties they naturally define continuous linear maps
\cite{BGP,S}
\[
G^{\pm}:\Gamma_0(\mathcal{V})\rightarrow\Gamma_{sc}(\mathcal{V}),\quad
G^{\pm}:\Gamma_{tc}(\mathcal{V})\rightarrow\Gamma(\mathcal{V}).
\]
If we let $(G^{\pm})^*$ denote the advanced and retarded fundamental solutions for $P^*$ we find
from a formal partial integration that
\begin{equation}\label{Eqn_Gpmdual}
((G^{\pm})^*\alpha,A)=(\alpha,G^{\mp}A),\quad \alpha\in\Gamma_0(\mathcal{V}^*),\
A\in\Gamma_0(\mathcal{V}),
\end{equation}
because the supports of $(G^{\pm})^*\alpha$ and $G^{\mp}A$ have compact intersection
(cf.\ \cite{BGP} Lemma 3.4.4). This equality remains true when either $\alpha$ or $A$ only has
timelike compact support.

The fundamental solutions can be used to find (distributional) solutions
$A\in\Gamma'(\mathcal{V})$ to the wave equation $PA=0$, simply by setting $A:=G\beta$ with
$\beta\in\Gamma'_{tc}(\mathcal{V})$ and $G:=G^--G^+$ and exploiting the duality to define the action
of $G$ on distribution densities. If $\beta$ is smooth, then so is $A$. We will see below that all
solutions are of this form and that for $G\beta\in\Gamma'_{sc}(\mathcal{V})$ we may choose $\beta$
to be compactly supported. Moreover, $G$ can be used to give a useful expression of a general
solution $A$ in terms of its initial data, as the next lemma shows.
\begin{lemma}\label{Lem_Asmeared}
If $A\in\Gamma'(\mathcal{V})$ satisfies $PA=j\in\Gamma(\mathcal{V})$ and
$\alpha\in\Gamma_0(\mathcal{V}^*)$, then
\[
(\alpha,A)=\sum_{\pm}\int_{J^{\pm}(\Sigma)}((G^{\mp})^*\alpha)(j)d\mathrm{vol}_g+
\int_{\Sigma} (G^*\alpha)_1(A_0)-(G^*\alpha)_0(A_1),
\]
where $\Sigma$ is a Cauchy surface with future pointing unit normal vector field $n^a$,
$G^*:=(G^-)^*-(G^+)^*$, $A_0:=A|_{\Sigma}$, $A_1:=n^a\nabla_aA|_{\Sigma}$,
$(G^*\alpha)_0:=G^*\alpha|_{\Sigma}$, $(G^*\alpha)_1:=n^a\nabla^*_aG^*\alpha|_{\Sigma}$, and
$\nabla_a$, resp.\ $\nabla^*_a$, are the $P$-compatible, resp.\ $P^*$-compatible, connections.
\end{lemma}
\begin{proof*}
In the smooth case this follows from Stokes' Theorem by a well-known computation:
\begin{eqnarray}
(\alpha,A)&=&\sum_{\pm}\int_{J^{\pm}(\Sigma)}
(P^*(G^{\mp})^*\alpha)(A)d\mathrm{vol}_g\nonumber\\
&=&\sum_{\pm}\int_{J^{\pm}(\Sigma)} ((G^{\mp})^*\alpha)(j)
-\nabla^a((\nabla_a^*(G^{\mp})^*\alpha)(A)
-((G^{\mp})^*\alpha)(\nabla_aA))d\mathrm{vol}_g\nonumber\\
&=&\sum_{\pm}\int_{J^{\pm}(\Sigma)} ((G^{\mp})^*\alpha)(j)d\mathrm{vol}_g+
\sum_{\pm}\int_{\Sigma}\pm n^a((\nabla_a^*(G^{\mp})^*\alpha)(A)
-((G^{\mp})^*\alpha)(\nabla_aA))d\mathrm{vol}_h\nonumber\\
&=&\sum_{\pm}\int_{J^{\pm}(\Sigma)} ((G^{\mp})^*\alpha)(j)d\mathrm{vol}_g+
\int_{\Sigma}(n^a\nabla_a^*G^*\alpha)(A)-(G^*\alpha)(n^a\nabla_aA))d\mathrm{vol}_h.\nonumber
\end{eqnarray}
For the distributional case we note that the initial data of $A$ are well-defined, because $PA=j$
is smooth, so $WF(A)$ only contains light-like vectors.\footnote{For a definition of the wave
front set $WF(A)$ and background material on microlocal analysis we refer to \cite{SV,H}.}
The computation above can then be performed in an analogous fashion, by multiplication with the
characteristic functions of the sets $J^{\pm}(\Sigma)$.
\end{proof*}

We are now ready to consider the well-posedness of the Cauchy problem in the distributional case.
\begin{theorem}\label{Thm_WaveCauchy'}
Given $j\in\Gamma(\mathcal{V})$ and $A_0,A_1\in\Gamma'(\mathcal{V}|_{\Sigma})$, there exists a
unique $A\in\Gamma'(\mathcal{V})$ such that
\[
PA=j,\quad A|_{\Sigma}=A_0,\quad n^a\nabla_aA|_{\Sigma}=A_1
\]
and $A$ depends continuously on the data $(j,A_0,A_1)$ (in the distributional topology on the
$A_i$). Furthermore, $A$ is supported in $J(K)$ with
$K:=\mathrm{supp}(j)\cup\mathrm{supp}(A_0)\cup\mathrm{supp}(A_1)$.
\end{theorem}
\begin{proof*}
The expression in Lemma \ref{Lem_Asmeared} proves that the solution $A$ is uniquely determined
in terms of the data $(j,A_0,A_1)$. Moreover, it proves the existence of a solution $A$, because
the right-hand side of that expression depends continuously and linearly on $\alpha$, as the maps
$\alpha\mapsto (G^{\pm})^*\alpha$ and the subsequent restriction to initial data are linear and
continuous. Furthermore, if we define $A$ by the right-hand side, then $A$ solves the desired
equation (as $\alpha=P^*\beta$ has $(G^{\pm})^*\alpha=\beta$) and one may check that it reproduces
the given initial data (by approximating the $A_i$ by smooth data). Finally we note that $A$
depends continuously on the data $(j,A_0,A_1)$, which again follows immediately from the
expression in Lemma \ref{Lem_Asmeared}, and that its support property follows from those of
$(G^{\pm})^*$.
\end{proof*}

If $j\in\Gamma'(\mathcal{V})$ with $WF(j)\cap N^*\Sigma=\emptyset$, where $N^*\Sigma$ is the
conormal bundle of $\Sigma\subset M$ (cf.\ \cite{H}), then the proof of existence and uniqueness
still works. To prove the continuous dependence of $A$ on $j$, however, one presumably
needs to impose some H\"ormander topology on $j$ near $\Sigma$, so that the products
$\chi^{\pm}j$ with the characteristic functions $\chi^{\pm}$ of $J^{\pm}(\Sigma)$ depend
continuously on $j$. In any case, in the remainder of our paper we will only be concerned with
smooth $j$, so that the wave front set of the solution $A$ only contains light-like covectors
and its initial data are well-defined on all spacelike Cauchy surfaces.

The fundamental solutions $G^{\pm}$ are also useful to characterise the freedom involved in
writing a solution $A$ of the homogeneous wave equation in the form $G\beta$
(cf.\ \cite{D} Prop.4):
\begin{proposition}\label{Prop_SCSoln}
In the notation of Theorem \ref{Thm_WaveCauchy'}, $A$ has spacelike compact support if and only
if all of $j,A_0,A_1$ have spacelike compact support. When $j=0$, $A=G\alpha$ for some
$\alpha\in\Gamma'_{tc}(\mathcal{V})$. If $A$ has spacelike compact support we can choose
$\alpha\in\Gamma'_0(\mathcal{V})$ and if $A$ is smooth, then $\alpha$ can be chosen smooth also.
Finally, if $G\alpha=0$ with $\alpha\in\Gamma'_{tc}(\mathcal{V})$, then $\alpha=P\beta$ for some
$\beta\in\Gamma'_{tc}(\mathcal{V})$, $\beta$ has compact support if and only if $\alpha$ does and
$\beta$ can be chosen smooth if and only if $\alpha$ is smooth.
\end{proposition}
\begin{proof*}
Suppose $A$ has spacelike compact support, $\mathrm{supp}(A)\subset J(K)$ with compact
$K\subset M$. The initial data on any Cauchy surface $\Sigma$ are compactly supported, because
$J(K)\cap\Sigma$ is compact, while the support of $j=\Box A$ is contained in that of $A$, so $j$
also has spacelike compact support. For the converse of the first claim we first consider
compactly supported $j$, in which case the result follows directly from Theorem
\ref{Thm_WaveCauchy'}. By linearity it then only remains to consider the case of vanishing
initial data on $\Sigma$ and, moreover, $j\equiv 0$ on a neighbourhood of $\Sigma$. Let
$K\subset M$ be a compact set such that $J(K)$ contains the support of $j$. $J(K)$ has a compact
intersection $L$ with $\Sigma$ and we note that
$\mathrm{supp}(j)\cap J^{\pm}(\Sigma)\subset J^{\pm}(L)$. Now let
$\phi_n\in\Omega^0_0(I^+(\Sigma))$ be a partition of unity of $I^+(\Sigma)$, with
$n\in\mathbb{N}$, and let $j_n:=\phi_nj$. We may then consider the solutions
$A_N:=\sum_{n=0}^NG^+j_n$ of $PA_N=\sum_{n=0}^Nj_n$, which have vanishing initial data on $\Sigma$
and their support is contained in $J^+(L)$. The
limit $A^+:=\lim_{N\rightarrow\infty}A_N$ is well-defined, because for every Cauchy surface
$\Sigma'$ the set $J^-(\Sigma')\cap J^+(\Sigma)\cap\mathrm{supp}(j)$ is compact in $I^+(\Sigma)$,
so for $M,N$ sufficiently large we have $A_N=A_M$ on $J^-(\Sigma')$. Moreover,
$\mathrm{supp}(A^+)\subset J^+(L)$, because this is true for all $N$. Constructing a solution
$A^-$ of $PA^-=j$ on $I^-(\Sigma)$ in a similar way we find $A:=A^++A^-$ with spacelike compact
support satisfying $PA=j$ on all of $M$ and with vanishing initial data. This completes the proof
of the first statement.

When $j=0$ it is clear that $G\alpha$ is a solution (with spacelike compact support, when $\alpha$
is compactly supported), by the properties of $G^{\pm}$. Conversely, given initial data
$A_i\in\Gamma'(\mathcal{V}|_{\Sigma})$ we may define $\beta\in\Gamma'_{tc}(\mathcal{V})$ by
\[
(\beta,\eta):=-\int_{\Sigma}A_0(n^b\nabla_b\eta)-A_1(\eta),\quad \eta\in\Gamma_0(\mathcal{V}).
\]
Because the identity (\ref{Eqn_Gpmdual}) can be extended to the case where one of the sections
is a distribution, we see that for any $D\in\Gamma_0(\mathcal{V})$
\[
(G\beta,\eta)=-(\beta,G\eta)=(A,\eta),
\]
where we used Lemma \ref{Lem_Asmeared} for the final equality. Therefore $A=G\beta$. Now let
$\chi\in\Omega^0(M)$ be identically $1$ to the future of some Cauchy surface $\Sigma^+$ and
identically $0$ to the past of some Cauchy surface $\Sigma^-$. We let
$\alpha:=-P\chi A\in\Gamma'_{tc}(\mathcal{V})$ and note that
$\alpha=\nabla^a((\nabla_a\chi)A)-(\nabla^a\chi)\nabla_aA$. The compact support,
resp.\ smoothness, of $\alpha$ follow from spacelike compact support, resp.\ smoothness, of $A$.
Because $\chi G^-\beta$ and $(1-\chi)G^+\beta$ are compactly supported too we have
\[
G^{\pm}\alpha=-G^{\pm}P(\chi G^-\beta)+G^{\pm}PG^+\beta-G^{\pm}P((1-\chi)G^+\beta)
=-\chi G^-\beta+G^{\pm}\beta-(1-\chi)G^+\beta
\]
and hence $G\alpha=G\beta=A$. This proves the second statement.

Finally, if $G\alpha=0$ for $\alpha\in\Gamma'_{tc}(\mathcal{V})$, then
$\beta:=G^-\alpha=G^+\alpha$ has timelike compact support and $\alpha=P\beta$. Moreover, $\beta$
is smooth, resp.\ compactly supported, if and only if $\alpha$ is smooth, resp.\ compactly
supported. This completes the proof.
\end{proof*}

\begin{remark}\label{Rem_Homeo}
The solution map
$\map{\mathcal{S}}{\Gamma'(\mathcal{V}|_{\Sigma})^{\oplus 2}}{\Gamma'(\mathcal{V})}$ is not only
continuous (by Theorem \ref{Thm_WaveCauchy'}), but also a homeomorphism onto its range. To see
why the inverse is continuous, fix two test-sections
$\alpha_0,\alpha_1\in\Gamma_0(\mathcal{V}^*|_{\Sigma})$. Using Theorem \ref{Thm_WaveCauchy'} and
Proposition \ref{Prop_SCSoln}, applied to $P^*$, we find a test-section
$\alpha\in\Gamma_0(\mathcal{V}^*)$ such that $G^*\alpha$ has initial data $\alpha_0,\alpha_1$.
Because of Lemma \ref{Lem_Asmeared} this means that the convergence of solutions $A$ on $M$
implies the convergence of their initial data.
\end{remark}

To conclude this subsection we return to the special case of the de Rham-Laplace-Beltrami
operator $\Box$. One easily verifies that
\[
d\Box=\Box d,\quad \delta\Box=\Box\delta,\quad *\Box=\Box*
\]
and $(\alpha,*\Box\beta)=(\Box\alpha,*\beta)$ when either $\alpha$ or $\beta$ has compact support.
We now let $G^{\pm}$ denote the advanced and retarded fundamental solutions for $\Box$ on
$\Omega(M)$ and we note that their restrictions to any $\Omega^p(M)$ are the corresponding
fundamental solutions for the restriction of $\Box$ to $\Omega^p(M)$. It follows that (\cite{P}
Prop.\ 2.1)
\[
dG^{\pm}=G^{\pm}d,\quad \delta G^{\pm}=G^{\pm}\delta,\quad
dG=Gd,\quad \delta G=G\delta,
\]
as may easily be checked by noticing that for any source $\alpha\in\Omega^p_{tc}(M)$ the
solutions $\beta=dG^{\pm}\alpha-G^{\pm}d\alpha$ and
$\beta=\delta G^{\pm}\alpha-G^{\pm}\delta\alpha$ of $\Box\beta=0$ vanish to the past or future of
some Cauchy surface and hence $\beta=0$.

\begin{corollary}\label{Cor_Exactness}
Let $\alpha\in\mathcal{D}^p_{tc}(M)$.
\begin{enumerate}
\item $dG\alpha=0$ if and only if $d\alpha=d\delta\beta$ for some
$\beta\in\mathcal{D}^{p+1}_{tc,d}(M)$. If, in addition, $\delta\alpha=0$, then
$\alpha=\delta\beta$.
\item $\delta G\alpha=0$ if and only if $\delta\alpha=\delta d\beta$ for some
$\beta\in\mathcal{D}^{p-1}_{tc,\delta}(M)$. If, in addition, $d\alpha=0$, then $\alpha=d\beta$.
\end{enumerate}
In both cases $\beta$ can be chosen smooth, resp.\ compactly supported, whenever $\alpha$ is
smooth, resp.\ compactly supported.
\end{corollary}
\begin{proof*}
We first note that if $0=dG\alpha=Gd\alpha$, then $d\alpha=\Box\beta$ for some
$\beta\in\mathcal{D}^{p+1}_{tc}(M)$, by Proposition \ref{Prop_SCSoln}. $\beta$ can be chosen
smooth, respectively compactly supported, whenever $\alpha$ is smooth, respectively compactly
supported, by the same proposition. Note that $0=dd\alpha=d\Box\beta=\Box d\beta$, so $d\beta=0$,
because it has timelike compact support. Thus we have $d\alpha=d\delta\beta$. Conversely, if
$d\alpha=d\delta\beta$ and $d\beta=0$, then $dG\alpha=Gd\delta\beta=-G\delta d\beta=0$. Now, if
in addition $\delta\alpha=0$, then $\Box(\alpha-\delta\beta)=\delta d(\alpha-\delta\beta)=0$, so
$\alpha=\delta\beta$ by the timelike compact support of $\alpha-\delta\beta$. The proof for the
case $\delta G\alpha=0$ is completely analogous.
\end{proof*}

\subsection{The Maxwell equations for $p$-form fields modulo gauge equivalence}\label{SSec_Maxwell}

We now turn to the Cauchy problem for the Maxwell equations. In Paragraph \ref{SSSec_Geo} we set
the scene by considering the geometric setting for electromagnetism and we discuss the Lagrangian
formulation for $p$-form fields. Paragraph \ref{SSSec_Data} establishes a parametrisation for the
initial data of $p$-form fields that is suitable for solving the Maxwell equations
\cite{P,Lang}. This leads to computations which are somewhat involved, because $[A]$ is most
easily described in terms of differential geometric notation, whereas the initial data are most
naturally described in terms of tensor calculus. Finally, in Paragraph \ref{SSSec_Maxwell}, we
discuss the Cauchy problem for the Maxwell equations.

\subsubsection{The geometric setting of the vector potential}\label{SSSec_Geo}

Let us then consider the physical situation of electromagnetism. A classical vector potential, in
the most general setting, is a principal connection on a principal $U(1)$-bundle $P$ over $M$
(see \cite{N} Ch.\ 10.1 or \cite{BDS} for more details). This $U(1)$-bundle arises as the
structure (gauge) group of matter fields that carry electric charge, but we will not need an
explicit description of these matter fields. We may identify the connection $A$ with a one-form
$A\in\Omega^1(M)$. (This one-form is not canonical. The space of all connections is an affine
space modeled over $\Omega^1(M)$, cf.\ \cite{BDS,BDS2}.)
A gauge transformation on $P$ can then be described by a $U(1)$-valued function $\lambda$ on $M$,
which changes the connection one-form $A$ into $A':=A-i\lambda^{-1}d\lambda$.\footnote{Here the
term $\lambda^{-1}d\lambda$ is to be interpreted in adapted local coordinates, viewing $U(1)$ as
a subset of $\mathbb{C}$. The gauge transformations are exactly all fibre bundle automorphisms of
$P$ covering the identity which preserve the Lagrangian (\ref{Eqn_Lagrangian}) of the theory,
assuming that the matter fields that give rise to $j$ also transform appropriately.}
We will denote the space of all $U(1)$-valued functions on $M$ by $\mathcal{G}(M)$. In particular
we may choose $\lambda=e^{i\chi}$ for any $\chi\in \Omega^0(M)$, so that $A':=A+d\chi$. This
means that $A$ is gauge equivalent to $A'$ whenever $A-A'\in d\Omega^0(M)$, but the converse is
not necessarily true, because not all $U(1)$-valued functions are necessarily of the exponential
form $e^{i\chi}$.

A generally covariant perspective brings to light a problem that indicates that the space
$\mathcal{G}(M)$ is too large to act as the physical gauge group. To exemplify this we consider
an embedding $\map{\psi}{M}{\tilde{M}}$ of two spacetimes together with two connection one-forms
$\tilde{A},\tilde{A}'$ on $\tilde{M}$ and their pull-backs $A:=\psi^*(\tilde{A})$,
$A':=\psi^*(\tilde{A}')$ to $M$. Now suppose that $A$ and $A'$ are gauge equivalent, which means
that we cannot distinguish between $A$ and $A'$ by performing measurements in $M$. Based on
general covariance one would expect that it follows that $\tilde{A}$ and $\tilde{A}'$ cannot be
distinguished by any measurements in $\psi(M)$. In other words, given a
$\lambda\in\mathcal{G}(M)$ one expects that there is a $\tilde{\lambda}\in\mathcal{G}(\tilde{M})$
such that $\psi^*(\tilde{\lambda}^{-1}d\tilde{\lambda})=\lambda^{-1}d\lambda^{-1}$. However, in
Example \ref{Ex_ABeffect} below, where we describe the Aharonov-Bohm effect, we will see
explicitly that this is not always true. This problem can even occur when $\psi$ is causal, so no
classical information from the spacelike complement $\psi(M)^{\perp}$ in $\tilde{M}$ should
influence the physical description in $\psi(M)$.

To resolve this issue we take the perspective of general covariance and show how it motivates us
to modify the gauge equivalence. In analogy to \cite{BFV} we introduce the following two
categories:
\begin{definition}\label{Def_Spac}
\begin{itemize}
\item $\mathfrak{Spac}$ is the category whose objects are globally hyperbolic spacetimes
$M=(\mathcal{M},g)$ and whose morphisms are orientation and time orientation preserving
embeddings $\map{\psi}{M}{\tilde{M}}$ such that $\psi^*\tilde{g}=g$ and
$\psi(M)\subset\tilde{M}$ is causally convex
(i.e.\ $\psi^*(\tilde{J}^{\pm}(\psi(p)))=J^{\pm}(p)$ for all $p\in M$).
\item $\mathfrak{Grp}$ is the category whose objects are groups and whose morphisms are
group homomorphisms.
\end{itemize}
\end{definition}
There is a functor $\map{\mathfrak{G}}{\mathfrak{Spac}}{\mathfrak{Grp}}$ such that
$\mathfrak{G}(M)=\mathcal{G}(M)$ and $\mathfrak{G}(\psi)=\psi^*$ is the pull-back. We will endow
the space $\mathcal{G}(M)$ with the topology of uniform convergence of all derivatives on all
compact sets of $M$ (cf.\ \cite{Hi}).
\begin{theorem}\label{Thm_PhysGauge}
There exists a unique functor $\map{\mathfrak{G}_0}{\mathfrak{Spac}}{\mathfrak{Grp}}$ such that
\begin{enumerate}
\item $\mathfrak{G}_0(M)\subset\mathcal{G}(M)$,
\item for any morphisms $\map{\psi}{M}{\tilde{M}}$ in $\mathfrak{Spac}$,
$\mathfrak{G}_0(\psi)=\mathfrak{G}(\psi)|_{\mathfrak{G}_0(\tilde{M})}=
\psi^*|_{\mathfrak{G}_0(\tilde{M})}$
has a dense range (in the relative topology induced by the $\mathcal{G}(M)$),
\item for any functor $\mathfrak{G}_0'$ satisfying the first two properties we have
$\mathfrak{G}_0'(M)\subset\mathfrak{G}_0(M)$.
\end{enumerate}
If the spacetime dimension $n\ge 3$, then
\[
\mathfrak{G}_0(M)=\left\{e^{i\chi}|\ \chi\in\Omega^0(M)\right\}.
\]
\end{theorem}
$\mathfrak{G}_0(M)$ is the largest subgroup of $\mathcal{G}(M)$ which avoids the problem
indicated above, up to a topological closure. For this reason we make the following definition:
\begin{definition}\label{Def_PhysGauge}
We call $\mathcal{G}_0(M):=\mathfrak{G}_0(M)$ the \emph{physical gauge group}.
\end{definition}
\begin{proof*}
Let $\mathscr{F}$ be the set of functors satisfying the first two conditions. $\mathscr{F}$ is
not empty, because it contains the trivial functor with $M\mapsto \left\{e\right\}$ where $e$ is
the identity element of $\mathcal{G}(M)$. Now define
\[
\mathfrak{G}_0(M):=\overline{\left\{\gamma_1\cdots\gamma_j|\ \gamma_i\in\mathfrak{F}_i(M), \mathfrak{F}_i\in\mathscr{F}\right\}}.
\]
Because $\mathcal{G}(M)$ is a commutative group, $\mathfrak{G}_0(M)$ is a subgroup. Because the
product in $\mathcal{G}(M)$ is jointly continuous one may verify directly that
$\mathfrak{G}_0\in\mathscr{F}$. Furthermore, for any $\mathfrak{F}\in\mathscr{F}$ we have
$\mathfrak{F}(M)\subset\mathfrak{G}_0(M)$ for all $M$, by construction. This maximality property
also entails the uniqueness of $\mathfrak{G}_0$.

Now consider the functor $\mathfrak{F}_0$ with
\[
\mathfrak{F}_0(M):=\left\{e^{i\chi}|\ \chi\in\Omega^0(M)\right\}
\]
and $\mathfrak{F}_0(\psi)=\psi^*$. Any morphism $\map{\psi}{M}{\tilde{M}}$ is an embedding, so
the push-forward $\map{\psi_*}{\Omega^0_0(M)}{\Omega^0_0(\tilde{M})}$ is well-defined. By
considering $\tilde{\lambda}=e^{i\psi_*\chi}$ with $\chi\in\Omega^0_0(M)$ we see that
$\psi^*(\tilde{\lambda})=e^{i\chi}$, so $\psi^*(\mathfrak{F}_0(\tilde{M}))$ contains $e^{i\chi}$
for all $\chi\in\Omega^0_0(M)$. This is already a dense set in $\mathfrak{F}_0(M)$, so
$\mathfrak{F}_0\in\mathscr{F}$.

Let $\lambda\in\mathcal{G}(M)$ be arbitrary. Locally $\lambda$ is always of exponential form,
$\lambda=e^{i\chi}$, where $\chi$ is unique up an additive constant in $2\pi\mathbb{Z}$. To see
if $\lambda$ is globally of exponential form we fix a base point $x_0\in M$ and a
$\chi_0\in\mathbb{R}$ such that $\lambda(x_0)=e^{i\chi_0}$. For each $x_1\in M$ we can find a
smooth curve $\map{\gamma}{[0,1]}{M}$ starting at $x_0$ and ending at $x_1$, because $M$ is
arcwise connected. For each such curve there is a unique smooth function $\xi_{\gamma}$ on
$\gamma$ such that $\lambda=e^{i\xi_{\gamma}}$ on $\gamma$ and $\xi_{\gamma}(x_0)=\chi_0$.
For $x_1=\gamma(1)$ we may try to define $\chi(x_1):=\xi_{\gamma}(\gamma(1))$ and the only
question is whether this is independent of the choice of $\gamma$. In other words, $\lambda$ is
globally of exponential form if and only if for each loop $\map{\gamma}{[0,1]}{M}$ starting and
ending at $x_0\in M$ we have $\xi_{\gamma}(\gamma(1))=\xi_{\gamma}(\gamma(0))$. (Using suitable
approximations in contractible neighbourhoods of the end points, the loop may always be chosen
smooth.) Note that this condition is invariant under homotopy, so the condition is equivalent to
the vanishing of all holonomies (cf.\ \cite{N,KoNo}). Also note that the holonomy along a curve
$\gamma$ depends continuously on $\lambda$.

If $M$ is simply connected, then $\mathcal{G}(M)=\mathfrak{F}_0(M)$ and hence
$\mathfrak{G}_0(M)=\mathfrak{F}_0(M)$. To prove this equality for general $M$ (and $n\ge 3$) we
proceed in several small steps. First we suppose that for some $M$ there is a
$\lambda\in\mathfrak{G}_0(M)$ which has a non-zero holonomy $a$ along a loop $\gamma$. We may
pick an arbitrary (smooth, spacelike) Cauchy surface $\Sigma_0\subset M$ and foliate
$M=(\mathcal{M},g)$ by Cauchy surfaces, such that there is a diffeomorphism
$\map{\psi}{\mathcal{M}}{\mathbb{R}\times\Sigma}$ for which the projection $T$ onto the first
coordinate yields a global time function $t=\psi^*T$ with $\Sigma_0=t^{-1}(0)$ \cite{BS2}. We
then write $\gamma(s)=(t(s),\rho(s))$ and we construct a homotopy $H$ between $\gamma$ and the
curve $\gamma_0(s):=(0,\rho(s))$, simply by setting $H(\tau,s):=((1-\tau)t(s),\rho(s))$. As
holonomies of $\lambda$ are homotopy invariant, we see that the holonomy along $\gamma_0$ is again
$a\not=0$. Thus we see that it suffices to consider loops in an arbitrary Cauchy surface of $M$.

As a second step we consider a morphism $\map{\psi}{M}{\tilde{M}}$. If there exists a
$\lambda\in\mathcal{G}_0(M)$ which has a non-zero holonomy $a$ along a loop $\gamma$ in $M$ and
$\epsilon>0$, then by assumption on the functor $\mathfrak{G}_0$ there exists a
$\tilde{\lambda}\in\mathcal{G}_0(\tilde{M})$ which has a holonomy along $\psi_*(\gamma)$ in
$(a-\epsilon,a+\epsilon)$. Choosing $\epsilon$ small enough we can arrange for this holonomy to
be non-zero, so the existence of non-zero holonomies for $\mathfrak{G}_0$ in $M$ implies the
existence of non-zero holonomies for $\mathfrak{G}_0$ in $\tilde{M}$. When the range of $\psi(M)$
contains a Cauchy surface for $\tilde{M}$, the converse is also true by the previous paragraph.
Using the functorial properties of $\mathfrak{G}_0$ and a spacetime deformation argument
\cite{F,FNW} we may then conclude that the existence of non-zero holonomies for $\mathfrak{G}_0$
in $M$ is equivalent to the existence of non-zero holonomies for $\mathfrak{G}_0$ in any
spacetime $\tilde{M}$ diffeomorphic to $M$. In particular we may choose $\tilde{M}$ to be
ultrastatic, by endowing $\Sigma$ with a complete Riemannian metric $h$ and setting
$\tilde{M}=(\mathbb{R}\times\Sigma,-dt^2+h)$. (Note that such a spacetime is always globally
hyperbolic \cite{Sanc}.)

For the third step we consider an embedding $\gamma$ of $\mathbb{S}^1$ into the Cauchy surface
$\Sigma_0=t^{-1}(0)$ of an ultrastatic spacetime $M=(\mathbb{R}\times\Sigma,-dt^2+h)$, where $t$
is the Killing time coordinate. With a slight abuse of notation we will denote the range of the
embedding again by $\gamma$. We may choose a tubular neighbourhood $V$ of $\gamma$ \cite{Hi}
Theorem 4.5.2, i.e.\ a vector bundle $V$ over $\mathbb{S}^1$ with an embedding $\map{\tau}{V}{M}$
such that $\tau(V)$ is an open neighbourhood of $\gamma$ in $M$ and the restriction of $\tau$ to
the zero section of $V$ coincides with the embedding $\gamma$. By construction the tubular
neighbourhood is diffeomorphic to the normal bundle $N\gamma$ of $\gamma$ in $\Sigma$, where we
use the Riemannian metric on $\Sigma$ to identify $N\gamma$ as a subbundle of
$T\Sigma|_{\gamma}$. Consider the short exact sequence of vector bundles
\[
0\rightarrow T\gamma\rightarrow T\Sigma|_{\gamma}\rightarrow N\gamma\rightarrow 0
\]
and note that both $T\gamma$ and $T\Sigma|_{\gamma}$ are orientable vector bundles, because both
$\mathbb{S}^1$ and $\Sigma$ are orientable. It follows from \cite{Hi} Lemma 4.4.1 that
$V\simeq N\gamma$ is also an orientable vector bundle. Now consider another embedding
$\tilde{\gamma}$ of $\mathbb{S}^1$ into $\tilde{\Sigma}:=\mathbb{R}^{n-1}$, viewed as a Euclidean
space. Such an embedding exists when $n\ge 3$. We may again choose a tubular neighbourhood
$\tilde{V}$ of $\tilde{\gamma}$, which is an orientable vector bundle by the same argument as for
$V$. Moreover, we may ensure that the range of $\tilde{\tau}$ is bounded. By \cite{Hi} Section
4.4 Exercise 2 there is a (vector bundle) isomorphism $\map{\psi_V}{V}{\tilde{V}}$, because $V$
and $\tilde{V}$ are both orientable and they are of the same dimension. This means that there is
a diffeomorphism $\psi:=\tilde{\tau}\circ\psi_V\circ\tau^{-1}$ between the tubular neighbourhoods
$\tau(V)$ of $\gamma$ in $\Sigma$ and $\tilde{\tau}(\tilde{V})$ in $\tilde{\Sigma}$.

By using a partition of unity on $\Sigma$ subordinate to the cover $\left\{\tilde{\tau}(\tilde{V}),\tilde{\Sigma}\setminus\tilde{\gamma}\right\}$ we may construct
a complete Riemannian metric $\tilde{h}$ on $\tilde{\Sigma}$ which coincides with $\psi_*h$ on a
neighbourhood $\tilde{U}$ of $\tilde{\gamma}$. (Here we use the fact that the range of
$\tilde{\tau}$ is bounded, so we may recover the usual Euclidean metric outside a bounded set and
thus ensure completeness of $\tilde{h}$.) We let $\tilde{M}$ be the ultrastatic spacetime
$\tilde{M}=(\mathbb{R}\times\tilde{\Sigma},-d\tilde{t}^2+\tilde{h})$. Note that there is an
isometric diffeomorphism $\map{\psi^{-1}}{\tilde{U}}{U}$ onto some neighbourhood $U\subset\Sigma$
of $\gamma$. We can extend this to an isometric diffeomorphism $\Psi$ of $D(\tilde{U})$ onto
$D(U)\subset M$ by setting $\Psi(\tilde{t},\psi(p)):=(t,p)$, where $t$ and $\tilde{t}$ are the
Killing time coordinates on $D(U)$ and $D(\tilde{U})$, which vanish on $U$ and $\tilde{U}$,
respectively. (Note that the range of $t$ with $(t,p)\in D(U)$ is exactly equal to the range of
$\tilde{t}$ with $(\tilde{t},\psi(p))\in D(\tilde{U})$.)

Because $\tilde{M}$ is simply connected, there is no $\tilde{\lambda}\in\mathcal{G}_0(\tilde{M})$
with a non-zero holonomy along $\tilde{\gamma}$. Hence the same is true for the subspacetime
$D(\tilde{U})\subset\tilde{M}$. Because $\gamma=\psi^*\circ\tilde{\gamma}$ we see that there
cannot be any $\lambda\in\mathcal{G}_0(D(U))$ with a non-zero holonomy. Moreover, as the loop
$\gamma$ was an arbitrary embedding into $\Sigma$, we see that there can be no non-zero
holonomies along any embedding $\map{\gamma}{\mathbb{S}^1}{\Sigma}$.

To complete the proof we note that for $n\ge4$, any loop $\gamma$ into $\Sigma$ can be
approximated arbitrarily closely by an embedding (\cite{Hi} Theorem 2.2.13), so there are no
non-zero holonomies. For $n=3$, $\gamma$ can be approximated by an immersion with clean double
points, i.e.\ when $\gamma(s_0)=\gamma(s_1)$ and $s_0\not=s_1$, then there are disjoint open
neighbourhoods $U_i\subset\mathbb{S}^1$ of $s_i$ such that the restrictions $\gamma|_{U_i}$ are
embeddings whose ranges are in general position (\cite{Hi} Theorem 2.2.12 and Exercise 1 of
Section 3.2). Note that for any $s_0\in\mathbb{S}^1$ there are at most finitely many points
$s_1,\ldots,s_k\in\mathbb{S}^1$ with $\gamma(s_i)=\gamma(s_0)$, because $\mathbb{S}^1$ is
compact. It follows that there is an open neighbourhood $U_0$ of $s_0$ such that $\gamma(U_1)$
contains at most one double point. Using compactness of $\mathbb{S}^1$ again there are at most
finitely many double points in the range of $\gamma$. We may now partition $\gamma$ into a finite
number of piecewise smooth loops $\gamma_j$ in $\Sigma$ without double points. The corners of
the $\gamma_j$ can be smoothed out within a contractible neighbourhood, without changing its
holonomy, so we may take the $\gamma_j$ to be embeddings. As before, all holonomies along the
$\gamma_j$ must now vanish. The holonomy of any $\lambda$ along $\gamma$ is the sum of the
holonomies along the $\gamma_i$, so it too must vanish. This completes the proof.
\end{proof*}

\begin{remark}\label{Rem_PhysGauge}
\begin{enumerate}
\item For $n=2$ one may show that $\mathfrak{G}_0=\mathfrak{G}$. Instead of giving this case
the separate treatment that it deserves, we will prefer to consider the ''unphysical'' gauge
group $\mathfrak{F}_0$ consisting of exponential type gauge transformations. This makes our
arguments more convenient, as it is in line with the higher dimensional case. Note that
$\mathfrak{F}_0(M)\not=\mathfrak{G}_0(M)$ if and only if $M$ has the topology of a cylinder
$\mathbb{R}\times\mathbb{S}^1$.
\item A remark on the general geometric situation is in order (see \cite{N} Ch.\ 10.1 or
\cite{BDS} for more details). In the presence of non-trivial principal $U(1)$-bundles $P$
the analog of Theorem \ref{Thm_PhysGauge} is less clear, because a morphism
$\map{\psi}{M}{\tilde{M}}$ may not necessarily admit a fibre bundle morphism
$\map{\Psi}{P}{\tilde{P}}$ covering $\psi$. We will ignore this issue for the time being,
because it is unclear whether it has any physical relevance. In fact, if the tangent bundle
$TM$ of $M$ is isomorphic to a trivial bundle, one may choose to describe spinors using the
Clifford algebra bundle over $TM$. In this way one may argue that, at least for
electrodynamics, all physically relevant bundles are trivial. Even though it is not
entirely clear whether this assumption holds for all four-dimensional globally hyperbolic
spacetimes,\footnote{The results of Geroch \cite{Ge70,Ge68} only hold for spatially compact
spacetimes. We are grateful to an anonymous referee for pointing this out to us.} we
should also note that even a non-trivial principle $U(1)$-bundle $P$ still has a trivial
adjoint bundle (cf.\ e.g.\ \cite{BDS2}). Because $A$ takes values in this adjoint bundle,
any physical effects would have to be very subtle.
\item The identification of the affine space of connections with sections of
$A\in\Omega^1(M)$ is not unique, as it uses a reference connection (cf.\ \cite{BDS}). Note,
however, that by considering the Maxwell equations with a source term, we will already
automatically end up with an affine Poisson space. A more proper treatment of the affine
space of connections will be given in \cite{BDS2}.
\end{enumerate}
\end{remark}

For general $p$-form fields we will consider the kinematic space of field configurations
\[
\mathcal{F}^p(M):=\mathcal{D}^p(M)/d\mathcal{D}^{p-1}(M),
\]
consisting of gauge equivalence classes of $p$-forms. For $p=1$ and $n\ge 3$ this is in
line with Theorem \ref{Thm_PhysGauge}. By Theorem \ref{Thm_dR}, the denominator is a
closed subspace (in the distributional topology), so we can endow $\mathcal{F}^p(M)$ with
the quotient topology, making it a Hausdorff locally convex topological vector space. The
space of continuous linear maps is then simply $\mathcal{F}^p(M)^*=\Omega^p_{0,\delta}(M)$,
under the duality $(.,*.)$ (cf.\ \cite{KN} 14.5).

We consider the dynamics for $[A]\in\mathcal{F}^p(M)$, $p<n$, against the background of a fixed
metric $g$ and electromagnetic current density $j\in\Omega^p(M)$. The equations of motion are
derived from the Lagrangian density
\begin{equation}\label{Eqn_Lagrangian}
\mathcal{L}:=\frac12 F\wedge *F+A\wedge *j,
\end{equation}
where $F:=dA$. The Euler-Lagrange equations are the Maxwell equations:
\begin{equation}\label{Eqn_Maxwell}
\delta dA=j.
\end{equation}
Note that this equation is well-defined for gauge equivalence classes, because $[A]=0$ entails
$dA=0$.

For $p=1$ the relation between equation (\ref{Eqn_Maxwell}) and the usual form of the Maxwell
equations can be seen by noting that $dF=0$ and $\delta F=j$ and writing out these equations in
terms of local Gaussian coordinates near a Cauchy surface $\Sigma$. We will do this in some
detail in the next paragraph, where we consider a suitable parametrisation of the initial data.

\subsubsection{Initial data for $p$-forms}\label{SSSec_Data}

If $\Sigma$ is a Cauchy surface with future pointing unit normal vector field $n^a$, we may
extend $n^a$ to a neighbourhood of $\Sigma$ by defining it as the coordinate vector field of a
Gaussian normal coordinate. The extended vector field satisfies
\begin{equation}\label{Eqn_na}
n^an_a\equiv 1,\quad n^a\nabla_an^b\equiv 0,\quad \nabla_{[a}n_{b]}=0,
\end{equation}
where the last equation can be derived using Frobenius' Theorem (e.g.\ \cite{W} Theorem B.3.2).
We let $P_a^{\phantom{a}b}:=\delta_a^{\phantom{a}b}-n_an^b$. On $\Sigma$,
$P_a^{\phantom{a}b}|_{\Sigma}$ is just the pointwise orthogonal projection of $TM|_{\Sigma}$ onto
$T\Sigma$.

Throughout this paragraph we will assume that $A\in\mathcal{D}^p(M)$ satisfies
$WF(A)\cap N^*\Sigma=\emptyset$, so that $A$ has well-defined initial data on $\Sigma$. We may
decompose these data as follows:
\[
A_0=a+n\wedge\phi,\quad A_1=\dot{a}+n\wedge\dot{\phi},
\]
where $n$ is viewed as a one-form $n_a$ and we introduced the tangential and normal components
of $A_i$, defined by
\begin{eqnarray}\label{Eqn_Def_aphi}
a_{a_1\cdots a_p}&:=&P_{a_1}^{\phantom{a_1}b_1}\cdots P_{a_p}^{\phantom{a_p}b_p}
A_{b_1\cdots b_p}|_{\Sigma}\nonumber\\
\dot{a}_{a_1\cdots a_p}&:=&n^a\nabla_a P_{a_1}^{\phantom{a_1}b_1}\cdots P_{a_p}^{\phantom{a_p}b_p}
A_{b_1\cdots b_p}|_{\Sigma}\nonumber\\
\phi_{a_2\cdots a_p}&:=&n^{a_1}A_{a_1\cdots a_p}|_{\Sigma}\nonumber\\
\dot{\phi}_{a_2\cdots a_p}&:=&n^b\nabla_bn^{a_1}A_{a_1\cdots a_p}|_{\Sigma}.\nonumber
\end{eqnarray}
Note that, by the properties of $n^a$, the normal derivative commutes with the contraction with
$n^a$ and with the projections $P_a^{\phantom{a}b}$. For $p=1$ we may interpret $a$ as the spatial
vector potential, $\phi$ as the scalar potential, and $\dot{\phi}$, $\dot{a}$ as their normal
derivatives. In further analogy to the $p=1$ case we may consider the field strength
$F=dA\in\mathcal{D}^{p+1}(M)$ (which for $p=1$ is the curvature of the connection,
$F=dA-A\wedge A=dA$). Decomposing this in a similar way yields
\[
F|_{\Sigma}=B+n\wedge E,
\]
where
\begin{eqnarray}\label{Eqn_Def_EB}
E_{a_1\cdots a_p}&:=&n^{a_0}(dA)_{a_0\cdots a_p}|_{\Sigma}\nonumber\\
B_{a_0\cdots a_p}&:=&P_{a_0}^{\phantom{a_0}b_0}\cdots P_{a_p}^{\phantom{a_p}b_p}
(dA)_{b_0\cdots b_p}|_{\Sigma}.\nonumber
\end{eqnarray}
The expression for $B$ entails that
$B=\iota^*_{\Sigma}(dA)=d^{\Sigma}\iota^*_{\Sigma}A=d^{\Sigma}a$, because the exterior derivation
commutes with the pull-back under the canonical embedding $\map{\iota_{\Sigma}}{\Sigma}{M}$.

In order to reparametrise the initial data in more differential geometric terms we need the
following
\begin{lemma}\label{Lem_EdeltaA}
If $A\in\mathcal{D}^p(M)$ satisfies $WF(A)\cap N^*\Sigma=\emptyset$ on a Cauchy surface
$\Sigma\subset M$ and its initial data are given by $(a,\dot{a},\phi,\dot{\phi})$, then
\begin{eqnarray}
E_{a_1\cdots a_p}|_{\Sigma}&=&\dot{a}_{a_1\cdots a_p}-(d^{\Sigma}\phi)_{a_1\cdots a_p}
+p(\nabla_{[a_1}n^c)a_{|c|a_2\cdots a_p]}\nonumber\\
(\delta A)_{a_2\cdots a_p}|_{\Sigma}&=&-\dot{\phi}_{a_2\cdots a_p}-
(\delta^{\Sigma}a)_{a_2\cdots a_p}-p(\nabla^{a_1}n_{[a_1})\phi_{a_2\cdots a_p]}
-(n\wedge\delta^{\Sigma}\phi)_{a_2\cdots a_p}.\nonumber
\end{eqnarray}
\end{lemma}
\begin{proof*}
The proof is a straightforward computation, using in particular \cite{W} Lemma 10.2.1, which
states that
\[
\nabla^{\Sigma}_cT^{a_1\cdots a_k}_{\phantom{a_1\cdots a_k}b_1\cdots b_l}=
P^{\phantom{d_1}a_1}_{d_1}\cdots P^{\phantom{d_k}a_k}_{d_k}P^{\phantom{b_1}e_1}_{b_1}\cdots
P^{\phantom{b_l}e_l}_{b_l}P_c^{\phantom{c}f}\nabla_f
T^{d_1\cdots d_k}_{\phantom{d_1\cdots d_k}e_1\cdots e_l},
\]
where $T$ is a tensor field on $\Sigma$ and $\nabla^{\Sigma}$ is the Levi-Civita derivative of
the induced metric $h$ on $\Sigma$. For the first expression we now expand the
anti-symmetrisation in $dA$, perform a partial integration and then pull back to $\Sigma$. For the
second expression we write
$(\delta A)_{a_2\cdots a_p}=-(n^{a_0}n^{a_1}-h^{a_0a_1})\nabla_{a_0}A_{a_1\cdots a_p}$ and then
insert a factor $\delta_a^{\phantom{a}b}=P_a^{\phantom{a}b}+n_an^b$ for each of the indices of
$A$, to the right of the derivative operator. We omit the details.
\end{proof*}

Introducing a notation for the pull-back of $\delta A$,
\[
\omega_{a_2\cdots a_p}:=P_{a_2}^{\phantom{a_2}b_2}\cdots P_{a_p}^{\phantom{a_p}b_p}
(\delta A)_{b_2\cdots b_p}|_{\Sigma},
\]
we can parametrise the initial data of $A$ as follows:
\begin{corollary}\label{Cor_NewData}
There is a linear homeomorphism on
$\mathcal{D}^p(\Sigma)^{\oplus 2}\oplus\mathcal{D}^{p-1}(\Sigma)^{\oplus 2}$ which maps the
initial data $(a,\dot{a},\phi,\dot{\phi})$ of $A\in\mathcal{D}^p(M)$ to
$(a,E,\phi,\omega)$.
\end{corollary}
The same statement is also valid for data in
$\Omega^p(\Sigma)^{\oplus 2}\oplus\Omega^{p-1}(\Sigma)^{\oplus 2}$, or in
$\Omega^p_0(\Sigma)^{\oplus 2}\oplus\Omega^{p-1}_0(\Sigma)^{\oplus 2}$.
\begin{proof*}
Lemma \ref{Lem_EdeltaA} shows how to express $E$ and $\omega$ in terms of the initial data
$(a,\dot{a},\phi,\dot{\phi})$. From these expressions we also see that $\dot{a}$ can be expressed
in terms of $(a,\phi,E)$ and $\dot{\phi}$ in terms of $(a,\phi,\omega)$ and the maps in both
directions are clearly continuous.
\end{proof*}

\begin{corollary}\label{Cor_Asmeared}
If $A\in\mathcal{D}^p(M)$ satisfies $\Box A=j\in\Omega^p(M)$ and $\alpha\in\Omega^p_0(M)$,
then
\[
(A,*\alpha)=\sum_{\pm}\int_{J^{\pm}(\Sigma)}j\wedge *G^{\mp}\alpha+
\int_{\Sigma} \phi\wedge *_{\Sigma}\omega_{\alpha}-a\wedge *_{\Sigma}E_{\alpha}
-\phi_{\alpha}\wedge *_{\Sigma}\omega+a_{\alpha}\wedge *_{\Sigma}E,
\]
where $\Sigma$ is a Cauchy surface with future pointing unit normal vector field $n^a$ and the
initial data $(a,E,\phi,\omega)$ refer to $A$, whereas
$(a_{\alpha},E_{\alpha},\phi_{\alpha},\omega_{\alpha})$ refer to $G\alpha$.
\end{corollary}
\begin{proof*}
The proof is similar to that of Lemma \ref{Lem_Asmeared}, but now performing partial integrations
using $d$ and $\delta$. We refer to \cite{P} Proposition 2.2 for more details on the proof, but we
note that this reference uses compactly supported $j$, so it gets away with using
$G^{\pm}\alpha$ on $J^{\pm}(\Sigma)$, rather than $G^{\mp}\alpha$. This causes an overall sign
difference for the integrations of the initial data.
\end{proof*}

In addition we will also make use of the following technical lemma:
\begin{lemma}\label{Lem_NormaldeltaA}
If $A\in\mathcal{D}^p(M)$ has initial data on $\Sigma$ such that $\phi=\omega=0$, then
$\delta A|_{\Sigma}=0$ and
\[
n^a\nabla_a\delta A|_{\Sigma}=pn^{a_1}(\Box A)_{a_1\cdots a_p}|_{\Sigma}
-p(\delta^{\Sigma}E)_{a_2\cdots a_p}|_{\Sigma}.
\]
\end{lemma}
\begin{proof*}
Note that for any $X\in\mathcal{D}^p(M)$ with $WF(X)\cap N^*\Sigma=\emptyset$ we have
\begin{eqnarray}\label{Eqn_ndelta}
n^{a_2}(\delta X)_{a_2\cdots a_p}|_{\Sigma}&=&
-\nabla^{a_1}n^{a_2}X_{a_1\cdots a_p}|_{\Sigma}
=h^{a_0a_1}\nabla_{a_0}n^{a_2}X_{a_1\cdots a_p}|_{\Sigma}\\
&=&-h^{a_0a_2}\nabla^{\Sigma}_{a_0}Y_{a_2\cdots a_p}|_{\Sigma}
=(\delta^{\Sigma}Y)_{a_3\cdots a_p}|_{\Sigma},\nonumber
\end{eqnarray}
where $Y_{a_2\cdots a_p}:=n^{a_1}X_{a1\cdots a_p}|_{\Sigma}$ and we used the antisymmetry of $X$
and the symmetry of $\nabla^an^b$. In case $X=A$ we have $Y=\phi=0$, so the equality above proves
that the normal component of $\delta A$ on $\Sigma$ vanishes. Together with $\omega=0$ this
implies $\delta A|_{\Sigma}=0$. Similarly we can consider the normal component of the normal
derivative:
\begin{eqnarray}
n^{a_1}n^{a_2}\nabla_{a_1}(\delta A)_{a_2\cdots a_p}|_{\Sigma}&=&
-n^{a_0}n^{a_2}\nabla_{a_0}\nabla^{a_1}A_{a_1\cdots a_p}|_{\Sigma}\nonumber\\
&=&-n^{a_0}n^{a_2}\nabla^{a_1}\nabla_{a_0}A_{a_1\cdots a_p}|_{\Sigma}\nonumber\\
&=&(\nabla^{a_1}n^{a_0})n^{a_2}\nabla_{a_0}A_{a_1\cdots a_p}|_{\Sigma}
-\nabla^{a_1}n^{a_0}n^{a_2}\nabla_{a_0}A_{a_1\cdots a_p}|_{\Sigma}\nonumber\\
&=&(\nabla^{a_0}n^{a_1})\nabla_{a_0}n^{a_2}A_{a_1\cdots a_p}|_{\Sigma}
+\nabla^{a_2}n^{a_0}\nabla_{a_0}n^{a_1}A_{a_1\cdots a_p}|_{\Sigma}\nonumber\\
&=&-(\nabla^{a_0}n^{a_2})\nabla^{\Sigma}_{a_0}\phi_{a_2\cdots a_p}|_{\Sigma}
+(\delta^{\Sigma}\dot{\phi})_{a_2\cdots a_p}|_{\Sigma},\nonumber
\end{eqnarray}
where the interchange of derivatives gives no curvature terms because of $\phi=0$ and
we repeatedly used $\nabla^{[a}n^{b]}=0$, the anti-symmetry of $A$ and the symmetry of
$(\nabla^an^b)(\nabla_an^c)$ in $(bc)$. Now note that $\phi=0$ and $\delta A|_{\Sigma}=0$, so
$\dot{\phi}=\delta^{\Sigma}a$ and hence $\delta^{\Sigma}\dot{\phi}=0$ too. For the spatial
component of the normal derivative of $\delta A$ we eliminate the second order derivative in the
normal direction in favour of $\Box A$ as follows:
\begin{eqnarray}
n^{a_1}\nabla_{a_1}(\delta A)_{a_2\cdots a_p}|_{\Sigma}&=&
pn^{a_1}(d\delta A)_{a_1\cdots a_p}|_{\Sigma}
+(p-1)n^{a_1}\nabla_{[a_2}(\delta A)_{|a_1|a_3\cdots a_p]}|_{\Sigma}\nonumber\\
&=&pn^{a_1}(\Box A)_{a_1\cdots a_p}|_{\Sigma}-pn^{a_1}(\delta dA)_{a_1\cdots a_p}|_{\Sigma}
\nonumber\\
&&+(p-1)n^{a_1}\nabla_{[a_2}(\delta A)_{|a_1|a_3\cdots a_p]}|_{\Sigma}.\nonumber
\end{eqnarray}
The normal component of the last term vanishes, as we have just seen. Furthermore, the
pull-back of the last term also vanishes, as this is just
$d^{\Sigma}\delta^{\Sigma}\phi$. Using $X=dA$ in the first paragraph we can rewrite the second
term on the right-hand side as $-p\delta^{\Sigma}E$, which completes the proof.
\end{proof*}

\subsubsection{The Cauchy problem for the Maxwell equations}\label{SSSec_Maxwell}

In order to solve the Maxwell equations, we first show that each equivalence class
$[A]\in\mathcal{F}^p(M)$ has sufficiently nice representatives:
\begin{lemma}[Lorenz gauge]\label{Lem_FixGauge}
For any $A\in\mathcal{D}^p(M)$, $[A]\in\mathcal{F}^p(M)$ has a representative $A'$ satisfying the
Lorenz gauge condition $\delta A'=0$. Furthermore, if $A\in\mathcal{D}^p_{sc}(M)$ we can choose
$A'$ such that $A'-A\in d\mathcal{D}^{p-1}_{sc}(M)$.
\end{lemma}
\begin{proof*}
Let $\phi^+\in\Omega^0(M)$ such that $\phi^+\equiv 0$ in a neighbourhood of a Cauchy surface
$\Sigma^+$ and such that $\phi^-:=1-\phi^+\equiv 0$ in a neighbourhood of another Cauchy surface
$\Sigma^-$. Given $A$, let $\chi^{\pm}$ be the unique solutions of
$\Box\chi^{\pm}=-\delta(\phi^{\pm}A)$ with vanishing initial data on $\Sigma^{\pm}$
(cf.\ Theorem \ref{Thm_WaveCauchy'}). Note that $\chi^{\pm}$ vanishes near $\Sigma^{\pm}$ and
that $\Box\delta\chi^{\pm}=\delta\Box\chi^{\pm}=0$, so $\delta\chi^{\pm}=0$. Furthermore,
$\chi^{\pm}$ has spacelike compact support if and only if $A$ does (cf.\ Proposition
\ref{Prop_SCSoln}). Hence $\chi:=\chi^++\chi^-$ satisfies $\delta\chi=0$, $\Box\chi=-\delta A$
and it has spacelike compact support if and only if $A$ does. Setting $A':=A+d\chi$ completes the
proof.
\end{proof*}

Note that the lemma does not require that $A$ has well-defined initial data on some Cauchy
surface. Also note that there is a residual gauge freedom: $A\sim A'$ and $\delta A=\delta A'=0$
hold if and only if $A-A'=d\chi$ with $\chi\in\mathcal{D}^{p-1}(M)$ such that
$\delta d\chi=0$. Interestingly, this is the homogeneous Maxwell equations for a $p-1$ form.
A further gauge fixing, which is often possible, is the temporal gauge, which consists in setting
$\phi=0$ on a given Cauchy surface:
\begin{lemma}[Temporal gauge]\label{Lem_Temporal}
Let $A\in\mathcal{D}^p(M)$ with $WF(A)\cap N^*\Sigma=\emptyset$ and initial data
$(a,E,\phi,\omega)$. Then there is a representative $A'\in[A]$ with initial data
$(a,E,\phi'=0,\omega'=0)$. In particular, $\delta A'=0$.
\end{lemma}
\begin{proof*}
We solve $\Box\chi=-\delta A$ with initial data $(0,-\phi,0,0)$ for $\chi$. By Lemma
\ref{Lem_NormaldeltaA} and equation (\ref{Eqn_ndelta}) we see that the initial data of
$\delta\chi$ on $\Sigma$ vanish. As $\Box \delta\chi=0$ we have $\delta\chi=0$, so $A':=A+d\chi$
has $\delta A'=0$. One may verify directly that $E'=E$, $\phi'=0$ and
$a'-a=\iota^*_{\Sigma}(d\chi)=d^{\Sigma}\iota^*_{\Sigma}\chi=0$.
\end{proof*}

\begin{remark}\label{Rem_Coulomb}
Lemma \ref{Lem_Temporal} implies that the Lorenz gauge, $\delta A=0$, and the temporal gauge,
$\phi=0$, can be achieved simultaneously. \cite{P} uses the term Coulomb gauge for this
combination of gauge conditions, but in the physics literature the term Coulomb gauge usually
refers to the gauge condition $\delta^{\Sigma}a=0$. If a given $[A]$ has any Coulomb gauge
representatives, then it has representatives that satisfy Lorenz, temporal and Coulomb gauge
simultaneously.
\end{remark}
Note that both the Coulomb and the temporal gauge are required to be valid only on the prescribed
Cauchy surface.

We now make the following fundamental observation:
\begin{lemma}\label{Lem_Lorenz}
Let $M=(\mathcal{M},g)$ be a globally hyperbolic spacetime, $\Sigma$ a Cauchy surface with future
pointing unit normal vector field $n^a$ and let $j\in\Omega^p(M)$ with $\delta j=0$. Any
$A\in\mathcal{D}^p(M)$ solves
\begin{equation}\label{Eqn_ProblemB}
\delta d A=j,\quad \delta A=0,
\end{equation}
if and only if it solves
\begin{equation}\label{Eqn_ProblemC}
\Box A=j,\quad \delta A|_{\Sigma}=0,\ n^a\nabla_a\delta A|_{\Sigma}=0,
\end{equation}
in which case it also solves
\begin{equation}\label{Eqn_ProblemA}
\delta dA=j.
\end{equation}
\end{lemma}
\begin{proof*}
If $A$ solves (\ref{Eqn_ProblemB}), it clearly also solves (\ref{Eqn_ProblemA}) and
(\ref{Eqn_ProblemC}). On the other hand, if $A$ solves (\ref{Eqn_ProblemC}), then
$\Box\delta A=\delta\Box A=\delta j=0$, so $\delta A$ satisfies a wave equation with vanishing
initial data. From Theorem \ref{Thm_WaveCauchy} we find $\delta A=0$, so $A$ solves
(\ref{Eqn_ProblemB}).
\end{proof*}

The requirement that the current $j$ is conserved, $\delta j=0$, is no real restriction, because
if $\delta j\not=0$ there can be no solutions to $\delta dA=j$, in view of $\delta^2=0$. In fact,
by the same reasoning we should even restrict attention to co-exact source terms $j$.

When considering gauge equivalence classes $[A]$ we encounter the problem that not all
representatives $A$ may have well-defined initial data on a given Cauchy surface $\Sigma$, due to
their distributional nature. We deal with this issue using the following definition:
\begin{definition}\label{Def_DataExist}
We say that an $[A]\in\mathcal{F}^p(M)$ has \emph{well-defined initial data on a Cauchy surface}
$\Sigma$ if and only if every Lorenz gauge representative $A$ has
$WF(A)\cap N^*\Sigma=\emptyset$. In this case we write $\iota_{\Sigma}^*([A])$ for
$[\iota_{\Sigma}^*(A)]$ with any Lorenz gauge representative.
\end{definition}
Note that it suffices to find one Lorenz gauge representative satisfying the wave front set
condition. Indeed, for any residual gauge term $d\chi$ we may use Lemma \ref{Lem_FixGauge},
with $\chi$ in the role of $A$, to write $d\chi=d\chi'$ where $\Box\chi'=\delta\chi'=0$, so
$WF(\chi')\cap N^*\Sigma=\emptyset$ and $WF(d\chi)=WF(d\chi')\subset WF(\chi')$.

Applying Lemma \ref{Lem_FixGauge} to $A$ in the same way we see that it suffices to study the
equation (\ref{Eqn_ProblemC}) instead of equation (\ref{Eqn_ProblemA}). Thus we obtain our main
result:
\begin{theorem}\label{Thm_MaxwellCauchy}
Given $j\in\Omega^p(M)$, $E\in\mathcal{D}^p(\Sigma)$ and
$[a]\in\mathcal{D}^p(\Sigma)/d\mathcal{D}^{p-1}(\Sigma)$, there is at most one
$[A]\in\mathcal{F}^p(M)$ with well-defined initial data on $\Sigma$, such that
\begin{equation}\label{Eqn_ProblemA'}
\delta dA=j,\quad \iota^*_{\Sigma}([A])=[a],\
n^{a_0}(dA)_{a_0\cdots a_p}|_{\Sigma}=E_{a_1\cdots a_p}.
\end{equation}
Such a solution exists if and only if $j$ is co-closed and
\begin{equation}\label{Eqn_Constraint}
(\delta^{\Sigma}E)_{a_2\cdots a_p}=n^{a_1}j_{a_1\cdots a_p}|_{\Sigma}.
\end{equation}
Moreover, if we define
\[
\mathscr{D}^p_j(\Sigma):=\frac{\mathcal{D}^p(\Sigma)}{d\mathcal{D}^{p-1}(\Sigma)}\oplus
\left\{E\in\mathcal{D}^p(\Sigma)|\ \delta^{\Sigma}E=n^aj_{a\cdots}|_{\Sigma}\right\},
\]
endowed with the topology that is obtained from the distributional topology by taking relative
topologies, quotients and direct sums, then $[A]$ depends continuously on
$([a],E)\in\mathscr{D}^p_j$ and on $j\in\Omega^p(M)$.
\end{theorem}

Note that for any Cauchy surface $\Sigma$, $\mathscr{D}^p_j(\Sigma)$ is the space of initial
data, modulo gauge equivalence, satisfying the constraint equation (\ref{Eqn_Constraint}). It is
empty when $j$ is not co-exact (because $j=\delta dA$), while it is otherwise an affine space
modeled over the linear space
\[
\mathscr{D}^p_0(\Sigma)=\frac{\mathcal{D}^p(\Sigma)}{d\mathcal{D}^{p-1}(\Sigma)}\oplus
\mathcal{D}^p_{\delta}(\Sigma).
\]

\begin{proof*}
We first prove existence of a solution. If $a$ is some representative of $[a]$, then there exists
a unique $A\in\mathcal{D}^p(M)$ which solves $\Box A=j$ with initial data
$(a,E,\phi=0,\omega=0)$, by Theorem \ref{Thm_WaveCauchy'} and Corollary \ref{Cor_NewData}.
Furthermore, $n^a\nabla_a(\delta A)|_{\Sigma}=0$, by Lemma \ref{Lem_NormaldeltaA}, so
$\delta A=0$ by Lemma \ref{Lem_Lorenz}. This implies that $A$ is a Lorenz gauge solution to
$\delta dA=j$ with the prescribed initial data. Any other Lorenz gauge solution in $[A]$ has the
same $E$ and $[a]$.

To prove uniqueness we let $A,A'\in\mathcal{D}^p(M)$ be two solutions to equation
(\ref{Eqn_ProblemA'}), both in Lorenz gauge. Then $B:=A-A'$ is in Lorenz gauge and satisfies
$\Box B=\delta B=0$ with $n^{a_0}(dB)_{a_0\cdots a_p}|_{\Sigma}=0$ and
$\iota^*_{\Sigma}(B)=d^{\Sigma}b$ for some $b\in\mathcal{D}^{p-1}(\Sigma)$. By the previous
paragraph we may solve $\delta d\chi=0$, with initial data such that $\iota^*_{\Sigma}\chi=b$ and
$n^{a_1}(d\chi)_{a_1\cdots a_p}|_{\Sigma}=n^{a_1}B_{a_1\cdots a_p}|_{\Sigma}$, because
(cf.\ equation (\ref{Eqn_ndelta}))
\[
-(\nabla^{\Sigma})^{a_2}n^{a_1}B_{a_1\cdots a_p}|_{\Sigma}=
n^{a_2}(\delta B)_{a_2\cdots a_p}|_{\Sigma}=0.
\]
Then, $C:=B-d\chi$ solves $\Box C=\delta C=0$ and the initial data, in the form of Corollary
\ref{Cor_NewData}, are easily seen to vanish, as
e.g.\ $\iota^*_{\Sigma}(C)=\iota^*_{\Sigma}(B)-d^{\Sigma}\iota^*_{\Sigma}(\chi)=0$. Hence, $C=0$
and $B=d\chi$, proving that $[A]=[A']$. The continuous dependence on that data follows by taking
a Lorenz gauge representative and using Corollary \ref{Cor_Asmeared}.
\end{proof*}

The statement of Theorem \ref{Thm_MaxwellCauchy} also holds if we assume
$j\in\Omega^p_{sc}(M)$ with data $E\in\mathcal{D}^p_0(\Sigma)$ and
$[a]\in\mathcal{D}^p_0(\Sigma)/d\mathcal{D}^{p-1}_0(\Sigma)$ and if we replace the gauge
equivalence by $A\sim A'$ iff $A-A'=d\chi$ with $\chi\in\mathcal{D}^{p-1}_{sc}(M)$. The proof
is completely analogous.

\begin{remark}\label{Rem_sj}
The solution map $\map{s^p_j}{\mathscr{D}^p_j}{\mathcal{F}^p(M)}$ is not only continuous, but it
is also a homeomorphism onto its range (when taken in the relative topology). This follows from
Corollary \ref{Cor_Asmeared}, using Lorenz and temporal gauge representatives of $[A]$ and the
following observation: for any test-forms $(a_{\alpha},E_{\alpha})$ on $\Sigma$ we can find a
solution $\chi\in\Omega^p_{sc}(M)$ to equation (\ref{Eqn_ProblemB}) with vanishing source term
and initial data $(a_{\alpha},E_{\alpha},0,0)$. (This follows e.g.\ from Theorem
\ref{Thm_MaxwellCauchy}.) By Proposition \ref{Prop_SCSoln} we may write $\chi=G\alpha$ with
$\alpha\in\Omega^p_0(M)$. When the pairing with $(a_{\alpha},E_{\alpha})$ in Corollary
\ref{Cor_Asmeared} is gauge invariant, so is the pairing with $A$. In that case the convergence
of $[A]$ on $M$ implies the convergence of its initial data.
\end{remark}

\begin{remark}
Let $\map{\iota_{\Sigma}}{\Sigma}{M}$ be the canonical embedding of a Cauchy surface $\Sigma$ in
a globally hyperbolic spacetime $M$. Theorem \ref{Thm_MaxwellCauchy} entails in particular that
if $j$ is co-closed and $n^{a_1}j_{a_1\ldots a_p}|_{\Sigma}$ is co-exact, then $j$ is co-exact.
This is the Hodge dual statement of the fact that the restriction of the pull-back map
$\map{\iota^*_{\Sigma}}{\Omega^p(M)}{\Omega^p(\Sigma)}$ to closed forms descends to an
isomorphism $\map{\iota^*_{\Sigma}}{H^p(M)}{H^p(\Sigma)}$. Similarly, a closed form
$\alpha\in\Omega^p(M)$ is of the form $\alpha=d\chi$ for some
$\chi\in\Omega^p_{sc}(M)$ if and only if
$[\iota^*_{\Sigma}\alpha]=0\in H^p_0(\Sigma)$.

To prove these statements we note that exterior derivatives commute with pull-backs, so the
restriction $\iota^*_{\Sigma}$ descends to a well-defined map between the cohomology groups.
By the K\"unneth formula, these groups are vector spaces of the same dimension, so it suffices
to show that $\iota^*_{\Sigma}$ is injective. For a given cohomology class $[\alpha]$ we may
define $j:=*\alpha\in\Omega^{n-p}_{\delta}(M)$. If $\iota_{\Sigma}^*\alpha=d^{\Sigma}\beta$,
then
$n^{a_1}j_{a_1\ldots a_{n-p}}|_{\Sigma}=(\delta^{\Sigma}*_{\Sigma}\beta)_{a_2\ldots a_{n-p}}$.
We may then consider the Maxwell equations $\delta dA=j$ with Cauchy data
$(a,E,\phi,\omega)=(0,*_{\Sigma}\beta,0,0)$. This satisfies the constraint equation
(\ref{Eqn_Constraint}), so by Theorem \ref{Thm_MaxwellCauchy} there exists a solution $A$, which
explicitly shows that $j=\delta dA$ is co-exact and hence $\alpha$ is exact. Similarly, when
$\alpha=d\chi$ with $\chi\in\Omega^p_{sc}(M)$, then $*_{\Sigma}\beta$ has compact support.
Conversely, when $*_{\Sigma}\beta$ has compact support, then
$A\in\Omega^{n-p}_{sc}(M)$ and hence
$\alpha=(-1)^{p(n-p)+1}d\delta A\in d\Omega^{p-1}_{sc}(M)$.
\end{remark}

\section{The Poisson Structure for $p$-Form Fields}\label{Sec_Poisson}

In this section we will consider the phase space of solutions to the Maxwell equations and we
will explain in some more detail how the Aharonov-Bohm effect is related to the choice of gauge
equivalence in Subsection \ref{SSec_Soln}. Next we endow the space of local, affine observables
with a Poisson bracket in Subsection \ref{SSec_Poisson}, which will be used to quantise the
theory in Section \ref{Sec_Quantum}. Moreover, we will compute the degeneracies of the Poisson
bracket in Subsection \ref{SSec_Degeneracies} and show how they may be interpreted in terms of
Gauss' law.

\subsection{Observables and the space of solutions}\label{SSec_Soln}

We have already introduced the kinematic space of field configurations
\[
\mathcal{F}^p(M)=\mathcal{D}^p(M)/d\mathcal{D}^{p-1}(M),
\]
and the continuous dual space $\mathcal{F}^p(M)^*=\Omega^p_{0,\delta}(M)$, under the duality
$(.,*.)$. We will interpret the elements of the dual space $\mathcal{F}^p(M)^*$ as local, linear
observables on $\mathcal{F}$ and we will write
\[
F_{\alpha}(A):=\int_Mf_{\alpha}(A),\qquad f_{\alpha}(A):=A\wedge*\alpha
\]
with $\alpha\in\mathcal{F}^p(M)^*$. As an illustration of these observables we will now elaborate
how the Aharonov-Bohm effect can be described within our mathematical framework.

\begin{figure}
\centering
\includegraphics[scale=0.4]{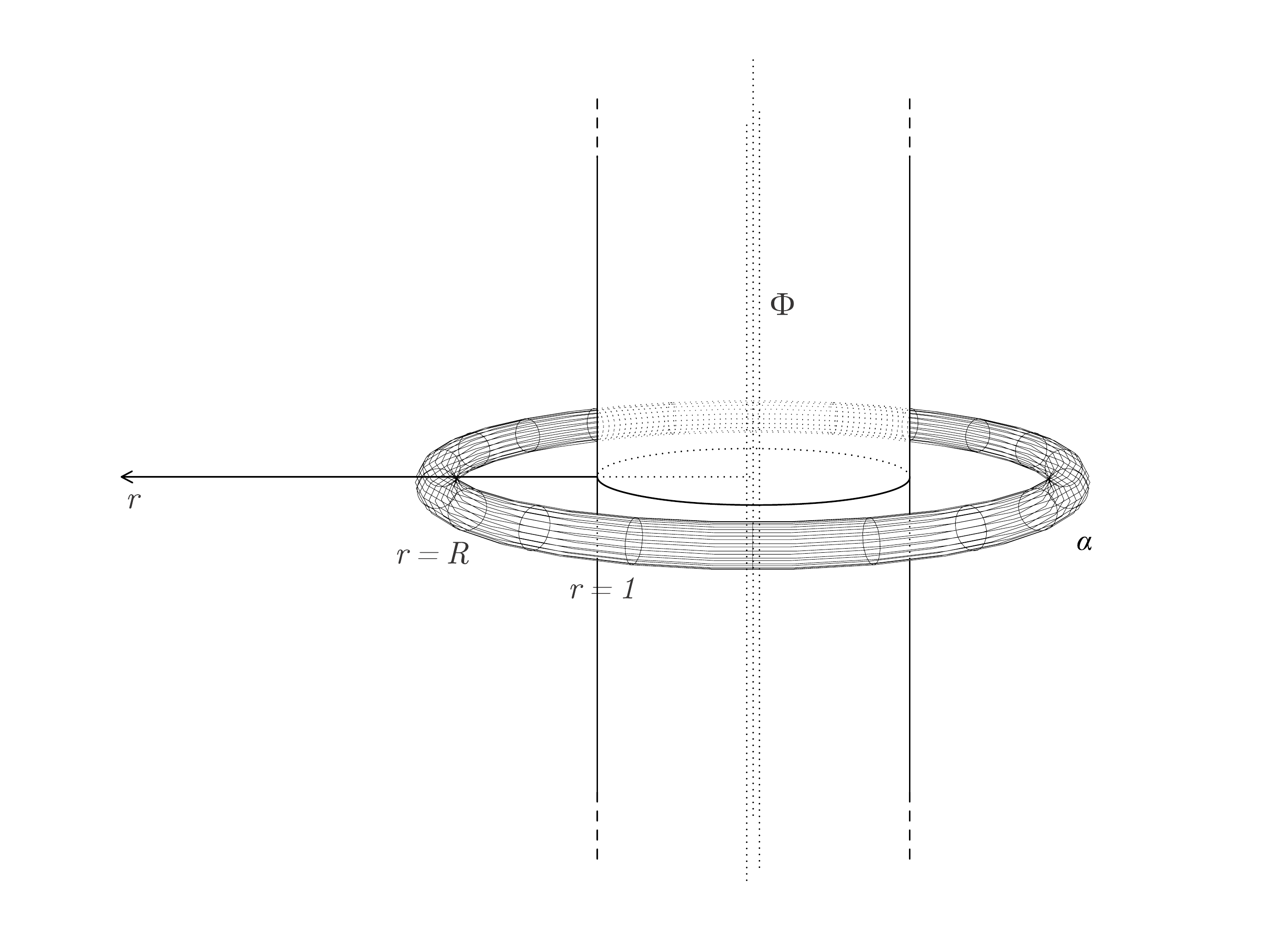}
\caption{An illustration of the Aharonov-Bohm effect, as described in Example \ref{Ex_ABeffect},
for $p=1$. The observable $\alpha$, proportional to the differential $d\varphi$ of the angular
coordinate, is supported in a ring-shaped region and essentially measures the magnetic flux
through a disc of radius $R$ intersecting the coil.}
\label{Fig_AB}
\end{figure}

\begin{example}\label{Ex_ABeffect}
The following example is illustrated in Figure \ref{Fig_AB}. Let $M_0$ denote Minkowski spacetime
and let $\zeta$ denote the solid cylinder along the $z$-axis, which is given in cylindrical
coordinates $(r,\varphi,z,t)$ by $r\le 1$. We suppose that the cylinder $\zeta$ contains a
conducting coil with a current running through it and we denote the current density 1-form by
$j$. The flux of the current through the $t=z=0$ plane will be denoted by $\Phi$. The current
generates a vector potential, which can be represented in very good approximation by
$A_{\Phi}=\frac{\phi(r)}{2\pi}d\varphi$, for some $\phi$ which equals $\Phi$ outside the coil.
(The approximation lies in the fact that in reality the current also has a small component along
the $z$-axis, which we ignored.)

In the Aharonov-Bohm experiment one uses quantum particles that, effectively, go around the coil
and measure a quantum phase shift that is proportional to the integral of $A$ along its circular
path. (For a more proper description see \cite{PT}.) We model this observable by the compactly
supported $1$-form $\alpha=r^{-1}f(r-R)f(z)f(t)d\varphi$, where $f\in C_0^{\infty}(\mathbb{R})$
has its support in $(-1,1)$ and $R>2$. One may verify that $\alpha$ is co-closed, so
$\alpha\in\mathcal{F}^1(M_0)^*$ does indeed define an observable, in our sense. A short
computation yields $F_{\alpha}(A_{\Phi})=\Phi\left(\int f\right)^3$. This observable is therefore
non-trivial, unless $\int f=0$.

We now focus on the region of spacetime
$M_1=M_0\setminus J(\zeta)\simeq \mathbb{R}^{\times 3}\times S^1$. In this region, $A_{\Phi}$ is
closed, but not exact. Nevertheless, $\alpha$ is supported only in $M_1$ and defines a
non-trivial observable there, as evidenced by the Aharonov-Bohm effect. (By Hodge duality,
$\alpha$ is not co-exact on $M_1$.) For this reason we cannot identify vector potentials whose
difference is not exact. Interpreting $A_{\Phi}$ as a connection one-form we see that
$A_{\Phi}=-i\lambda^{-1}d\lambda$ with $\lambda:=\exp(i\frac{\Phi}{2\pi}\varphi)\in\mathcal{G}(M_1)$.
However, $\varphi$ is not a well-defined smooth function on $M_1$ and $\lambda$ is not in
$\mathfrak{G}_0(M_1)$, because it has a non-trivial holonomy. (Cf.\ Subsection \ref{SSSec_Geo}.)

It is not difficult to construct higher dimensional analogues of this example. Indeed, let
$M_0$ be the $n$-dimensional Minkowski space and choose $p$ such that $1\le p\le n-2$. In the
time zero Cauchy surface $\Sigma_0$ we now remove a hyperplane $H$ of codimension $p+1$ to
obtain a surface
$\Sigma:=\Sigma_0\setminus H\simeq\mathbb{R}^{n-2-p}\times\mathbb{S}^p\times\mathbb{R}_{>0}$.
We let $M:=D(\Sigma)\subset M_0$, so that
$M\simeq\mathbb{R}^{n-1-p}\times\mathbb{S}^p\times\mathbb{R}_{>0}$. Let $\omega$ denote the
volume form on $\mathbb{S}^p$ and define the $p$-form potential
$A_{\Phi}:=\frac{\phi(r)}{\int\omega}\omega$, where $\phi(r)$ takes some constant value
$\Phi$ on $r>1$. One may verify by direct computation that $dA_{\Phi}=0$ on the region where
$\phi\equiv\Phi$ and in particular it solves the homogeneous analogue of Maxwell's equation
there. Choosing $f$ and $R$ as above we may define the compactly supported $p$-form
$\alpha:=r^{-p}f(x^1)\cdots f(x^{n-p-2})f(t)\omega$, which is again co-closed, so it
defines an observable. We then find $F_{\alpha}(A_{\Phi})=\Phi(\int f)^{n-p}$. For a
physical interpretation analogous to the Aharonov-Bohm effect we note that the observable
$\alpha$ must now describe an experiment involving $(p-1)$-dimensional objects, rather than
particles. (For $p=0$ the spacetime $M$ would no longer be connected. For $p=n-1$ one could
still take a Cauchy surface $\Sigma\simeq\mathbb{S}^{n-1}$ and find a closed $A$ which is
not exact, but this can no longer be obtained by removing a hyperplane from $M_0$, so an
interpretation analogous to the Aharonov-Bohm experiment seems less obvious.)
\end{example}

Example \ref{Ex_ABeffect} motivates us to make the following definition, for $p$-forms:
\begin{definition}\label{Def_ABConf}
A field configuration $[A]\in\mathcal{F}^p(M)$ is called an \emph{Aharonov-Bohm configuration} if
and only if $0\not=[A]\in H^p(M)$.
Furthermore, we call an observable $\alpha\in\mathcal{F}^p(M)^*$ a
\emph{field strength observable} if and only if $\alpha\in\delta\Omega^{p+1}_0(M)$.
\end{definition}
The definition of field strength observables is motivated by the fact that an observable
$\alpha=\delta\beta$ is only sensitive to the field strength $dA$, because
$F_{\alpha}(A)=\int_MdA\wedge *\beta$. Equivalently, they are the observables that vanish on all
Aharonov-Bohm field configurations.

For any given background current $j\in\Omega^p(M)$ we let $\mathcal{S}^p_j$ denote the phase space
of solutions of (\ref{Eqn_ProblemA}) :
\[
\mathcal{S}^p_j:=\left\{[A]\in\mathcal{F}^p(M)|\ \delta d[A]=j\right\}
=\left\{A\in\mathcal{D}^p(M)|\ \delta dA=j\right\}/\sim,
\]
where we divide out the gauge equivalence relation. For any $[A']\in\mathcal{S}^p_j$, the map
\[
e_{[A']}:\mathcal{S}^p_0\ni [A]\mapsto [A+A']\in\mathcal{S}^p_j
\]
is a well-defined bijection. Furthermore,
\[
e_{[A']}(0)=[A'],\quad e_{[A']}(B)=e_{[A]}(B+e_{[A]}^{-1}([A'])).
\]
This means that $\mathcal{S}^p_j$ is an affine space modeled over the vector space
$\mathcal{S}^p_0$.

If we equip $\mathcal{D}^p(M)$ with the usual distribution topology, then the set
$\mathcal{S}^p_0$ is a quotient of closed linear subspaces. (That the denominator is closed
follows from Theorem \ref{Thm_dR}.) $\mathcal{S}^p_0$ will be equipped with the quotient topology
of the relative topology and we equip $\mathcal{S}^p_j$ with the unique topology that makes all
the maps $e_{[A']}$ homeomorphisms.

The affine space $\mathcal{S}^p_j$ can be identified with the space $\mathscr{D}^p_j$ of initial
data satisfying the constraint equation, via the map
$\map{s^p_j}{\mathscr{D}^p_j}{\mathcal{S}^p_j}$, which sends (equivalence classes of) initial
data to the corresponding (equivalence classes of) solutions to equation (\ref{Eqn_ProblemA'}).
Note that the $s^p_j$ are affine bijections, by the well-posedness of the initial value problem,
Theorem \ref{Thm_MaxwellCauchy}. Moreover, if we endow $\mathscr{D}^p_j$ with the topology that
is obtained from the distributional topology by taking quotients and direct sums, then all
$s^p_j$ are homeomorphisms (cf.\ Remark \ref{Rem_sj}).

We may view $\mathcal{S}^p_j$ as an infinite dimensional manifold, where we use $e_{[A']}^{-1}$
as a single coordinate chart. By considering smooth curves into $\mathcal{S}^p_j$ one finds that
the (kinematic) tangent bundle of this manifold is given by
\[
T\mathcal{S}^p_j\simeq \mathcal{S}^p_j\times\mathcal{S}^p_0.
\]
For the (kinematic) cotangent bundle we consider the continuous linear maps on each tangent
space, so we define
\[
T^*\mathcal{S}^p_j\simeq \mathcal{S}^p_j\times(\mathcal{S}^p_0)^*.
\]
We now establish an explicit representation for $(\mathcal{S}^p_0)^*$:
\begin{proposition}\label{Prop_S0*}
We have
\[
(\mathcal{S}^p_0)^*=\frac{\Omega^p_{0,\delta}(M)}{\delta d\Omega^p_0(M)}\qquad
(\mathscr{D}^p_0)^*=\frac{\Omega^p_0(\Sigma)}{d^{\Sigma}\Omega^{p-1}_0(\Sigma)}\oplus
\Omega^p_{0,\delta}(\Sigma).
\]
There are isomorphisms $\map{G}{(\mathcal{S}^p_0)^*}{\mathcal{S}^{p,\infty}_{sc}}$ and
$\map{\rho}{\mathcal{S}^{p,\infty}_{sc}}{(\mathscr{D}^p_0)^*}$ with
\[
\mathcal{S}^{p,\infty}_{sc}:=\left\{A\in\Omega^p_{sc}(M)|\ \Box A=\delta A=0\right\}
/d\Omega^{p-1}_{sc}(M)
\]
and $\rho(A):=(\iota^*_{\Sigma}([A]),n^a(dA)_{a\cdots}|_{\Sigma})$.
\end{proposition}
\begin{proof*}
Because there is a homeomorphism $\map{s^p_0}{\mathscr{D}^p_0}{\mathcal{S}^p_0}$ it is clear that
$(s^p_0)^*$ is a homeomorphism of $(\mathcal{S}^p_0)^*$ to $(\mathscr{D}^p_0)^*$ (cf.\ Remark
\ref{Rem_sj}). The expression for $(\mathscr{D}^p_0)^*$ follows immediately from Theorem
\ref{Thm_dR}, keeping in mind the general facts that for a closed subspace $L\subset T$ of a
locally convex topological vector space $T$ we have $(T/L)^*=L^{\perp}$ and $L^*=T^*/L^{\perp}$,
where $L^{\perp}\subset T^*$ is the subspace that annihilates $L$ (cf.\ \cite{KN} 14.5).

The map $\rho$ is a well-defined linear isomorphism, by the smooth version of Theorem
\ref{Thm_MaxwellCauchy}. We now show that $G$ descends to an isomorphism from
\[
V:=\Omega^p_{0,\delta}(M)/\delta d\Omega^p_0(M)
\]
to $\mathcal{S}^{p,\infty}_{sc}$. For any $\alpha\in\Omega^p_{0,\delta}(M)$ we have
$\Box G\alpha=0$ and $\delta G\alpha=G\delta\alpha=0$. Furthermore, if $\alpha=\delta d\beta$ for
some $\beta\in\Omega^p_0(M)$, then $G\alpha=G\delta d\beta=-Gd\delta\beta=d\chi$ with
$\chi:=-G\delta\beta\in\Omega^{p-1}_{sc}(M)$. This means that $G$ descends to a well-defined
linear map from $V$ to $\mathcal{S}^{p,\infty}_{sc}$. Now suppose that
$A\in\mathcal{S}^{p,\infty}_{sc}$. By Proposition \ref{Prop_SCSoln} there is an
$\alpha\in\Omega^p_0(M)$ such that $A=G\alpha$. As $\delta A=0$ we conclude from Corollary
\ref{Cor_Exactness} that $\delta\alpha=\delta d\beta$ for some $\beta\in\Omega^{p-1}_0(M)$.
Defining $\gamma:=\alpha-d\beta$ we have $\delta\gamma=0$, so $[\gamma]\in V$, while
$G\gamma=A-dG\beta\sim A$ in $\mathcal{S}^{p,\infty}_{sc}$. This means that $G$ is surjective.

To prove injectivity of $G$ we choose $\alpha\in\Omega^p_{0,\delta}(M)$ and we assume that
$G\alpha=d\chi$ for some $\chi\in\Omega^{p-1}_{sc}(M)$. Without loss of generality we may assume
that $\delta\chi=0$ (cf.\ Lemma \ref{Lem_FixGauge}). Note that $dG\alpha=0$, so by Corollary
\ref{Cor_Exactness} that $\alpha=\delta\beta$ for some $\beta\in\Omega^{p+1}_{0,d}(M)$.
Furthermore, $\Box\chi =\delta d\chi=\delta G\delta\beta=0$, so that $\chi=G\eta$ for some
$\eta\in\Omega^{p-1}_0(M)$. As $0=\delta\chi=\delta G\eta$ we have $\delta\eta=\delta d\sigma$
for some $\sigma\in\Omega^{p-2}_{0,\delta}(M)$, by Corollary \ref{Cor_Exactness}. Putting
everything together we have $G(\delta\beta-d\eta)=G\alpha-d\chi=0$, so we can find
$\tau\in\Omega^p_0(M)$ with $\delta\beta=d\eta+\Box\tau$ (cf.\ Proposition \ref{Prop_SCSoln}).
Note that $\Box(\beta-d\tau)=d(\delta\beta-\Box\tau)=d^2\eta=0$, which means that $\beta=d\tau$,
by the compact supports. Hence, $\alpha=\delta\beta=\delta d\tau$, so $[\alpha]=0$ in $V$ and $G$
is injective.

From Corollary \ref{Cor_Asmeared}, taking $[A]$ in Lorenz and temporal gauge, we see that the
composition $\rho\circ G$ on $V$ is just the dual map to the solution map $s^p_0$. In particular,
$(\mathcal{S}^p_0)^*=V$.
\end{proof*}

\subsection{The Poisson structure}\label{SSec_Poisson}

Our next goal is to deduce a Poisson structure on the phase space $\mathcal{S}^p_j$, using
Peierls' method (cf.\ \cite{Pei}, or \cite{Ha} Section I.4). For this purpose we consider for any
observable $F_{\alpha}$, $\alpha\in\mathcal{F}^p(M)^*$, and for any $\epsilon>0$ the modified
Lagrangian
\[
\mathcal{L}_{\epsilon}:=\mathcal{L}+\epsilon f_{\alpha}(A).
\]
This gives rise to the equations of motion
\[
\delta dA=j+\epsilon\alpha.
\]
Given $[A]\in\mathcal{S}^p_j$ we let $[A^{\pm}_{\epsilon,\alpha}]$ denote the gauge equivalence
class of solutions to the modified equation which coincide (up to gauge equivalence) with $[A]$ in
the past ($+$), resp.\ future ($-$), of the support of $\alpha$. Due to the affine structure of
the equation of motion these solutions are uniquely defined, up to gauge equivalence, and they are
represented by
\[
A^{\pm}_{\epsilon,\alpha}=A+\epsilon G^{\pm}\alpha.
\]
The function $F_{\alpha}$ then defines a vector field on $\mathcal{S}^p_j$ by setting
\[
[A]\mapsto\delta_{F_{\alpha}}[A]:=\partial_{\epsilon}(A^+_{\epsilon,\alpha}
-A^-_{\epsilon,\alpha})|_{\epsilon=0}=-G\alpha.
\]
For another function $F_{\beta}$ on $\mathcal{F}^p(M)$, $\beta\in\Omega^p_{0,\delta}(M)$, we then
define the anti-symmetric bilinear map
\begin{equation}\label{PB}
\left\{F_{\alpha},F_{\beta}\right\}:=\delta_{F_{\alpha}}(F_{\beta})
=-(G\alpha,*\beta)=(\alpha,*G\beta).
\end{equation}
This map descends to an anti-symmetric bilinear map on $(\mathcal{S}^p_0)^*$, by similar
computations as in the proof of Proposition \ref{Prop_S0*}. We define the quotient map to be the
Poisson bracket, which is an anti-symmetric bilinear map on each cotangent space
$T^*_{[A]}\mathcal{S}^p_j\simeq(\mathcal{S}^p_0)^*$ (it defines a $2$-vectorfield). The canonical
trivialisation of the cotangent bundle ensures that the Poisson bracket takes a form which is
independent of the base point $[A]\in\mathcal{S}^p_j$.

In terms of the initial data $(a_{\alpha},E_{\alpha},\phi_{\alpha},0)$ of $G\alpha$ and
$(a_{\beta},E_{\beta},\phi_{\beta},0)$ of $G\beta$ on a Cauchy surface $\Sigma$ we have:
\begin{equation}\label{Eq_PBracket}
\left\{F_{\alpha},F_{\beta}\right\}(A)\equiv\int_{M}\alpha\wedge* G\beta
=\int_{\Sigma}E_{\beta}\wedge *_{\Sigma}a_{\alpha}-E_{\alpha}\wedge *_{\Sigma}a_{\beta},
\end{equation}
as may be seen directly from Corollary \ref{Cor_Asmeared}. Also note that the constraint
equation $\delta^{\Sigma}E_{\alpha}=\delta^{\Sigma}E_{\beta}=0$ is satisfied, by Theorem
\ref{Thm_MaxwellCauchy}.

\begin{remark}\label{Rem_WrongQ}
Our proof of Proposition \ref{Prop_S0*} also shows the rather remarkable fact that
$(\mathcal{S}^p_0)^*$ is isomorphic to $\mathcal{S}^{p,\infty}_{sc}$, which is a space of
spacelike compact, smooth solutions to the Maxwell equations, in Lorenz gauge
(see \cite{K12} Sec.5.3 for similar comments). Furthermore, under this identification the
Poisson bracket that we have derived, using Peierls' method, takes the same form as the
usual (pre-)symplectic form, or Lichnerowicz propagator, on $\mathcal{S}^{p,\infty}_{sc}$.
This makes it tempting to believe that the two quantisation schemes, using the Poisson
bracket on observables or using the (pre-)symplectic structure on the spacelike compact
solutions, are equivalent.

However, there is an important, but subtle difference: the gauge equivalence on
$\mathcal{S}^{p,\infty}_{sc}$ is defined by $d\Omega^{p-1}_{sc}(M)$, so it differs from the
original gauge equivalence $d\Omega^{p-1}(M)$. Using the wrong gauge equivalence would lead to a
theory without non-local behaviour \cite{Lang}, which, however, does not behave well under
embeddings.\footnote{To prove these claims one uses arguments as in the proof of Proposition
\ref{Prop_S0*} to find the space of observables
\[
\Omega^p_{0,\delta}(M)/\delta d\Omega^p_{tc}(M)\simeq
\Omega^p_0(\Sigma)/d\Omega^{p-1}(\Sigma)\oplus\Omega^p_{0,\delta}(\Sigma).
\]
By Corollary \ref{Cor_Asmeared} one sees that the Poisson bracket has no degeneracies. However,
item 2 of Remark \ref{Rem_Embed} below gives an example where this theory behaves badly under
embeddings, because the usual push-forward on $\Omega^p_{0,\delta}(M)$ would map a trivial
observable to a non-trivial one.} The previous literature has dealt with this subtle difference
in various ways: it was either evaded, by considering only compact Cauchy surfaces \cite{D,FP,P},
the field strength tensor \cite{DL} or a different choice of gauge equivalence \cite{DS}, or at
best it was taken into account in an ad hoc fashion (cf.\ \cite{FeH} for the case of linearised
gravity, or \cite{HS} for a more axiomatic approach). The point of our paper is that the origin
of this subtle difference can be understood: it stems from taking the dual space of
$\mathcal{S}^p_0$ as the space of observables, in line with Peierls' method.
\end{remark}

\subsection{Degeneracies of the Poisson bracket}\label{SSec_Degeneracies}

In general the Poisson bracket is degenerate, which means that there can be degenerate elements
$\alpha\in(\mathcal{S}^p_0)^*$, i.e.\ elements such that $\left\{F_{\alpha},F_{\beta}\right\}=0$
for all $\beta\in(\mathcal{S}^p_0)^*$. The subspace $\mathcal{C}^p(M)$ of degenerate elements of
$(\mathcal{S}^p_0)^*$ can be fully characterised in terms of the topology (and causal structure)
of $M$:
\begin{proposition}\label{Prop_Degeneracy}
We have
\begin{equation}
\mathcal{C}^p(M):=\frac{\delta(\Omega^{p+1}_0(M)\cap d\Omega^p_{tc}(M))}{\delta d\Omega^p_0(M)}
\simeq\frac{\Omega^p_0(\Sigma)\cap d^{\Sigma}\Omega^{p-1}(\Sigma)}
{d^{\Sigma}\Omega^{p-1}_0(\Sigma)}=:\mathcal{C}^p(\Sigma).\nonumber
\end{equation}
\end{proposition}
\begin{proof*}
We start with the expression for $\mathcal{C}^p(\Sigma)$, which is easiest to obtain. Using the
formula for $(\mathscr{D}^p_0)^*$ (Proposition \ref{Prop_S0*}), the Poisson bracket as given in
equation (\ref{Eq_PBracket}) and Poincar\'e duality one sees that $([a],E)\in(\mathscr{D}^p_0)^*$
is degenerate if and only if $E=0$ and any representative $a$ of $[a]$ is exact,
$a\in d^{\Sigma}\Omega^{p-1}(\Sigma)$, but not necessarily in $d^{\Sigma}\Omega^{p-1}_0(\Sigma)$.
The expression for $\mathcal{C}^p(\Sigma)$ then follows.

Next we note that any $\alpha\in\delta(\Omega^{p+1}_0(M)\cap d\Omega^p_{tc}(M))$ does define a
degenerate observable in $(\mathcal{S}^p_0)^*$, because if $\alpha=\delta d\beta$ for some
$\beta\in\Omega^p_{tc}(M)$, then every $\gamma\in\Omega^p_{0,\delta}(M)$ satisfies
$(\alpha,*G\gamma)=-(G\delta d\beta,*\gamma)=(Gd\delta\beta,*\gamma)
=-(G\delta\beta,*\delta\gamma)=0$. It follows from Proposition \ref{Prop_S0*} that there is a
linear injection from the second expression into $\mathcal{C}^p(M)$. To prove that this map is
surjective we suppose that $\alpha\in(\mathcal{S}^p_0)^*$ is degenerate. For any
$\gamma\in\Omega^p_{0,\delta}(M)$ we must then have $(G\alpha,*\gamma)=0$. This implies firstly
that $G\alpha$ is closed (by choosing $\gamma\in\delta\Omega^{p+1}_0(M)$) and even that $G\alpha$
is exact, by Poincar\'e duality. Thus $G\alpha=d\chi$ for some $\chi\in\Omega^{p-1}(M)$. Now,
following the penultimate paragraph of the proof of Proposition \ref{Prop_S0*}, but allowing
supports to be timelike compact only, where needed, we find that $\alpha=\delta d\tau$ for some
$\tau\in\Omega^p_{tc}(M)$ with $\beta:=d\tau\in\Omega^{p+1}_0(M)$. This completes the proof.
\end{proof*}

\begin{remark}
Since the moment we made our choice of gauge equivalence, we have only followed standard
procedures to find $\mathcal{C}^p(M)$. It is therefore tempting to think that observables in
$\mathcal{C}^p(M)$ are related to the Aharonov-Bohm effect, which motivated our choice of gauge
equivalence. However, $\mathcal{C}^p(M)$ consists entirely of field strength observables
(cf.\ Def.\ \ref{Def_ABConf}), which are not sensitive to the Aharonov-Bohm effect. (This is in
accord with the observations of \cite{Ashtekar}.)
\end{remark}

It is clear that $\mathcal{C}^p(M)$ is trivial whenever $\Sigma$ is compact, or whenever
$H^p_0(\Sigma)\simeq H^{p+1}_0(M)$ is trivial. Furthermore, if $H^p(M)\simeq H^p(\Sigma)$ is
trivial, then $\mathcal{C}^p(M)=\delta H^{p+1}_0(M)$ and $\mathcal{C}^p(\Sigma)=H^p_0(\Sigma)$.
To close this section we will give an alternative description of $\mathcal{C}^p(\Sigma)$ for
general $p$ and general spacetimes $M$. This will allow us to physically interpret the
degeneracies in the case of electromagnetism, $p=1$.

Suppose, then, that $[\alpha]\in\mathcal{C}^p(\Sigma)$, and let $\beta\in\Omega^{p-1}(\Sigma)$
be such that $\alpha=d^{\Sigma}\beta$. Note that $\beta$ is unique up to a closed form and
therefore $[\alpha]=0$ if and only if there is a closed form $\gamma\in\Omega^{p-1}_d(\Sigma)$
such that $\beta-\gamma$ has compact support. We will first argue that it suffices to find
$\gamma$ such that $\beta-\gamma$ is exact outside a compact set. Indeed, if $\beta-\gamma=0$
in some region, then it is certainly exact there. Conversely, if $\beta-\gamma$ is exact on
the complement $K^c:=\Sigma\setminus K$ of some compact set $K$, $\beta-\gamma=d^{\Sigma}\zeta$,
then we may use a partition of unity subordinate to $K^c$ and some relatively compact
$V\supset K$ to find a $\zeta'\in\Omega^{p-2}(\Sigma)$ which coincides with $\zeta$ outside
$K':=\overline{V}$. Hence, $\gamma':=\gamma+d^{\Sigma}\zeta'$ is closed and $\beta-\gamma'$
vanishes outside $K'$. This means that $[\alpha]=0$ as an observable if and only if we can
find $[\gamma]\in H^{p-1}(\Sigma)$ and a compact $K\subset\Sigma$ such that
$[(\beta-\gamma)|_{K^c}]=0$ in $H^{p-1}(K^c)$.

For any compact $K$ the canonical embedding $\map{\iota}{K^c}{\Sigma}$ gives rise to the
linear restriction map $\map{\iota^*}{H^{p-1}(\Sigma)}{H^{p-1}(K^c)}$. If $K$ contains
the support of $\alpha$, then $\beta$ is closed on $K^c$ and determines an element
$[\beta|_{K^c}]\in H^{p-1}(K^c)$. We then see that $[\alpha]=0$ if and only if $[\beta|_{K^c}]$
is in the range of $\iota^*$, for some compact $K\supset\mathrm{supp}(\alpha)$. By
Poincar\'e duality this is equivalent to the fact that $[\beta|_{K^c}]$, interpreted as a
linear map on $H_0^{n-p}(K^c)$, vanishes on the kernel of the push-forward map
$\map{\iota_*}{H^{n-p}_0(K^c)}{H^{n-p}_0(\Sigma)}$.

When $p=1$ this situation simplifies, because $H^0(\Sigma)$ consists only of all constant
functions and $H^0(K^c)$ of locally constant functions. Let us decompose
$\mathrm{supp}(\alpha)^c$ into connected components $V_i$, $i\in\mathcal{I}$ some index
set, and assume that $\mathcal{I}$ is finite. Let $\mathcal{I}'\subset\mathcal{I}$ be the
subset of indices for which $V_i$ has a non-compact closure in $\Sigma$. Note that
$[\alpha]=0$ if and only if $\beta$ takes the same constant value on all regions $V_i$
with $i\in\mathcal{I}'$. Indeed, if this is not the case, then $\beta-\gamma$ can never
have compact support for any constant $\gamma$. Conversely, if $\beta|_{V_i}=\gamma$ for
all $i\in\mathcal{I}'$, then $\beta-\gamma$ has support in the compact set
$K:=\mathrm{supp}(\alpha)\bigcup_{i\in\mathcal{I}\setminus\mathcal{I}'}\overline{V_i}$,
because it vanishes on $K^c=\bigcup_{i\in\mathcal{I}'}V_i$.

Note that in order to have $[\alpha]\not=0$ in $\mathcal{C}^1(M)$, $\mathcal{I}'$ must
contain at least two distinct indices. As a physical interpretation, one may think of
one of the regions $V_i$, $i\in\mathcal{I}'$, as a neighbourhood of infinity, whereas the
others may be seen as regions which are influenced by some electric charges (which
themselves lie outside of the spacetime). The support of $\alpha$ separates all these
regions, and we may interpret $F_{\alpha}$ as an observable which exploits Gauss' law to
measure the electromagnetic flux through a surface that separates the regions with charge
from the neighbourhood of infinity.

\begin{figure}
\centering
\includegraphics[scale=0.4]{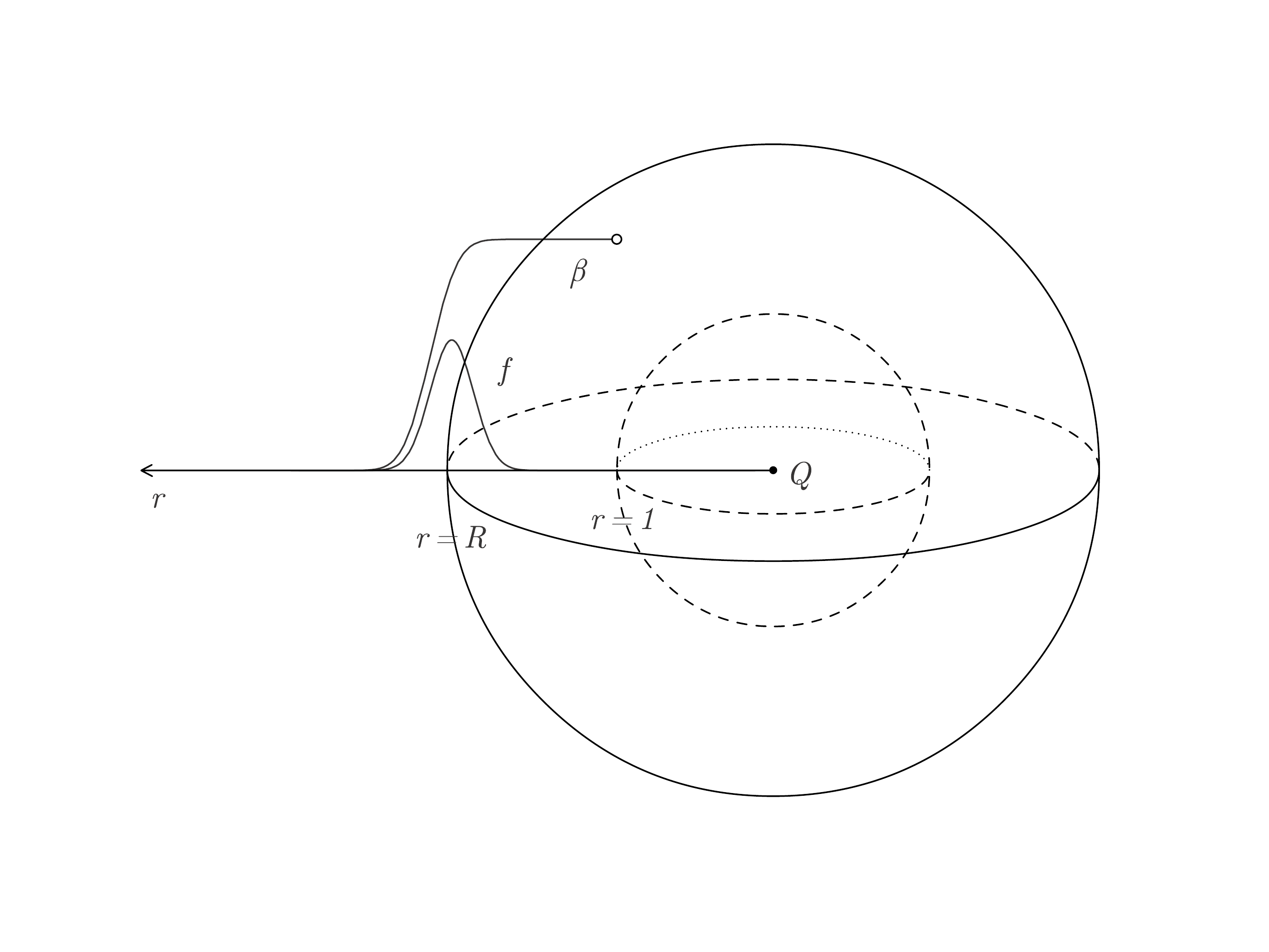}
\caption{An illustration of Gauss' law, as described in Example \ref{Ex_Gauss}. The observable
$\alpha=d\beta=f(r)dr$, supported near a sphere of radius $R$ around the locus of the electric
point charge $Q$, essentially measures the electric flux through the surface of that sphere.}
\label{Fig_Gauss}
\end{figure}

We have already seen a concrete example of a non-trivial $\mathcal{C}^1(\Sigma)$ in Example
\ref{Ex_Exactness}, where $\Sigma\simeq\mathbb{R}$ and the observable is given by $\alpha=f(r)dr$
for some $f\in C_0^{\infty}(\mathbb{R})$. We now elaborate the relationship between this example
and Gauss' law:
\begin{example}\label{Ex_Gauss}
The following example is illustrated in Figure \ref{Fig_Gauss}. Let $M_0$ be $n$-dimensional
Minkoswki spacetime, $n\ge 3$, and let $\Sigma_0:=\left\{t=0\right\}$ for some inertial time
coordinate $t$. Define $M:=D(\Sigma)\subset M_0$, with $\Sigma:=\Sigma_0\setminus B_1$, where
$B_1$ is the closed unit ball. (The case $n=2$ requires slight modifications, as $\Sigma$ would
be disconnected.) Then $\Sigma$ is a Cauchy surface for $M$ and
$\Sigma\simeq\mathbb{R}_{>1}\times\mathbb{S}^{n-2}$. We consider the $1$-form $\alpha=f(r)dr$,
where $f\in C_0^{\infty}((1,\infty))$ and $r\in\mathbb{R}_{>1}$ is the radial coordinate on
$\Sigma\subset\Sigma_0$. In analogy to Example \ref{Ex_Exactness}, $\alpha\in\Omega^1_0(\Sigma)$
and $\alpha=-d\beta$ with $\beta(r):=\int_r^{\infty}f(s)ds$. $\beta$ vanishes near $r=\infty$
and it is constant in a neighbourhood of the (removed) unit ball $B_1$. $\beta$ is compactly
supported if and only if $\int f=0$.

This example can be extended from the Cauchy surface $\Sigma$ to the spacetime $M$ as follows.
Let $\tau\in\Omega^0_0(\mathbb{R})$ with $\int\tau=1$ and support sufficiently close to the
origin to ensure that $\alpha'(r,t):=\alpha(r)\wedge\tau(t)dt$ is in $\Omega^2_0(M)$. We have
$\alpha'=d\beta'$ with $\beta'(r,t):=-\beta(r)\wedge\tau(t)dt$ and $\beta'$ is compactly supported
if and only if $\int f=0$. Now consider the observable $\nu:=-\delta\alpha'=-\delta d\beta'$.
Note that $\nu\in\delta d\Omega^1_{tc}(M)$, so $\nu\in\mathcal{C}^1(M)$. We will show that $\nu$
is not trivial, i.e.\ $\nu\not\in\delta d\Omega^1_0(M)$. For this purpose we consider the field
configuration $A_Q:=\frac{Q}{(3-n)c_{n-2}r^{n-3}}dt$ when $n\ge 4$, or
$A_Q:=\frac{Q}{2\pi}\log(r)dt$ when $n=3$, with $Q\in\mathbb{R}$ and $c_{n-2}$ is the volume of
the unit sphere $\mathbb{S}^{n-2}$. One may verify that $A_Q\in\mathcal{S}^p_0$ is a (Lorenz
gauge) solution to the Maxwell equations without source. In fact, it is the field generated by a
point charge at the origin in $M_0$, but the region of charge has been removed from $M$. Direct
computations now show that $dA_Q=\frac{Q}{c_{n-2}r^{n-2}}dr\wedge dt$ and
\[
(\nu,*A_Q)=(\alpha',*dA_Q)=\int_M f(r)\tau(t)\frac{Q}{c_{n-2}r^{n-2}}d\mathrm{vol}_M
=\int_{\Sigma} f(r)\frac{Q}{c_{n-2}r^{n-2}}d\mathrm{vol}_{\Sigma}=Q\int f.
\]
This proves that $\nu\not=0\in(\mathcal{S}^p_0)^*$, by Proposition \ref{Prop_S0*}. Moreover, the
final equation exhibits the relation between the form $\nu$ and Gauss' law.
\end{example}

The discussion above and these examples motivate us to make the following definition
\begin{definition}\label{Def_GaussObs}
An observable $\alpha\in\mathcal{C}^1(M)$ is called an \emph{(external) electric monopole
observable}. We call a field configuration $A\in\mathcal{F}^1(M)$ \emph{free of (external)
electric monopoles} if and only if $F_{\alpha}(A)=0$ for all $\alpha\in\mathcal{C}^1(M)$.
\end{definition}

\begin{remark}\label{Rem_MagMonopoles}
Our theory does not contain any magnetic monopole observables, because $F=dA$ is always
exact. To obtain such magnetic monopoles one could e.g.\ directly quantise the theory for
$F$ (see \cite{DL} and Remark \ref{Rem_F} below), or one can use a $3$-form field $B$
such that $F=\delta B$ and with a gauge equivalence based on co-closed or co-exact forms.
In these cases $F$ can be closed without being exact and magnetic monopoles can occur. On
the other hand, the theory would no longer be able to describe the Aharonov-Bohm effect.
Alternatively, one may obtain magnetic monopoles by quantising a theory of principal
$U(1)$-connections \cite{BDS2}, or by adding by hand a space of central magnetic monopole
observables, indexed e.g.\ by a basis of a suitable cohomology group. The latter approach,
however, is somewhat ad hoc and it does not seem amenable to the geometric techniques that
we advocate, using Lagrangians and Peierls' method. This means in particular that any
choice of a space of extra central observables cannot be motivated from a geometric
analysis similar to the space of electric monopole observables (at least not without
reverting to other theories, e.g.\ the one based on $F$).
\end{remark}

Our interpretation offers a nice explanation for the fact that $\mathcal{C}^p(M)$ is trivial when
$M$ has compact Cauchy surfaces. Namely, such a spacetime can only be isometrically embedded in
one with a diffeomorphic Cauchy surface. Thus, in particular, it is not possible to embed $M$
into a spacetime with an electric charge located outside of the image of the embedding.

For future convenience we make here a remark, which is closely related to the previous example of
Gauss' law:
\begin{remark}\label{Rem_Embed}
Consider two embeddings of $\Sigma:=\mathbb{R}\times\mathbb{S}^{n-2}$, $n\ge 3$, into other
manifolds. The first is an embedding $\map{\psi_1}{\Sigma}{\mathbb{R}^{n-1}}$, which is defined
as the identity in polar coordinates. The second is an embedding
$\map{\psi_2}{\Sigma}{\mathbb{S}^1\times\mathbb{S}^{n-2}}$, where we used an embedding
$\mathbb{R}\rightarrow\mathbb{S}^1$ in the first factor. Now consider the compactly supported
1-form $\alpha$ of Example \ref{Ex_Gauss} on $\Sigma$ with $\int f\not=0$, and its push-forwards
$\alpha_i:=(\psi_i)_*\alpha$. Because the exterior derivative commutes with the push-forward, both
$\alpha_i$ are closed. Recall that $\alpha\in d\Omega^0(\Sigma)$, but
$\alpha\not\in d\Omega^0_0(\Sigma)$. For the $\alpha_i$ this is different:
\begin{enumerate}
\item $\alpha_1\in d\Omega^0_0(\mathbb{R}^{n-1})$.
\item $\alpha_2\not\in d\Omega^0(\mathbb{S}^1\times\mathbb{S}^{n-2})$.
\end{enumerate}
The first statement follows from the fact that $H^1_0(\mathbb{R}^{n-1})$ is trivial for $n\ge 3$.
For the second statement we argue by contradiction and suppose that $\alpha_2=d\beta_2$ for some
function $\beta_2$ on $\mathbb{S}^1\times\mathbb{S}^{n-2}$. Note that the complement of the range
of $\psi_2$ is connected, and that $\beta_2$ is constant there. Without loss of generality we may
assume that $\beta_2$ vanishes there, so it follows that $\beta:=\psi_2^*(\beta_2)$ satisfies
$d\beta=\alpha$ and $\beta$ vanishes outside a compact set. However, we know from the Examples
\ref{Ex_Exactness} and \ref{Ex_Gauss} (and from the discussion above Example \ref{Ex_Gauss}) that
this cannot be the case, as $\int f\not=0$.
\end{remark}

To close this subsection we provide an example concerning degenerate observables for $p=2$,
indicating why an interpretation in terms of charge is more complicated in that case.
\begin{example}
For $p>1$ one may easily find degenerate observables by generalising Example \ref{Ex_Gauss},
taking $\Sigma=\mathbb{R}_{>0}\times\mathbb{R}^{\times p-1}\times\mathbb{S}^{n-1-p}$ and
$\alpha=d^{\Sigma}\beta$ with
$\beta =(\int_1^rf)f(x^2)\cdots f(x^p)dx^2\wedge\ldots\wedge dx^p$, where the
$x^i$ are Cartesian coordinates on $\mathbb{R}^{p-1}$. Then $\alpha$ has compact support and
$\beta$ has compact support if and only if $\int_1^{\infty} f=0$.

Perhaps a more interesting example for $p=2$ can be obtained by taking
$\Sigma:=\mathbb{R}^2\setminus\left\{x_1,x_2\right\}$, where the $x_i$ are two distinct
points. We let $V_0\subset\mathbb{R}^2$ denote the complement of a closed ball which contains
both $x_i$ and we let $V_1,V_2\subset\mathbb{R}^2$ be punctured open balls around $x_1,x_2$,
respectively, such that the closures of $V_0,V_1,V_2$ are pairwise disjoint. We will view
$V_0,V_1,V_2$ as subsets of $\Sigma$ and we note that their topologies are all equal to
$\mathbb{R}\times\mathbb{S}^1$. For $i=0,1,2$ we may choose $\eta_i\in\Omega^1_{0,d}(V_i)$
which is not exact, so that $[\eta_i]$ generates $H_0^1(V_i)$, and we may choose
$\beta_i\in\Omega^1(\Sigma)$ such that $(\eta_i,\beta_j|_{V_i})=\delta_{ij}$, the
Kronecker delta.

For any given constants $b_i\in\mathbb{R}$ we can now construct a one-form
$\beta\in\Omega^1(\Sigma)$ such that $\beta|_{V_i}=b_i\beta_i$, simply by using a
suitable partition of unity. Setting $\alpha:=d^{\Sigma}\beta$ we see that
$\alpha\in\Omega^2_0(\Sigma)$ has support in the compact set $V_0^c$. We may wonder
whether $\alpha=d^{\Sigma}\beta'$ for some $\beta'\in\Omega^1_0(\Sigma)$ with compact
support. Now note that $[\gamma]\in H^1(\Sigma)$ is uniquely determined by
$c_i:=(\eta_i,\gamma)$ with $i=1,2$ only. Moreover, the complement of any compact set
$K\subset\Sigma$ will contain representatives of all three $[\eta_i]$. Hence, for
$\beta-\gamma$ to be exact on $K^c$ we would need $(\eta_i,\beta-\gamma)=0$ for
$i=0,1,2$. We can choose $\gamma$ to ensure this equality for $i=1,2$, but the remaining
equality puts a necessary restriction on $\beta$. Indeed, in $\Sigma$ the three
$[\eta_i]$ are linearly dependent, say $[\eta_2]=[\eta_0]-[\eta_1]$. A short computation
then shows that the necessary condition for $[\alpha]=0$ is $b_2=b_0-b_1$. This
condition is also sufficient.

Note that the equation $b_2=b_0-b_1$ involves all three constants $b_i$, which changes
the interpretation somewhat. If we identify $V_0$ as a neighbourhood of infinity we
may choose $\gamma$ such that $(\eta_0,\beta-\gamma)=0$, meaning that there is no
charge at infinity. Replacing $\beta$ by $\tilde{\beta}:=\beta-\gamma$ we are left with
the condition that $\tilde{b}_2=-\tilde{b}_1$, where $\tilde{b}_i=(\eta_i,\tilde{\beta})$.
In the analogous case for $p=1$ we would find conditions involving only one constant
$\tilde{b}_i$, which we could then interpret as a charge located at $x_i$, $i=1,2$.
In the present case, however, it seems we must attribute the charge instead to the union
of the two points.

Note that the situation above can also be formulated in $\mathbb{R}^3$, simply by adding
an extra dimension, removing two parallel lines, choosing $[\eta_i]\in H^2_0(V_i)$, etc.
If we would instead remove a circle and a line from $\mathbb{R}^3$, then the linear
dependence between the $[\eta_i]$ would only involve two of these classes and we would
obtain an interpretation in terms of a charge located on the line. On the other hand, if
we remove two circles, then the three $[\eta_i]$ would be linearly independent, so no
charge is present. This seems to be independent of whether the removed circles are
knotted or linked in any way.
\end{example}

\section{The Quantised $p$-Form Field and Field Strength}\label{Sec_Quantum}

After studying the classical dynamics of the $p$-form field in the presence of a given
background current $j$, we now discuss the corresponding quantum theory. In the case where $j=0$
we can directly quantise the linear Poisson space and let the Poisson bracket correspond to a
commutator between operators in the usual way. In the general case, however, we only have an
affine Poisson space. For this reason we will first discuss a general method for quantising
affine Poisson spaces, which can be viewed as a special case of Fedosov's quantisation scheme
from the theory of deformation quantisation \cite{Wa}.

\subsection{The quantisation of affine Poisson spaces}\label{SSec_QAffine}

In this subsection we consider a real affine space $V$, modeled over a real linear topological
vector space $V_0$. (Requiring a topology is no loss of generality, because one may always choose
the discrete topology.) This means that for every $x\in V$ there is an affine bijection
$\map{e_x}{V_0}{V}$ such that
\[
e_x(0)=x,\qquad e_x(v)=e_y(v+e_y^{-1}(x))\ \forall y\in V.
\]
We equip $V$ with the unique topology that makes each $e_x$ a homeomorphism, so we may view
$e_x$ as a coordinate chart covering the affine manifold $V$. We denote by $V^*_0$ the
topological dual of $V_0$ and we denote the space of continuous affine maps on $V_0$ by
$V'_0\simeq\mathbb{R}\oplus V_0^*$. We may identify the tangent and cotangent bundles of $V$ with
\[
TV\simeq V\times V_0,\quad T^*V\simeq V\times V^*_0.
\]
In particular we can use the derivative $D_0e_x$ at $0$ to identify $T_0V_0\simeq V_0$ with
$T_xV$ in a canonical way, so there is an isomorphism $\map{e_x\circ(D_0e_x)^{-1}}{T_xV}{V}$. We
define a canonical affine connection on $TV$ using the maps $\map{\sigma_{yx}}{T_xV}{T_yV}$
defined by $\sigma_{yx}:=D_0e_y\circ e_y^{-1}\circ e_x\circ(D_0e_x)^{-1}$. This affine connection
is characterised by $(x,D_0e_xv)\mapsto (y,D_0e_y(v+e_y^{-1}x))$.
Because the connection is only affine and not linear it will be convenient to introduce the bundle
$T'V\simeq V\times V'_0$, so that $T'_xV$ is the space of continuous affine maps on $T_xV$.

Now assume that there is a Poisson structure on $V$, i.e.\ at each $x\in V$, $P_x$ is an
antisymmetric, bilinear map on $T^*_xV$. Equivalently, $P_x$ can be viewed as a bilinear map on
$T'_xV$ which vanishes when at least one of the arguments is a constant map. For any $x\in V$ we
can pull-back $P_x$ to a bilinear map $\pi_x$ on $V'_0$ (or on $V^*_0$),
$\pi_x(\xi,\eta):=P_x(\xi\circ (D_0e_x)^{-1},\eta\circ (D_0e_x)^{-1})$ for all $\xi,\eta\in V'_0$.
We assume in addition that the Poisson structure is covariantly constant with respect to the
canonical affine connection on $TV$,
i.e.\ $P_y(\xi,\eta)=P_x(\xi\circ\sigma_{yx},\eta\circ\sigma_{yx})$ for all $\xi,\eta\in T'_yV$.
The covariant constancy is equivalent to
$\pi_x(\xi,\eta)=\pi_y(\xi\circ e_x^{-1}\circ e_y,\eta\circ e_x^{-1}\circ e_y)=\pi_y(\xi,\eta)$
for all $x,y\in V$ and all $\xi,\eta \in V'_0$, where we used the fact that $\pi_y$ is bilinear
and vanishes when one of the two arguments is a constant. In other words, $\pi_x=\pi_y$ for all
$x,y\in V$. A further equivalent formulation uses the identification of $T_xV$ with $V$ by
$e_x\circ (D_0e_x)^{-1}$ to define an antisymmetric, bilinear map $P$ on $V'$ which vanishes when
one of the arguments is a constant map. Note that $P(\xi,\eta)=\pi(\xi\circ e_x,\eta\circ e_x)$
for any $x\in V$.

\begin{definition}\label{Def_AffPoisson}
By an \emph{affine Poisson space} we mean a pair $(V,P)$, where $V$ is an affine space, modeled
over a topological vector space $V_0$, and $P$ is an antisymmetric bilinear map on $V'$. We say
that $(P,V)$ is modeled over the linear Poisson space $(V_0,\pi)$, where $\pi$ is the
push-forward of $P$ by any of the canonical isomorphisms $\map{e_x}{V_0}{V}$, $x\in V$.
\end{definition}

\begin{remark}
If $V_0$ is finite-dimensional, the specification of a non-degenerate antisymmetric bilinear map
$\pi$ on $V^*_0$ is equivalent to the specification of a (non-degenerate) symplectic form $\sigma$
on $V_0$, using the linear isomorphism
$V^*_0\rightarrow V_0^{**}\simeq V_0: \xi\mapsto \pi(\xi,.)$. However, when $\sigma$ or $\pi$ is
degenerate, or when the space $V_0$ is infinite dimensional, the situation is more complicated.
Both complications apply for the vector potential and its $p$-form generalisations.
\end{remark}

For the linear Poisson space $(V_0,\pi)$ it is well-known how to construct the corresponding
quantum theory. One can construct a Weyl $C^*$-algebra $\mathscr{W}_0$, which is generated by
linearly independent Weyl operators $W(\xi)$, $\xi\in V_0^*$ satisfying the Weyl relations
\[
W(\xi)W(\eta)=e^{-\frac{i}{2}\pi(\xi,\eta)}W(\xi+\eta),\quad W(\xi)^*=W(-\xi).
\]
Note that this also works in the case where $\pi$ is degenerate \cite{BHR}. The operators
$W(\xi)$ are unitary and degenerate elements $\xi\in V_0^*$ generate the centre of
$\mathscr{W}_0$. We will mostly be interested in the infinitesimal Weyl algebra,
$\mathscr{A}_0$, which is the $^*$-algebra generated by an identity operator $I$ and the
operators $\Phi(\xi)$ satisfying the relation
\[
[\Phi(\xi),\Phi(\eta)]:=\Phi(\xi)\Phi(\eta)-\Phi(\eta)\Phi(\xi)=i\pi(\xi,\eta)I.
\]
It may equivalently be described as a $^*$-algebra of functions on $V_0$, which are symmetric
polynomials in elements of $V_0^*$, using a deformed (Weyl-Moyal) $\star$-product. For
monomials $F:=\xi^l$ and $G:=\eta^m$ with $\xi,\eta\in V_0^*$ this product is given by
(cf.\ \cite{Wa})
\[
F\star G=\sum_{n=0}^{\mathrm{min}(l,m)}\frac{i^n}{n!2^n}\pi(\xi,\eta)^n
\frac{l!m!}{(l-n)!(m-n)!}\xi^{l-n}\circ_s\eta^{m-n},
\]
where $\circ_s$ denotes the symmetrised product. Using $\xi\star\eta-\eta\star\xi=i\pi(\xi,\eta)$
and identifying $\Phi(\xi)$ with the linear map $\xi$ we recover
$[\Phi(\xi),\Phi(\eta)]=i\pi(\xi,\eta)$. Note that symmetric monomials of the form $\xi^l$
generate the linear space of all symmetric polynomials.

For the affine Poisson space $(V,P)$ we will construct an analogous algebra, using a prescription
that can be viewed as an application of Fedosov's quantisation technique from the theory of
deformation quantisation \cite{Wa}. For any $x\in V$ we let $\mathscr{A}_x$ be the infinitesimal
Weyl algebra of the cotangent space $T_x^*V$, in the Poisson bracket $P_x$ at $x\in V$. It may
be interpreted as a quantisation of the perturbations of the system, around the fixed point
$x\in V$. Taking the algebras for all $x\in V$ together, one may form a bundle $\mathcal{A}$ of
$^*$-algebras over $V$ and then consider the algebra $\tilde{\mathscr{A}}$ of sections of
$\mathcal{A}$, where algebraic operations are preformed pointwise in $\mathscr{A}_x$. Of course
the algebra $\tilde{\mathscr{A}}$ is too big for physical purposes, because it does not take into
account the relations between the algebras $\mathscr{A}_x$ at different points $x\in V$. To
remedy this defect, Fedosov's construction suggests to find a connection on the bundle
$\mathcal{A}$, suitably adapted to the Poisson structure $P$, and to consider the subalgebra
$\mathscr{A}$ of $\tilde{\mathscr{A}}$ generated by covariantly constant sections. We will now
show that for an affine Poisson space such a connection on $\mathcal{A}$ can be found, using the
canonical affine connection of $V$.

Consider two points, $x,y\in V$, and the affine isomorphism $\sigma_{yx}$ between $T_xV$ and
$T_yV$. We view $\mathscr{A}_y$ as the space of functions on $T_yV$ which can be written as
symmetric linear polynomials in $V_0^*$, endowed with a $\star$-product. Pulling back these
functions by $e_y^{-1}e_x$ yields a linear isomorphism onto $\mathscr{A}_x$, which we denote by
$\alpha_{xy}$. Under this isomorphism, a generator $\Phi_y((D_0e_y)_*\xi)$ at $y\in V$, with
$\xi\in V^*_0$, gets mapped to
\[
\alpha_{xy}(\Phi_y((D_0e_y)_*\xi))=\Phi_x((D_0e_x)_*\xi)+\xi(e_y^{-1}(x))I,
\]
because $((D_0e_y)_*\xi)\circ\sigma_{yx}=(D_0e_x)_*\xi+\xi(e_y^{-1}(x))$. One may verify that
$e_x^{-1}(y)=-e_y^{-1}(x)$, so that
$\sigma_{yx}=\sigma_{xy}^{-1}$ and $\alpha_{yx}=\alpha_{xy}^{-1}$. Furthermore,
$\alpha_{xy}\alpha_{yz}=\alpha_{xz}$ for all $x,y,z\in V$ and, using the explicit form for the
$\star$-product,
\[
\alpha_{xy}(F)\star\alpha_{xy}(G)=\alpha_{xy}(F\star G).
\]
As $x,y$ range over $V$, the $^*$-isomorphisms $\alpha_{xy}$ piece together the desired connection
$\alpha$ of the algebra bundle $\mathcal{A}$. A section of this bundle,
$F(x)\in\tilde{\mathscr{A}}$, is covariantly constant with respect to $\alpha$ if and only if
$F(x)=\alpha_{xy}(F(y))$ for all $x,y\in V$. The subalgebra $\mathscr{A}$ that we obtain from
Fedosov's quantisation method is therefore isomorphic to any $\mathscr{A}_x$.

The algebra $\mathscr{A}$ can also be constructed in a more direct and invariant way as follows.
We consider the algebra $\mathscr{A}$ of functions on $V$ that can be written as antisymmetric
polynomials in elements of $V'$, with a $\star$-product based on the antisymmetric bilinear map
$P$ on $V'$. Equivalently, we may define $\mathscr{A}$ as follows:
\begin{definition}\label{Def_PoissAlg}
For an affine Poisson space $(V,P)$ we define the algebra $\mathscr{A}(V,P)$, generated by
operators $\Phi(\chi)$, $\chi\in V'$, satisfying
\begin{enumerate}
\item $\Phi(1)=I$, where $1$ is the constant function on $V$ and $I$ the identity operator,
\item $\chi\mapsto\Phi(\chi)$ is linear,
\item $\Phi(\chi)^*=\Phi(\chi)$ (as $\chi$ is real-valued), and
\item $[\Phi(\chi),\Phi(\psi)]=iP(\chi,\psi)I$.
\end{enumerate}
\end{definition}

To see the relation between the generators $\Phi(\chi)$ of $\mathscr{A}(V,P)$ and $\Phi_x(\xi)$
of $\mathscr{A}_x$ for any $x\in V$, we view these algebras as linear spaces of functions on $V$,
resp.\ $T_xV$. The pull-back under $\map{e_x\circ (D_0e_x)^{-1}}{T_xV}{V}$ provides a linear
isomorphism $\map{\alpha_x}{\mathscr{A}(V,P)}{\mathscr{A}_x}$. For any $\chi\in V'$ we may define
the linear part $\xi_{\chi}$ of $\chi$ by $\xi_{\chi}(v):=(e_x^*\chi)(v)-(e_x^*\chi)(0)$. This is
a continuous linear map on $V_0$, which is independent of the choice of $x\in V$, because
(indicating the possible dependence on $x$ in the subscript)
\begin{eqnarray}
\xi_{\chi,x}(v)&=&\chi(e_x(v))-\chi(x)=(e_y^*\chi)(v+e_y^{-1}(x))-(e_y^*\chi)(e_y^{-1}(x))
\nonumber\\
&=&\xi_{\chi,y}(v+e_y^{-1}(x))-\xi_{\chi,y}(e_y^{-1}(x))=\xi_{\chi,y}(v).\nonumber
\end{eqnarray}
The map $\alpha_x$ then satisfies
\[
\alpha_x(\Phi(\chi))=\Phi_x((D_0e_x)_*\xi_{\chi})+\chi(x)I.
\]
One may directly verify that $\alpha_x$ is a $^*$-algebra isomorphism, because
$\alpha_x(F)\star\alpha_x(G)=\alpha_x(F\star G)$. Also note that
$\alpha_{xy}=\alpha_x\circ\alpha_y^{-1}$ by construction.

\begin{remark}
A $C^*$-algebraic formulation exists along similar lines, because the maps $\sigma_{xy}$ give
rise to Bogolyubov transformations of the second kind on the Weyl $C^*$-algebras $\mathscr{W}_x$
of each tangent space $T_xV\simeq V_0$, cf.\ \cite{BHR}.
\end{remark}

\subsection{Quantising the $p$-form fields}

The procedure described in the previous subsection applies in particular to the $p$-form fields,
for which we have seen that $\mathcal{S}^p_j$ is an affine Poisson space, modeled on the linear
Poisson space $\mathcal{S}^p_0$ with the Poisson bracket $\left\{,\right\}$.

Note that the notation can be made a bit more concrete by realising that the affine space
$\mathcal{S}^p_j$ is a subspace of the linear space $\mathcal{F}^p(M)$ and by using the maps
$e_{[A]}$ (cf.\ Subsection \ref{SSec_Soln}). Any continuous affine map $\chi$ in
$(\mathcal{S}^p_j)'$ can be extended to a continuous affine map on $\mathcal{F}^p(M)$ as follows.
First we may extend the linear part $\xi_{\chi}$ from $\mathcal{S}^p_0$ to $\mathcal{F}^p(M)$, so
there is an $\alpha\in\mathcal{F}^p(M)^*$ such that $\xi_{\chi}([A])=F_{\alpha}(A)$ for all
$[A]\in\mathcal{S}^p_0$. Then we may define the affine map $\chi_{\alpha}$ on $\mathcal{F}^p(M)$ by
\[
\chi_{\alpha}([A']):=\chi([A])+F_{\alpha}(A'-A)=(\chi([A])-F_{\alpha}(A))+F_{\alpha}(A')
\]
for some given $[A]\in\mathcal{S}^p_j$. The extension is not unique, because
$\chi_{\alpha}=\chi_{\alpha'}$ if and only if $\alpha-\alpha'\in\delta d\Omega^p_0(M)$
(cf.\ Proposition \ref{Prop_S0*}).

Following the results of Subsection \ref{SSec_QAffine} we may now describe the quantum $p$-form
field. Instead of the space $(\mathcal{S}^p_j)'$ we may consider
$\mathcal{F}^p(M)'\simeq\mathbb{R}\oplus\mathcal{F}^p(M)^*$, but we must divide out the
equivalence relation
\[
(c,\alpha)\sim 0\quad\Leftrightarrow\quad \alpha=\delta d\beta,\ \beta\in\Omega^p_0(M),\
c=-F_{\alpha}(A)=-(j,*\beta),
\]
where $A\in\mathcal{S}^p_j$ is arbitrary. This leads to the following
\begin{definition}\label{Def_Aj}
The (on-shell) $p$-form \emph{field algebra} is the $^*$-algebra $\mathscr{A}_j(M)$ generated by
a unit $I$ and the symbols $\mathbf{A}(\alpha)$, $\alpha\in\mathcal{F}^p(M)^*$, subject to the
relations
\begin{enumerate}
\item $\alpha\mapsto\mathbf{A}(\alpha)$ is linear,
\item $\mathbf{A}(\alpha)^*=\mathbf{A}(\alpha)$, (as $\alpha$ and $\mathbf{A}$ are real-valued),
\item $\left[\mathbf{A}(\alpha),\mathbf{A}(\beta)\right]=i\left\{F_{\alpha},F_{\beta}\right\}I$,
\item $(\delta d\mathbf{A})(\beta)=(j,*\beta)I$ for all $\beta\in\Omega^p_0(M)$.
\end{enumerate}
\end{definition}
Note that the constant maps $(c,0)$ give rise to multiples of the identity, whereas the linear
maps $(0,\alpha)$ give rise to the fields $\mathbf{A}(\alpha)$. The equivalence relation
$(c,\alpha)\sim 0$ gives rise to the equation of motion. For the homogeneous case, where $j=0$,
this simply reduces to the usual infinitesimal Weyl algebra $\mathscr{A}_0$.

Just like in Subsection \ref{SSec_QAffine} we may construct for each $[A]\in\mathcal{S}^p_j$ a
unique $^*$-isomorphism $\map{\alpha_{[A]}}{\mathscr{A}_j}{\mathscr{A}_0}$ which preserves the
unit and satisfies
\[
\alpha_{[A]}(\mathbf{A}(\alpha))=F_{\alpha}(A)I+\mathbf{A}_0(\alpha),
\]
i.e.\ $\alpha_{[A]}(\mathbf{A})(x):=A(x)I+\mathbf{A}_0(x)$, where we wrote the field operators
that generate $\mathscr{A}_0$ with a subscript $0$ for distinction. This $^*$-isomorphism can be
used to pull back states on $\mathscr{A}_0$ to states on $\mathscr{A}_j$. When $A$ is smooth, the
microlocal spectrum condition is preserved under pull-back by $e_{[A]}$ (cf.\ \cite{BDS}).

One may define the field strength $\mathbf{F}:=d\mathbf{A}$ also in the quantum case. More
precisely, we consider the field strength observables $\alpha=\delta\beta$ with
$\beta\in\Omega^{p+1}_0(M)$ and $\mathbf{F}(\beta):=-\mathbf{A}(\delta\beta)$. Together with the
unit $I$ these operators generate a subalgebra $\mathscr{F}_j(M)$ of $\mathscr{A}_j(M)$ such that
\begin{enumerate}
\item $\beta\mapsto\mathbf{F}(\beta)$ is linear,
\item $\mathbf{F}(\beta)^*=\mathbf{F}(\beta)$, (as $\beta$ is real-valued),
\item $\left[\mathbf{F}(\beta),\mathbf{F}(\gamma)\right]=
i\left\{F_{\delta\beta},F_{\delta\gamma}\right\}I$,
\item $(\delta\mathbf{F})(\gamma)=j(\gamma)I$ for all $\gamma\in\Omega^p_0(M)$.
\end{enumerate}
The homogeneous Maxwell equations $d\mathbf{F}=0$ already follow from the definition
$\mathbf{F}=d\mathbf{A}$.

\begin{remark}\label{Rem_F}
Note that the algebra $\mathscr{F}_j(M)$ differs from the field strength algebra considered in
\cite{DL}, even for $j=0$ and $p=1$, because our approach only considers observables on field
strength configurations that are derived from a vector potential, whereas \cite{DL} constructs an
algebra directly for $F$. This is related to our ability to accommodate the Aharonov-Bohm
effect and to our interpretation of electromagnetism as a theory for the connections of a trivial
principal $U(1)$-bundle. We rely heavily on the existence of a Lagrangian formulation and on
Peierls' method to derive the Poisson bracket. For this reason a direct application of our
results to analyse the centre of the field strength algebra of \cite{DL} is not straightforward.
\end{remark}

Proposition \ref{Prop_Degeneracy} implies the following result on the centre of $\mathscr{A}_j(M)$:
\begin{corollary}\label{Cor_centreA}
The centre of $\mathscr{A}_j(M)$ is generated by $I$ and
$\mathbf{F}(\beta)=-\mathbf{A}(\delta\beta)$ with $-\delta\beta\in\mathcal{C}^p(M)$.
\end{corollary}
A similar statement can be derived for the field strength algebra, whose centre is generated by
$I$ and $\mathbf{F}(\beta)$ with $\beta\in H^{p+1}_0(M)$.
This may be compared to \cite{DL}, who quantised the field strength tensor for $p=2$ directly
and whose results imply that the centre can be written as
\[
\frac{\Omega^2_{0,d}(M)+\Omega^2_{0,\delta}(M)}{d\Omega^1_0(M)+\delta\Omega^3_0(M)}
\simeq H^2_0(M)+H^2_{0,\delta}(M).
\]
Note that a globally hyperbolic spacetime has no $\gamma\in\Omega^2_0(M)$ which is
simultaneously closed and co-closed, since $\Box\gamma=0$ implies $\gamma=0$. Hence, any
$\omega\in\Omega^2_{0,d}+\Omega^2_{0,\delta}$ can be written uniquely as $\omega=\alpha+\beta$
with $\alpha\in\Omega^2_{0,d}(M)$ and $\beta\in\Omega^2_{0,\delta}(M)$. Using this
decomposition in the numerator and denominator we see that the space of degeneracies found by
\cite{DL} is isomorphic to $H^2_0(M)+H^2_{0,\delta}(M)$.

\subsection{General covariance and locality}\label{SSec_LC}

The occurrence of a non-trivial centre as in Corollary \ref{Cor_centreA} was first established by
\cite{DL} in their model of the field strength tensor, although the centre found in \cite{DL} is
larger than ours, as it also contains observables of the field strength tensor which are not
observables of the vector potential. Those authors also realised that the presence of this
non-trivial centre
implies that local covariance, in the sense of \cite{BFV}, fails. As we will see in this
subsection, this lack of local covariance is really only a lack of locality. Moreover, the lack
of locality is not surprising, because it already occurs at the classical level in the form of
Gauss' law. We will now use our careful computation of the centre to show that the lack of
locality at the quantum level can be traced back to the same source, thereby achieving the main
goal of this paper, namely to clarify the topological origins of the non-local (Gauss' law)
observables in the quantum theory.

Following the seminal paper \cite{BFV} we may analyse the general covariance of our theory in
a categorical framework. For this purpose we introduce the following categories:
\begin{definition}
\begin{itemize}
\item $\mathfrak{SpacCurr}$ is the category whose objects are triples $(\mathcal{M},g,j)$,
where
$M=(\mathcal{M},g)$ is a globally hyperbolic spacetime and $j\in\Omega^p_{\delta}(M)$, and whose
morphisms are orientation and time orientation preserving embeddings $\map{\psi}{M}{\tilde{M}}$
such that $\psi^*\tilde{\jmath}=j$, $\psi^*\tilde{g}=g$ and $\psi(M)\subset\tilde{M}$ is causally
convex (i.e.\ $\psi^{-1}(\tilde{J}^{\pm}(\psi(p)))=J^{\pm}(p)$ for all $p\in M$).
\item $\mathfrak{Alg}$ is the category whose objects are unital $^*$-algebras and whose
morphisms are unit preserving $^*$-homomorphisms.
\item $\mathfrak{Alg}'$ is the subcategory of $\mathfrak{Alg}$ with the same objects and only
injective morphisms.
\end{itemize}
\end{definition}
The inclusion of $j$ in the background structure is a natural modification of the framework of
\cite{BFV}. The original considered the category $\mathfrak{Spac}$ (cf.\ Definition
\ref{Def_Spac}), which is the full subcategory of $\mathfrak{SpacCurr}$ consisting of objects
with $j=0$. (Alternatively, there is a forgetful functor from $\mathfrak{SpacCurr}$ to
$\mathfrak{Spac}$.) Instead of $\mathfrak{Alg}$ one may consider topological $^*$-algebras (with
continuous morphisms) or $C^*$-algebras, but a choice of topology is not very relevant for our
current investigation and has been omitted.

The idea that physical theories should depend in a generally covariant way on the background
structure can now be stated in a concise way as follows:
\begin{definition}
A \emph{generally covariant quantum field theory with background current} is a covariant functor
$\mathfrak{A}:\mathfrak{SpacCurr}\to\mathfrak{Alg}$. This theory is \emph{locally covariant} if
and only if its range is contained in $\mathfrak{Alg}'$.
\end{definition}
The definition of locally covariant quantum field theory is directly analogous to that of
\cite{BFV} (which has no background current and uses $\mathfrak{Spac}$ instead of
$\mathfrak{SpacCurr}$). The slightly more general notion of generally covariant quantum field
theory is new. It has been introduced to accommodate the quantum $p$-form fields, as we will see
shortly.

Let us first investigate the functorial behaviour of our quantisation scheme of affine Poisson
spaces. (This is in analogy to the functorial behaviour of the infinitesimal Weyl quantisation of
(pre)-symplectic spaces, cf.\ \cite{FV,BGP}.) For this purpose we need the following additional
category (cf.\ Def.\ \ref{Def_AffPoisson}):
\begin{definition}
$\mathfrak{AffPoiss}$ is the category whose objects are affine Poisson spaces $(V,P)$ and whose
morphisms are continuous affine maps $\map{L}{V}{\tilde{V}}$ such that $\tilde{P}=L^*P$,
i.e.\ $\tilde{P}(\xi,\eta)=P(\xi\circ L,\eta\circ L)$ for all $\xi,\eta\in \tilde{V}'$.
\end{definition}
We can then prove:
\begin{lemma}\label{Lem_Q}
There is a contravariant functor $\map{\mathfrak{Q}}{\mathfrak{AffPoiss}}{\mathfrak{Alg}}$ such
that $\mathfrak{Q}(V,P)$ is the algebra $\mathscr{A}(V,P)$ of Definition \ref{Def_PoissAlg} and
such that for any morphism $\map{L}{V}{\tilde{V}}$,
$\map{\mathfrak{Q}(L)}{\mathscr{A}(\tilde{V},\tilde{P})}{\mathscr{A}(V,P)}$ is the $^*$-algebraic
homomorphism determined by $\mathfrak{Q}(L)(\tilde{\Phi}(\chi)):=\Phi(\chi\circ L)$ for all
$\chi\in\tilde{V}$. Furthermore, $\mathfrak{Q}(L)$ is injective if and only if
$\map{L^*}{\tilde{V}'}{V'}$ is injective.
\end{lemma}
\begin{proof*}
The non-trivial part is to show that $\mathfrak{Q}(L)$ is well-defined, i.e.\ that it
respects the relations of the algebras $\mathscr{A}(V,P)$ and $\mathscr{A}(\tilde{V},\tilde{P})$.
This follows from the fact that the pull-back $L^*$ is linear, and preserves the Poisson
structure. (Alternatively one may show that $(V,P)\mapsto (V',P)$ is a contravariant functor from
$\mathfrak{AffPoiss}$ to a category of pre-symplectic spaces and compose this functor with the
quantisation functor of \cite{FV}.) For the last statement we deduce from
Def.\ \ref{Def_PoissAlg} that $\mathfrak{Q}(L)$ is injective if and only if it is injective on
the generators $\tilde{\Phi}(\chi)$, $\chi\in\tilde{V}'$, which is equivalent to injectivity of
$L^*$.
\end{proof*}

In order to describe the $p$-form theory as a generally covariant quantum field theory, it remains
to show that there is a contravariant functor
$\map{\mathfrak{P}}{\mathfrak{SpacCurr}}{\mathfrak{AffPoiss}}$, so that the theory can be
described by the composition $\mathfrak{Q}\circ\mathfrak{P}$. This is the subject of the following
result:
\begin{proposition}\label{Prop_GCQFT}
There is a contravariant functor $\map{\mathfrak{P}}{\mathfrak{SpacCurr}}{\mathfrak{AffPoiss}}$,
which is defined as follows. To each object $(\mathcal{M},g,j)$, $\mathfrak{P}$ assigns the affine
Poisson space $\mathfrak{P}(M,j):=\mathcal{S}^p_j(M)$ (cf.\ Section \ref{Sec_Poisson}). To each
morphism $\map{\psi}{(\mathcal{M},g,j)}{(\tilde{\mathcal{M}},\tilde{g},\tilde{\jmath})}$,
$\mathfrak{P}$ assigns the pull-back
$\map{\mathfrak{P}(\psi)}{\mathcal{S}^p_{\tilde{\jmath}}(\tilde{M})}{\mathcal{S}^p_j(M)}$, so that
$\mathfrak{P}(\psi)(\tilde{A}):=\psi^*\tilde{A}$.
\end{proposition}
\begin{proof*}
The non-trivial part is to show that the morphisms are well-defined. To see this we first note
that each isometric embedding $\map{\psi}{M}{\tilde{M}}$ yields a well-defined push-forward map
$\map{\psi_*}{\Omega^p_0(M)}{\Omega^p_0(\tilde{M})}$ so, by duality, there are well-defined
pull-back maps $\map{\psi^*}{\mathcal{D}^p(\tilde{M})}{\mathcal{D}^p(M)}$. Furthermore, $\psi^*$
commutes with exterior derivatives, so it maps exact forms to exact forms and hence descends to a
well-defined linear map $\map{\psi^*}{\mathcal{F}^p(\tilde{M})}{\mathcal{F}^p(M)}$. For a
morphism in $\mathfrak{SpacCurr}$, $\psi^*$ also respects the equations of motion, so it
restricts to a map $\map{\psi^*}{\mathcal{S}^p_j(\tilde{M})}{\mathcal{S}^p_j(M)}$.

It remains to show that $\psi^*$ intertwines the Poisson structures of the two affine spaces. At
an abstract level this follows from the fact that $\psi^*$ preserves the Lagrangian densities and
from the well-definedness of Peierls' method. In more detail we note that a continuous affine map
$F_{\alpha}$ on $\mathcal{S}^p_j(M)$, with $\alpha\in\Omega^p_{0,\delta}(M)$, gets mapped to
$F_{\alpha}\circ\psi^*=F_{\psi_*\alpha}$ with $\psi_*\alpha\in\Omega^p_{0,\delta}(\tilde{M})$.
(A general continuous affine map on $\mathcal{S}^p_j(M)$ is of the form $F_{\alpha}$, up to a
constant. Because the Poisson structures vanish on constants, it suffices to consider only the
$F_{\alpha}$.) Expressing the Poisson structure in terms of the advanced-minus-retarded
fundamental solution $G$ we need to verify that
\[
(\psi_*\alpha,*\tilde{G}\psi_*\beta)=(\alpha,*G\beta).
\]
It is well-known that this follows from the uniqueness of the advanced and retarded fundamental
solutions, together with the fact that the range $\psi(M)$ is causally convex in $\tilde{M}$ (see for example \cite[Ch. 3]{BGP}).
Together these imply that $\psi^*(\tilde{G}^{\pm}\psi_*\beta)=G^{\pm}\beta$ and using the fact
that $\psi$ is isometric the result easily follows.
\end{proof*}

Putting things together we find a functor $\mathfrak{A}:=\mathfrak{Q}\circ\mathfrak{P}$ from
$\mathfrak{SpacCurr}$ to $\mathfrak{Alg}$ which describes the $p$-form field in a generally
covariant way. Furthermore, one may show that $\mathfrak{A}$ is causal and satisfies the
time-slice axiom, by which we mean the following:
\begin{definition}
We say that a generally covariant quantum field theory $\mathfrak{A}$ with background current is
\emph{causal}, if and only if for every pair of morphisms
$\psi_i:(\mathcal{M}_i,g_i,j_i)\to(\mathcal{M},g,j)$, $i=1,2$, in $\mathfrak{SpacCurr}$, whose
ranges $\psi_1(\mathcal{M}_1)$ and $\psi_2(\mathcal{M}_2)$ are causally disjoint in $M$,
$\left[\alpha_1(\mathscr{A}_1),\alpha_2(\mathscr{A}_2)\right]=0$ in
$\mathfrak{A}(\mathcal{M},g,j)$, where we have written $\alpha_i:=\mathfrak{A}(\psi_i)$ and
$\mathscr{A}_i:=\mathfrak{A}(\mathcal{M}_i,g_i,j_i)$ for brevity.

We say that a generally covariant quantum field theory $\mathfrak{A}$ with background current
satisfies the \emph{time-slice axiom}, if and only if for every morphism
$\psi:(\mathcal{M},g,j)\to(\tilde{\mathcal{M}},\tilde{g},\tilde{\jmath})$ in
$\mathfrak{SpacCurr}$, whose range $\psi(\mathcal{M})$ contains a Cauchy surface for $\tilde{M}$,
$\mathfrak{A}(\psi)$ is an isomorphism.
\end{definition}
The causality follows directly from the form of the Poisson structure, in particular from the
support properties of $G$. The time-slice axiom essentially follows from Corollary
\ref{Cor_Asmeared}. (For $j\not=0$ we note that $j(x)I$ is contained in every local algebra,
so it does not spoil the time-slice axiom.) We omit the details of the proofs, because they
proceed along familiar lines (cf.\ \cite{Dim} for the scalar field case.)

We now turn to the issue of locality. It was already remarked in \cite{DL} that locality may
fail, due to the presence of a gauge equivalence. The following result shows that, when treated
correctly, the lack of locality may be interpreted in terms of Gauss' law, also at the quantum
level. For completeness we also prove some no-go results for attempts to restore locality.
\begin{theorem}\label{Thm_QGauss}
The generally covariant theory $\mathfrak{A}$ is not locally covariant. For any morphism
$\map{\psi}{(\mathcal{M},g,j)}{(\tilde{\mathcal{M}},\tilde{g},\tilde{\jmath})}$ the kernel of
$\mathfrak{A}(\psi)$ is contained in the algebra generated by $\mathbf{A}(\alpha)$ with
$\alpha\in\mathcal{C}^p(M)$.
\end{theorem}
\begin{proof*}
Quite generally, the kernel of $\mathfrak{A}(\psi)$ is generated by the operators
$\mathbf{A}(\alpha)$ which $\mathfrak{A}(\psi)$ maps to $0$ (cf.\ Lemma \ref{Lem_Q}). This can
only occur if $\alpha\in\mathcal{C}^p(M)$, by the canonical commutation relations (and the
fact that morphisms preserve the unit $I\not=0$). To show that $\mathfrak{A}$ is not locally
covariant, we need to prove that non-injective morphisms do indeed exist. For $p=1$ this can be
seen from Example \ref{Ex_Gauss} and Remark \ref{Rem_Embed}, where we have a canonical injection
$\iota$ of a spacetime $M$ into Minkowski spacetime $M_0$, both with vanishing currents $j$. $M$
is constructed in such a way that $\mathcal{C}^1(M)$ is non-trivial (cf.\ Proposition
\ref{Prop_Degeneracy}), so the algebra $\mathfrak{A}(M)$ has a non-trivial centre. However,
$\mathcal{C}^1(M_0)$ is trivial, and hence so is the centre of $\mathfrak{A}(M_0)$. Consequently,
the $\mathfrak{A}(\iota)$ cannot be injective.

For general $p$ we proceed as follows. Let $\mathbb{S}^{p-1}$ be the unit sphere in
$\mathbb{R}^p$, with the induced Riemannian metric. We consider the embedding of the spacetime
$M\times\mathbb{S}^{p-1}$ into $M_0\times\mathbb{S}^{p-1}$ which is trivial in the second factor.
Now, if $\omega$ is the volume form on $\mathbb{S}^{p-1}$, then $\alpha\wedge\omega$ is a
compactly supported $p$-form on the Cauchy surface $\Sigma\times\mathbb{S}^{p-1}$, where
$\Sigma$ is the Cauchy surface for $M$ used in Example \ref{Ex_Gauss}. Note that
$\alpha\wedge\omega=d(\beta\wedge\omega)$, but $\beta\wedge\omega$ is not compactly supported.
Indeed, if there were some compactly supported $\gamma$ with $\alpha\wedge\omega=d\gamma$, then
$0=\int_1^{\infty}\int_{\mathbb{S}^{p-1}}\alpha\wedge\omega=\int_1^{\infty}f\times
\int_{\mathbb{S}^{p-1}}\omega$, which contradicts the choices of $f$ and $\omega$.
\end{proof*}

For $p=1$ the interpretation of Section \ref{SSec_Degeneracies} shows that the lack of
injectivity is due to electric monopole observables, which is in line with Gauss' law.

Because the spaces $\mathcal{S}^p_0(M)$ are locally convex, the injectivity of $\psi_*$ on
$\mathcal{S}^p_0(M)^*$ is equivalent to the surjectivity of $\psi^*$ on $\mathcal{S}^p_j(M)$ (by
the Hahn-Banach Theorem). If $p<n-1$ this leads to another proof that injectivity fails, as
follows. Consider any source $j\in\Omega^p_{\delta}(M_0)$ with support in $J(B_1)$, where $B_1$
is the unit ball of the Cauchy surface $\Sigma_0$. Let $A\in\Omega^p(M_0)$ be any on-shell
configuration, solving $\delta dA=j$ and $\delta A=0$ (which exists as per Theorem
\ref{Thm_MaxwellCauchy}). Let $\Sigma\subset\Sigma_0\setminus B_1$ be any connected open set and
define $M:=D(\Sigma)$, with canonical embedding $\map{\iota}{M}{M_0}$. The pull-back of $A$ to
$M$ solves the homogeneous Maxwell equations and generalises the electric monopole configuration
of Example \ref{Ex_Gauss}. Now, if $\alpha\in\mathcal{C}^p(M)$, then $\alpha=\delta\nu$ for some
$\nu\in\Omega^{p+1}_{0,d}(M)$. As $M_0$ is topologically trivial and $p<n-1$ we may write
$\iota_*\nu=d\beta$ for some $\beta\in\Omega^p_0(M_0)$. Hence,
$(\alpha,*\iota^*A)=(\delta d\beta,*A)=(\beta,*j)=:c$ with $c\in\mathbb{R}$ and therefore
\[
\alpha_{\iota}(\mathbf{A}(\alpha)-cI)=0.
\]
If non-trivial $\alpha$ exist, it follows that the kernel is non-trivial. The existence of
non-trivial $\alpha$ is equivalent to the fact that $\iota^*$ is not surjective.

By introducing background currents one can ensure that the electric monopole observables are
mapped to a non-zero constant, rather than to $0$. However, one can still construct linear
combinations of electric monopole observables and the unit which are in the kernel of the
embedding. For this reason the inclusion of background currents does not help to restore local
covariance.

\begin{remark}\label{Rem_A-alg-off}
We can also define an off-shell algebra for $p$-form fields, by dropping the equation of motion
from Definition \ref{Def_Aj}. In other words, for each globally hyperbolic spacetime $M$
(independently of any background current) we quantise the (linear) Poisson space
$\mathcal{F}^p(M)$, where the Poisson structure on $\mathcal{F}^p(M)^*$ is still given by
$\left\{F_{\alpha},F_{\beta}\right\}:=(\alpha,*G\beta)$. For morphisms $\psi$ we use once more
the pull-back to obtain a morphism of Poisson spaces. In this case there is no relation that
forces us to divide out a subspace of $\Omega^p_{0,\delta}(M)$, so that the push-forward map on
observables $F_{\alpha}$ is now injective. This entails that the off-shell theory does abide to
the principle of local covariance and it is also causal. However, it contains no information on
the dynamics, since no equation of motion is imposed, and hence the time-slice axiom fails.

Alternatively, one might say that for the off-shell algebra the external current is left
arbitrary. Indeed, defining the current as a quantum operator $\mathbf{j}:=\delta d\mathbf{A}$ it
is unrestricted. Thus it seems that considering arbitrary background currents should restore
local covariance and one may argue \cite{FH} that a fully interacting theory like QED, where
$\mathbf{j}$ is the usual normal ordered Dirac current, would give rise to a locally covariant
theory satisfying the time-slice axiom.
\end{remark}

To conclude this section we consider an attempt to restore local covariance by dividing out the
centre of the algebras $\mathfrak{A}(\mathcal{M},g,j)$. Although this procedure is well-defined for
each individual algebra, we now prove that it cannot be done in a functorial way, at least not
for $p=1$ and $n\ge 3$:
\begin{theorem}
Consider the case $p=1$ and $n\ge 3$. For each spacetime $M$ with background current $j$, let
$\mathcal{Z}(\mathfrak{A}(M))$ denote the centre of the algebra $\mathfrak{A}(M)$ and let
$\mathfrak{A}'(M):=\mathfrak{A}(M)/\mathcal{Z}(\mathfrak{A}(M))$. Not all morphisms
$\mathfrak{A}(\psi)$, where $\psi$ is a morphism in $\mathfrak{SpacCurr}$, descend to morphisms
of the algebras $\mathfrak{A}'(M)$.
\end{theorem}
\begin{proof*}
It suffices to provide one counterexample, which follows from the second item of Remark
\ref{Rem_Embed}. A spacetime formulation of that remark provides an example of a morphism
$\map{\psi}{(\mathcal{M},g,j)}{(\tilde{\mathcal{M}},\tilde{g},\tilde{j})}$ and a central operator
$\Phi(\alpha)$ in $\mathfrak{A}(M)$ for which $\alpha_{\psi}(\Phi(\alpha))$ is no longer in the
centre. Clearly $\alpha_{\psi}$ cannot descend to the quotient algebras.
\end{proof*}

\begin{remark}
In analogy to $\mathfrak{A}$ one may construct a generally covariant theory $\mathfrak{F}$ for
the field strength tensor $\mathbf{F}:=d\mathbf{A}$ (cf.\  Remark \ref{Rem_F} and \cite{DL}).
This theory is causal and satisfies the time-slice axiom, but similar arguments as for
$\mathfrak{A}$ show that $\mathfrak{F}$ is not local either. In fact, we have an even larger
space of degeneracies than before, namely $H^{p+1}_0(M)$ (cf.\ Corollary \ref{Cor_centreA} and
the remarks below it). Unlike for $\mathfrak{A}$, however, $\mathfrak{F}$ does allow the centre
of the algebra to be divided out in a functorial way, leading to a locally covariant theory.
Indeed, the observables are generated by $I$ and $F(\beta)$ with $\beta\in\frac{\Omega^{p+1}_0(M)}
{\Omega^{p+1}_{0,d}(M)}$, where we note that the latter space is well behaved under embeddings.
Unfortunately, the resulting theory is based on a classical phase space consisting of field
configurations $F$ which are co-exact, $F=dA=\delta B$. Consequently, the theory cannot measure
any electric charges or electromagnetic currents whatsoever.
\end{remark}

\section{Conclusions}\label{Sec_Conclusions}

In this paper, we have described the dynamics of the vector potential and its $p$-form
generalisations, both at a classical and at a quantum level, on a generic globally hyperbolic
spacetime and in the presence of (classical) background currents. At the classical level we have
recollected mostly well-known results, which are based on imposing the Lorenz gauge, and extended
them to a distributional setting. Like \cite{DL} we found a quantum theory which fails to be
generally locally covariant. However, unlike \cite{DL} we were able to ascribe the failure to the
lack of locality at the quantum level. At least for the vector potential ($p=1$), the source of
such a lack is the same as at the classical level: Gauss' law.

In order to accomplish this new understanding of the lack of locality, we made essential use of
three crucial ingredients. Firstly, we made a judicious choice of the gauge equivalence for the
vector potential. This choice was motivated by the geometric perspective of the standard model,
combined with insights from general covariance. It is experimentally justified by the
Aharonov-Bohm effect. Secondly, we used a quantisation scheme that improves on the ones that were
used in most of the previous literature by avoiding the use of (pre)-symplectic spaces. Instead
of using a space of classical solutions with spacelike compact support and using $G$ to define a
(pre-)symplectic form, we followed the general geometric method of Peierls', which automatically
leads to a Poisson structure on the space of local observables. Peierls' method applies to any
Lagrangian theory, even if the equation of motion is non-linear. Because we were studying a
linear equation (with source term), the extension of the classical space of solutions to include
distributional solutions allowed us to view the space of observables as the continuous dual
space, consisting of smooth test-forms. In this way the Poisson bracket that we found was
automatically well-defined on all observables. Thirdly, we continued to use a geometric
perspective by quantising the affine Poisson space using ideas from deformation quantisation, in
particular Fedosov's method. It is not clear if (or how) these methods can be extended to general
Lagrangian theories, but in our case they allowed us to interpret the quantum field operators in
terms of observables, in direct analogy to the classical case.

In addition to these general choices of procedure, our arguments relied on a somewhat involved
computation of the degeneracies of the Poisson bracket. Both at the classical and at the quantum
level we were able to relate these degeneracies to Gauss' law (at least for $p=1$) and they have
lead to a formula for the space of electric monopoles that can influence a spacetime. For the
field strength tensor associated to the vector potential we have mentioned that the degeneracies
are ruled by $H^2(M)$. These explanations vindicate the combination of the three procedural
ingredients mentioned above. In addition we would like to point out that the degeneracies are not
directly due to the Aharonov-Bohm effect, but our computation does establish that the
Aharonov-Bohm effect and Gauss' law are related, via the Poisson bracket.

The quantum theory $\mathfrak{A}$ that we constructed is not locally covariant. From a
mathematical point of view the reason can be ascribed to the degeneracy of the underlying Poisson
structure, which entails that the on-shell algebra possesses a non trivial centre, depending on
the topology of the underlying spacetime. Even though we motivated the gauge equivalence by
viewing electromagnetism as a $U(1)$ Yang-Mills theory in a general covariant setting, we did not
analyse in detail the effect of non-trivial principal $U(1)$-bundles on our results. Such a
formulation may shed further light on the geometric (or topological) causes of the failure of
locality \cite{BDS2}. We also omitted a detailed study of the special case $n=2$, where a
different choice of gauge equivalence would be appropriate. Another potentially interesting
research topic is the inclusion of Dirac fields and their interaction with the vector potential
at a perturbative level. In this framework, one may expect that the on-shell algebra would behave
well under general embeddings, leading to a locally covariant theory (cf.\ Remark
\ref{Rem_A-alg-off}).

Although the quantum theory $\mathfrak{A}$ is not locally covariant, we have argued that it is
generally covariant and that the failure of locality is a well-understood consequence of the
gauge invariance. Besides, we have shown that it is impossible to recover a locally covariant
theory by dividing out the degeneracies of the Poisson bracket, at least for the vector
potential. For the field strength such a procedure is possible, and the off-shell theory is
locally covariant from the start, but these theories are rather unphysical or have a limited
physical applicability. Finally, the idealisation of classical background currents is so useful
that it should be considered as a perfectly satisfactory model. In our opinion these are
sufficient grounds to consider modifying the framework of locally covariant quantum field theory,
so as to accommodate physical theories with gauge symmetries and non-local observables. Our
notion of generally covariant quantum field theory is clearly general enough for this purpose,
but it is not clear whether it is too general.

In the spirit of \cite{FH} one could argue that the principle of local covariance wants to stress
that algebras of arbitrarily small neighbourhoods of a point should depend only on the germ of
the metric at this point. This is in line with the fact that for the $p$-form theories under
consideration, the morphisms $\mathfrak{A}(\psi)$ are injective as soon as the source spacetime
of $\psi$ has a trivial topology. Such spacetimes could be used as building blocks for the entire
theory, in such a way that topological effects can be inferred out of a failure of Haag duality.
(How this works in detail, and how these ideas relate to the universal field strength algebra of
\cite{DL}, will hopefully be addressed by some of us in a future investigation.) On the other
hand, one could argue that the topology of the underlying spacetimes is only relevant because the
gauge symmetry of the $p$-forms fields is related to the topology. Other linear gauge theories,
in particular linearised general relativity, may have central elements in their algebras which
are sensitive to the background metric $g$ as well as (or instead of) the background topology of
$\mathcal{M}$. This could force us to search for other remedies, which are not of a purely
topological nature. (This issue too will hopefully be addressed in a future investigation
involving one of us.)

\section*{Acknowledgements}
C.D.\ and T.-P.H.\ are indebted to Chris Fewster and Klaus Fredenhagen for illuminating
discussions on the spirit of locally covariant quantum field theory. C.D.\ wishes to thank Chris
Fewster and Marco Benini for the useful discussions. K.S.\ is indebted to Bob Wald for useful
discussions and for drawing his attention to Fedosov's quantisation method and to Stefan Hollands
for making available his notes on this topic. We would also like to thank Chris Fewster and
Alexander Schenkel for contacting us about an error in an earlier version of Definition
\ref{Def_PoissAlg}.

The work of C.D.\ is supported partly by the University of Pavia and by the Indam-GNFM project
{``Effetti topologici e struttura della teoria di campo interagente''}. T.-P.H. gratefully
acknowledges financial support from the Hamburg research cluster LEXI ``Connecting Particles with
the Cosmos''. We are indebted to the the II.\ Institut f\"ur Theoretische Physik of the University
of Hamburg for the kind hospitality during the realization of part of this work.


\begin{thebibliography}{15}
\bibitem{AshIsh92}
A.\ Ashtekar and C.\ J.\ Isham, \emph{Inequivalent observable algebras. Another ambiguity in field
quantisation}, Phys.\ Lett.\ B {\bf 274} (1992) 393--398

\bibitem{Ashtekar}
A.\ Ashtekar and A.\ Sen, \emph{On the role of space-time topology in quantum phenomena:
superselection of charge and emergence of nontrivial vacua},
J.\ Math.\ Phys.\ {\bf 21} (1980) 526--533

\bibitem{BGP}
C.\ B\"ar, N.\ Ginoux and F.\ Pf\"affle, \emph{Wave equations on Lorentzian manifolds and
quantization},
EMS, Z\"urich (2007)

\bibitem{BDS}
M.\ Benini, C.\ Dappiaggi, A.\ Schenkel, \emph{Quantum field theory on affine bundles},
Ann.\ Henri Poincar\'e doi:10.1007/s00023-013-0234-z (2012)

\bibitem{BDS2}
M.\ Benini, C.\ Dappiaggi, A.\ Schenkel, \emph{Quantized Abelian principal connections on Lorentzian manifolds},
arXiv:1303.2515 [math-ph] (2013)

\bibitem{BS}
A.\ N.\ Bernal and M.\ S\'anchez, \emph{Smoothness of time functions and the metric splitting of
globally hyperbolic spacetimes},
Commun.\ Math.\ Phys.\ {\bf 257} (2005) 43--50

\bibitem{BS2}
A.\ N.\ Bernal and M.\ S\'anchez, \emph{Further results on the smoothability of Cauchy
hypersurfaces and Cauchy time functions},
Lett.\ Math.\ Phys.\ {\bf 77} (2006) 183--197

\bibitem{BHR}
E.\ Binz, R.\ Honegger and A.\ Rieckers, \emph{Construction and uniqueness of the $C^*$-Weyl
algebra over a general pre-symplectic space},
J.\ Math.\ Phys.\ {\bf 45} (2004), no.\ 7, 2885--2907

\bibitem{B}
P.\ J.\ M.\ Bongaarts, \emph{Maxwell's equations in axiomatic quantum field theory. I. Field
tensor and potentials},
J.\ Math.\ Phys.\ {\bf 18} (1977), no.\ 7, 1510--1516

\bibitem{BT}
R.\ Bott and L.\ W.\ Tu, \emph{Differential forms in algebraic topology},
Graduate texts in mathematics {\bf 82}, Springer, New York (1982)

\bibitem{BFV}
R.\ Brunetti, K.\ Fredenhagen and R.\ Verch, \emph{The generally covariant locality principle:
a new paradigm for local quantum field theory},
Commun.\ Math.\ Phys.\ {\bf 237} (2003) 31--68

\bibitem{DL}
C.\ Dappiaggi and B.\ Lang, \emph{Quantization of Maxwell's equations on curved backgrounds and
general local covariance},
Lett.\ Math.\ Phys.\ {\bf 101} (2012) 265--287

\bibitem{DS}
C.\ Dappiaggi and D.\ Siemssen, \emph{Hadamard states for the vector potential on asymptotically
flat spacetimes},
Rev.\ Math.\ Phys.\ {\bf 25} 13500025 (2013)

\bibitem{dR}
G.\ de Rham, \emph{Differentiable manifolds},
translated by F.\ R.\ Smith, Springer, Berlin (1984)

\bibitem{Dim}
J.\ Dimock, \emph{Algebras of local observables on a manifold},
Commun.\ Math.\ Phys.\ {\bf 77} (1980) 219--228

\bibitem{D}
J.\ Dimock, \emph{Quantized electromagnetic field on a manifold},
Rev.\ Math.\ Phys.\ {\bf 4} (1992) 223--233

\bibitem{DW}
J.\ P.\ Dowling and C.\ P.\ Williams, \emph{Maxwell duality, Lorentz invariance, and topological
phase},
Phys.\ Rev.\ Lett.\ {\bf 83} (1999) 2486--2489

\bibitem{F}
C.\ Fewster, \emph{On the notion of ''the same physics in all spacetimes''},
to appear in: Proceedings ''Quantum field theory and gravity'', Regensburg (2010)

\bibitem{FeH}
C.\ J.\ Fewster and D.\ S.\ Hunt, \emph{Quantization of linearized gravity in cosmological vacuum
spacetimes}, Rev.\ Math.\ Phys.\ {\bf 25} 1330003 (2013)

\bibitem{FP}
C.\ J.\ Fewster and M.\ J.\ Pfenning, \emph{A quantum weak energy inequality for spin-one fields
in curved spacetime},
J.\ Math.\ Phys.\ {\bf 44} (2003) 4480--4513

\bibitem{FV}
C.\ J.\ Fewster and R.\ Verch, \emph{Dynamical locality of the free scalar field},
Ann.\ Henri Poincar\'e {\bf 13} (2012) 1675--1709

\bibitem{FS}
F.\ Finster and A.\ Strohmaier, \emph{Gupta-Bleuler Quantization of the Maxwell Field in Globally Hyperbolic Space-Times},
arXiv:1307.1632 [math-ph] (2013)

\bibitem{FH}
K.\ Fredenhagen, \emph{Structural aspects of gauge theories in the algebraic framework of quantum
field theory},
FREIBURG-THEP 82/9, Talk presented at the Colloqium in honour of Prof.\ Rudolf Haag at the
occasion of his 60th birthday, Hamburg (1982)

\bibitem{FR12}
K.\ Fredenhagen and K.\ Rejzner, \emph{Batalin-Vilkovisky formalism in perturbative algebraic quantum field theory},
Commun.\ Math.\ Phys.\ {\bf 317} 697--725 (2013)

\bibitem{Fri}
F.\ G.\ Friedlander, \emph{The wave equation on a curved space-time},
Cambridge University Press (1975)

\bibitem{FNW}
S.\ A.\ Fulling, F.\ J.\ Narcowich and R.\ M.\ Wald, \emph{Singularity structure of the two-point
function in quantum field theory in curved spacetime},
Ann.\ Phys.\ (N.Y.) {\bf 136}, 243--272 (1981)

\bibitem{Ge68}
R.\ Geroch, \emph{Spinor structure of space-times in general relativity. I},
J.\ Math.\ Phys.\ {\bf 9}, 1739--1744 (1968)

\bibitem{Ge70}
R.\ Geroch, \emph{Spinor structure of space-times in general relativity. II},
J.\ Math.\ Phys.\ {\bf 11}, 343--348 (1970)

\bibitem{G}
N.\ Ginoux, \emph{Linear wave equations}, in: C.\ B\"ar and K.\ Fredenhagen (eds.)
\emph{Quantum field theory on curved spacetimes},
Springer, Berlin Heidelberg (2009)

\bibitem{Ha}
R.\ Haag, \emph{Local quantum physics},
Springer, Berlin Heidelberg (1996)

\bibitem{HS}
T.-P.\ Hack and A.\ Schenkel, \emph{Linear bosonic and fermionic quantum gauge theories on curved spacetimes},
Gen.\ Rel.\ Grav.\ {\bf 45} (5) 877--910 (2013)

\bibitem{Hi}
M.\ W.\ Hirsch, \emph{Differential topology}, Sprigner, New York (1994)

\bibitem{Hol}
S.\ Hollands, \emph{Renormalized Quantum Yang-Mills Fields in Curved Spacetime},
Rev.\ Math.\ Phys.\ {\bf 20} (2008) 1033--1172.

\bibitem{H}
L.\ H\"ormander, \emph{The analysis of linear partial differential operators I},
Springer, Berlin Heidelberg (1990)

\bibitem{KN}
J.\ L.\ Kelley and I.\ Namioka, \emph{Linear topological spaces},
Van Nostrand, Princeton, N.J. (1963)

\bibitem{K12}
I.\ Khavkine, \emph{Characteristics, conal geometry and causality in locally covariant field
theory},
arXiv:1211.1914 [gr-qc] (2012)

\bibitem{KoNo}
S.\ Kobayashi and K.\ Nomizu, \emph{Foundations of differential geometry},
Interscience, New York (1963-1969)

\bibitem{Lang}
B.\ Lang, \emph{Homologie und die Feldalgebra des quantisierten Maxwellfeldes},
Diplomarbeit - (2010) Universit\"at Freiburg, available at
www.desy.de/uni-th/theses/Dipl\_Lang.pdf

\bibitem{N}
M.\ Nakahara, \emph{Geometry, topology and physics},
Hilger, Bristol New York (1990)

\bibitem{PT}
M.\ Peshkin and A.\ Tonomura, \emph{The Aharonov-Bohm effect},
Lecture notes in physics {\bf 340}, Springer, Berlin New York (1989)

\bibitem{Pei}
R.\ E.\ Peierls, \emph{The commutation laws of relativistic field theory},
Proc.\ Roy.\ Soc.\ London.\ Ser.\ A.\ {\bf 214}, (1952) 143--157

\bibitem{P}
M.\ J.\ Pfenning, \emph{Quantization of the Maxwell field in curved spacetimes of arbitrary
dimension},
Class.\ Quantum Grav.\ {\bf 26} (2009) 135017

\bibitem{SV}
H.\ Sahlmann and R.\ Verch, \emph{Microlocal spectrum condition and Hadamard form for
vector-valued quantum fields in curved spacetime},
Rev.\ Math.\ Phys.\ {\bf 13} (2001), no.\ 10, 1203--1246

\bibitem{Sanc}
M.\ S\'anchez, \emph{On the geometry of static spacetimes}, Nonlinear Analysis {\bf 63}
e455--e463 (2005)

\bibitem{S}
K.\ Sanders, \emph{A note on spacelike and timelike compactness},
Class.\ Quantum Grav.\ {\bf 30} (2013) 115014

\bibitem{W}
R.\ M.\ Wald, \emph{General relativity},
University of Chicago Press, Chicago (1984)

\bibitem{Wa}
S.\ Waldmann, \emph{Poisson-Geometrie und Deformationsquantisierung},
Springer, Berlin Heidelberg New York (2007)
\end{thebibliography}
\end{document}